\documentclass[pra,aps,twocolumn,10pt,superscriptaddress,notitlepage,longbibliography,nofootinbib]{revtex4-2}

\usepackage[dvipsnames]{xcolor}
\usepackage[english]{babel} 
\usepackage{graphicx}
\usepackage{float}
\usepackage{braket}
\usepackage{epstopdf}
\usepackage{mathtools}
\usepackage{amsmath}
\usepackage{amssymb}
\usepackage[caption=false]{subfig}
\usepackage{pythonhighlight}
\usepackage{hyperref}
\usepackage{float}
\usepackage{bbold}
\usepackage[T1]{fontenc}
\usepackage{MnSymbol}
\usepackage{graphicx}
\usepackage{amsmath,amssymb,amsfonts}
\usepackage{mathrsfs}

\usepackage{booktabs}

\usepackage{tikz}
\usetikzlibrary{
    arrows.meta,
    positioning,
    calc,
    fit,
    backgrounds,
    decorations.pathmorphing,
    shapes.misc
}

\newcommand{\Tr}[1]{\mathrm{Tr}[#1]} 

\graphicspath{{./Figures/}{./Figures_thesis/}}

\usepackage[markup=underlined]{changes}
\makeatletter
\@namedef{Changes@AuthorColor}{magenta}
\colorlet{Changes@Color}{magenta}
\makeatother
\usepackage{hyperref}
 \hypersetup{
     colorlinks=true,
     linkcolor=blue,
     filecolor=blue,
     citecolor = magenta,      
     urlcolor=red,
     }

\usepackage{braket}

\usepackage{xargs}
\newcommandx{\greencom}[2][1=]
{\todo[inline, color=green!40,#1]{#2}}
\newcommandx{\bluecom}[2][1=]
{\todo[inline, color=blue!40,#1]{#2}}
\newcommandx{\bluemargin}[2][1=]
{\todo[color=blue!40,#1]{#2}}

\usepackage{letltxmacro}
\LetLtxMacro{\ORIGselectlanguage}{\selectlanguage}
\makeatletter
\DeclareRobustCommand{\selectlanguage}[1]{%
  \@ifundefined{alias@\string#1}
    {\ORIGselectlanguage{#1}}
    {\begingroup\edef\x{\endgroup
       \noexpand\ORIGselectlanguage{\@nameuse{alias@#1}}}\x}%
}
\newcommand{\definelanguagealias}[2]{%
  \@namedef{alias@#1}{#2}%
}

\makeatother

\definelanguagealias{en}{english}
\definelanguagealias{EN}{english}
\definelanguagealias{eng}{english}
\definelanguagealias{de}{ngerman}

\newcommand{\ii}{\mathrm{i}}
\newcommand{\dd}{\mathrm{d}}

\newcommand{\lsb}{\left[}
\newcommand{\rsb}{\right]}
\newcommand{\lcb}{\left\{}
\newcommand{\rcb}{\right\}}

\newcommand{\ee}{\mathrm{e}}

\newcommand{\tket}[1]{\lvert #1)}
\newcommand{\tbra}[1]{(#1\rvert}

\newcommand{\Lket}[1]{\lvert #1\rrangle}
\newcommand{\Lbra}[1]{\llangle #1\rvert}
\newcommand{\Lbraket}[2]{\llangle #1\vert #2 \rrangle}
\newcommand{\Fket}[1]{\lvert #1)\!\rangle}
\newcommand{\Fbra}[1]{\langle\!( #1\rvert}
\newcommand{\Fbraket}[2]{\langle\!( #1\vert #2 )\!\rangle}
\newcommand{\FLket}[1]{\lvert #1)\!\rrangle}
\newcommand{\FLbra}[1]{\llangle\!( #1\rvert}
\newcommand{\FLbraket}[2]{\llangle\!( #1\vert #2 )\!\rrangle}

\usepackage{pifont}

\begin{document}

\title{Floquet-Liouville Theory for Strongly Driven Open Quantum Systems}



\author{Kamran~Akbari}
\email[]{kamran.akbari@queensu.ca}
\affiliation{Department of Physics, Engineering Physics and Astronomy, Queen's University, Kingston ON K7L 3N6, Canada}
\author{Stephen~Hughes}
\email[]{shughes@queensu.ca}
\affiliation{Department of Physics, Engineering Physics and Astronomy, Queen's University, Kingston ON K7L 3N6, Canada}
\date{\today}

\begin{abstract}

Periodically driven quantum systems are commonly modeled using master equations constructed in the eigenbasis of an undriven Hamiltonian, implicitly assuming that environmental dissipation couples to static energy transitions even under strong time-periodic driving. The validity of this approximation beyond weak or near-resonant driving remains poorly understood.
To address the need for a more self-consistent quantum theory approach, we formulate a nonsecular Floquet--Markov generalized master equation (F-GME) in the quasienergy basis, treating interaction-induced (internal) and drive-induced (external) nonperturbative dressing on an equal footing. We subsequently investigate dissipation in two minimal driven open quantum systems---a harmonically driven two-level system and a harmonically driven coupled-two-level-system---each weakly coupled to a Markovian bath. Comparing the F-GME to a time-independent dressed-basis master equation, we show that even for
a flat-bath spectral density and weak dissipation, the two approaches can yield qualitatively different steady-state populations and emission spectra.
We resolve dissipation into drive-assisted sideband processes decaying via Floquet extended-space quasienergy channels, and show these channels can hybridize through nonsecular couplings into collective Floquet--Liouville modes governing observable spectral resonances. This analysis demonstrates that time-independent dissipative descriptions can incorrectly weight multiphoton Floquet transitions by collapsing quasienergy-resolved decay pathways into static energy gaps. The F-GME framework provides a systematic diagnostic for identifying regimes where Floquet-consistent dissipation is essential and clarifies the physical origin of discrepancies between commonly used master-equation approaches.

\end{abstract}

\maketitle
\section{Introduction}
\label{sec:Intro}
Periodically driven quantum systems play a central role across quantum optics, condensed-matter physics, and quantum technologies, enabling control protocols such as coherent population transfer~\cite{Bergmann_Coherent_1998,Mallavarapu_Population_2021}, dynamical decoupling~\cite{Viola1999DynamicalDecoupling,Viola1999UniversalControl,Szczygielski_Markovian_2015}, Floquet engineering of band structures~\cite{Castro_Floquet_2022}, phase transition~\cite{mercurio2026floquetdissipativephasetransitions,Wu_Floquet_2026} and synthetic gauge fields~\cite{Creffield_Generation_2014,Greschner_Density-Dependent_2014,Roushan2017ChiralCurrents,Bazavan_Synthetic_2024}. 
Moreover, periodic driving can serve as an engineering resource, enabling control of quasienergies, activation of off-resonant transitions, synthesis of effective interactions, and encoding in time-dependent dressed states, particularly pivotal in quantum information processing~\cite{Caldwell2018ParametricallyActivated,Viola1999DynamicalDecoupling,Viola1999UniversalControl,Huang2021DynamicalSweetSpots,Roushan2017ChiralCurrents,Mi2022TimeCrystallineOrder,Hastings2021DynamicallyGenerated,Davydova2023FloquetCodes,Jiang2011MajoranaFermions,Bomantara2020MeasurementOnly,Bai2023FloquetMetrology,Chen2025FloquetLiouvillians}.
In realistic systems, however, driven systems are inevitably coupled to environmental degrees of freedom (baths), making dissipation and decoherence essential ingredients for their dynamics~\cite{Grifoni_Driven_1998,Hone_Statistical_2009,Kohn_Periodic_2001,Kohler_Driven_2005,Hausinger_Dissipative_2010,Mori_Floquet_2023}.

A widely used strategy for modeling dissipation in periodically driven quantum systems is to construct the dissipative dynamics from the eigenstates of a 
{\it static system Hamiltonian}---either in the bare subsystem basis or in the eigenbasis of a time-independent dressed Hamiltonian---while treating the drive as an explicitly time-dependent perturbation~\cite{Breuer_Theory_2002,Settineri_Dissipation_2018,Salmon_Gauge-independent_2022}. This approach underlies many descriptions of driven quantum systems, particularly driven two-level systems (TLSs), including resonance fluorescence, dressed-state dynamics, and multiphoton processes. Although computationally convenient, such treatments implicitly assume that the environment couples primarily to transitions associated with static eigenfrequencies of the     undriven or time-independent dressed system~\cite{Settineri_Dissipation_2018,Kowalewska-kudlaszyk_Generalized_2001}.

For hybridized quantum systems---where subsystems interact strongly, such as in dipole--dipole-coupled quantum emitters or cavity quantum electrodynamics (cavity-QED) in the ultrastrong-coupling regime~\cite{FriskKockum_Ultrastrong_2019,Forn-Diaz_Ultrastrong_2019}---the situation becomes substantially more intricate~\cite{Carmichael_Master_1973,Beaudoin_Dissipation_2011,Settineri_Dissipation_2018,Salmon_Gauge-independent_2022,Akbari_Generalized_2023}. In such systems, two distinct dressing mechanisms may coexist simultaneously: (i) interaction-induced dressing arising from coherent subsystem coupling, which we refer to here as the `internal' dressing; and (ii) drive-induced dressing arising from the time-periodic field, which we refer to here as the `external' dressing. 

A consistent description of 
{\it driven} hybrid systems therefore requires both forms of dressing to be treated on
an equal footing. For closed hybrid systems, this interplay has recently been investigated in the context of periodically driven cavity-QED systems in the ultrastrong-coupling regime, where it was shown to produce substantial Floquet-state restructuring and nontrivial sideband dynamics~\cite{Akbari_Floquet_2025}. In open hybrid systems, however, most existing approaches in the ultrastrong-coupling regime primarily incorporate only the interaction-induced dressing through dressed-state dissipative formalisms~\cite{Beaudoin_Dissipation_2011,Settineri_Gauge_2021,Salmon_Gauge-independent_2022}. In contrast, outside the ultrastrong-coupling regime, the role of drive-induced dressing alone has a long history within Floquet treatments of periodically driven closed and open quantum systems~\cite{Shirley_Solution_1965,Sambe_Steady_1973,Breuer_Floquet_2000,Grifoni_Driven_1998,Kohler_Driven_2005,Shirai_Condition_2015,Restrepo_Quantum_2018,Restrepo_Driven_2019}. For noninteracting standalone systems, interaction-induced dressing is absent, thus only the drive-induced Floquet dressing remains relevant.

In an early foundational work on 
modeling dissipation in strongly interacting quantum systems~\cite{Carmichael_Master_1973}, Carmichael and Walls showed that, when subsystems interact strongly, the dissipative dynamics cannot generally be constructed from uncoupled subsystem transitions in the bare-state representation. They demonstrated that such treatments may produce unphysical relaxation dynamics and incorrect steady states. Instead, the master equation must be formulated in the eigenbasis of the interacting Hamiltonian, i.e., after diagonalizing the coherent coupled system. In modern terminology, this corresponds to a dressed-state master-equation approach. This principle later became a cornerstone of the theoretical description of open quantum systems, including 
the light--matter regime of  ultrastrong-coupling
regime~\cite{DeLiberato_Light-Matter_2009,Beaudoin_Dissipation_2011,Ridolfo_Saving_2012,Settineri_Dissipation_2018,Salmon_Gauge-independent_2022,Akbari_Generalized_2023}.

Motivated by this 
{\it dressed-state dissipation philosophy}, it was later recognized that, under sufficiently strong coherent driving, the external drive itself must also be incorporated into the dressing procedure before coupling the system to the environment~\cite{Kowalewska-kudlaszyk_Generalized_2001}. In the terminology adopted in our work here, this corresponds to the {\it external} dressing of the system. Within such strongly driven dressed-state descriptions, the environment effectively couples to drive-modified transition channels and sidebands rather than solely to the static undriven energy spectrum. Although formulated semiclassically, such driven dressed-state approaches already captured several key features  understood more systematically within Floquet theory, including sideband-resolved transitions and drive-modified dissipative channels.

A more systematic formulation of periodically driven open-system dynamics is provided by Floquet theory, where the dynamics is described in terms of {\it quasienergies and Floquet sidebands}~\cite{Shirley_Solution_1965,Sambe_Steady_1973,Grifoni_Driven_1998,Breuer_Floquet_2000,Kohler_Driven_2005,Shirai_Condition_2015,Restrepo_Driven_2019}. Within this powerful framework, transitions between Floquet states may occur not only at quasienergy differences, but also at frequencies shifted by integer multiples of the drive frequency. Consequently, dissipation generally proceeds through a hierarchy of drive-assisted multiphoton channels whose frequencies are determined by the quasienergy spectrum rather than by static energy eigenvalues. Within Floquet-based open-system descriptions, dissipation therefore becomes quasienergy resolved, allowing the bath to couple differently to distinct Floquet sidebands and multiphoton channels. This framework naturally incorporates sideband-dependent damping, structured reservoirs, and nonsecular dissipative couplings, and has been widely employed to describe strong-field fluorescence spectra, driven open-system dynamics, and the breakdown of naive time-independent dissipative descriptions~\cite{Grifoni_Driven_1998,Kohler_Driven_2005,Shirai_Condition_2015,Restrepo_Driven_2019}.

Related ideas were later extended to a broader class of driven open quantum systems, including pulsed excitation protocols and engineered environments designed to modify effective dissipative pathways. Representative examples include polaron master-equation approaches for strongly driven semiconductor quantum emitters interacting with 
structured phonon reservoirs~\cite{McCutcheon_Quantum_2010,Roy-Choudhury_Influence_2015,Gustin_Efficient_2020}, cavity-assisted single-photon generation in phonon environments~\cite{Manson_Polaron_2016,PhysRevB.96.085305}, and studies of coupled photon--phonon dissipative processes in semiconductor and cavity-QED platforms~\cite{Roy-Choudhury_Resonance_2016,Hargart_Cavity-enhanced_2016}. These works further demonstrated that strong driving can substantially reshape how environmental reservoirs sample the system spectrum and mediate dissipative dynamics.

Several works have developed Floquet-based master equations for periodically driven open quantum systems, typically within a Floquet--Born--Markov framework~\cite{Breuer_Adiabatic_1989,Grifoni_Driven_1998,Talkner_Failure_1986,Kohler_Driven_2005,deVega_Dynamics_2017,Hausinger_Dissipative_2010,Hausinger_Dissipative_2010PhDThesis,Restrepo_Driven_2019,Restrepo_Quantum_2018}. In many implementations, additional secular approximations in the quasienergy basis are introduced in order to obtain generators of Lindblad form~\cite{Talkner_Failure_1986,Grifoni_Driven_1998,Kohler_Driven_2005,deVega_Dynamics_2017}. Although such approaches successfully capture many qualitative aspects of driven dissipative dynamics, it is now understood that nonsecular terms---particularly those associated with near-degenerate quasienergy gaps---can become important and may qualitatively alter long-time observables and spectral properties in a system-dependent manner~\cite{Shirai_Condition_2015,Mori_Rigorous_2016,Restrepo_Quantum_2018}. Consequently, it remains unclear under what conditions time-independent dissipative descriptions provide reliable approximations to the full Floquet-consistent dynamics, especially when multiple quasienergy channels compete or when the Floquet spectrum reorganizes strongly with drive amplitude or detuning~\cite{Sieberer_Dynamical_2013}.

To our knowledge, previous Floquet master-equation approaches and generalized dressed-state master equations have typically emphasized either drive-induced (external) dressing or interaction-induced (internal) dressing separately, while a unified nonsecular treatment of both effects within a generalized Floquet dissipative framework remains comparatively unexplored. Conventional Floquet-Redfield and Floquet--Born--Markov approaches typically incorporate drive-induced quasienergy dressing for weakly interacting or effectively noninteracting systems, whereas generalized dressed-state master equations developed for strongly interacting hybrid systems primarily account for interaction-induced dressing alone. The present work combines {\it both} within a unified nonsecular Floquet generalized master-equation (F-GME) framework, enabling a quasienergy-resolved dissipative description of strongly driven hybrid open quantum systems beyond rotating-wave and secular approximations. As it nonperturbatively applies the Floquet theory for an open strongly-interacting quantum system, it also complements the approach based on the first-order Floquet perturbation theory of Raman scattering in ultrastrong cavity-QED~\cite{Macri_Spontaneous_2022}.

The purpose of our paper is
not to simply rederive the general principle that strong driving modifies dissipative bath sampling, which is already well known in the literature. Rather, our focus is on regimes where the drive {\it must} be treated nonperturbatively, where {\it rotating-wave and secular approximations with respect to the pump become insufficient}, and where {\it multiple quasienergy channels interact through nonsecular dissipative couplings}. In such situations, it becomes nontrivial to determine when a time-independent dissipator remains reliable, when a conventional Floquet master equation lacking internal dressing becomes insufficient, and when a full Floquet-consistent generalized dissipative treatment becomes necessary.

\begin{figure*}[!htpb]
    \centering
    \resizebox{\textwidth}{!}{
        \input{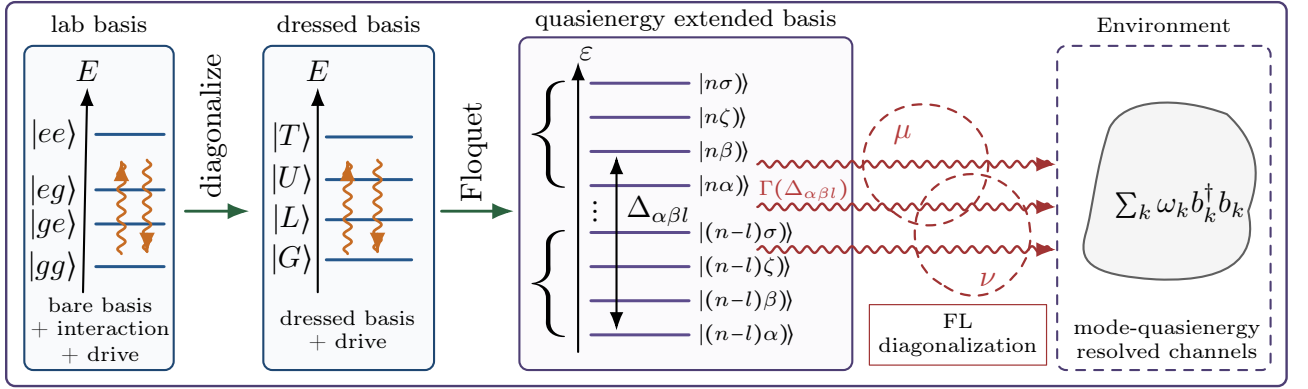}
    }
    \caption{
    Schematic illustration of the driven-dissipative single-TLS (a) and 
    coupled-TLS models (b), for example, two
    dipole-dipole coupled TLSs.
    In panel (a), the periodically driven two-level system is mapped to the Floquet extended basis, where dissipative processes are resolved through rates $\Gamma(\Delta_{\alpha\beta l})$, and then spectrally decomposed into FL modes.
    In panel (b), the coupled-TLS model (two interacting TLSs) is first transformed from the bare basis
    $\{|gg\rangle,|ge\rangle,|eg\rangle,|ee\rangle\}$, where e.g., $\ket{ge}=\ket{g}_1\otimes\ket{e}_2$,  to the dressed basis
    $\{|G\rangle,|L\rangle,|U\rangle,|T\rangle\}$ and then lifted to the Floquet extended basis, with $\alpha,\beta,\zeta,\sigma\in\{G,L,U,T\}$, where dissipation is resolved in terms of quasienergy transitions, and then spectrally decomposed into FL modes. 
    }
    \label{fig:schematics}
\end{figure*}

Specifically, we will investigate these questions in two minimal yet broadly relevant 
and pedagogically-important driven open quantum systems. The first example is  a harmonically driven-dissipative TLS [Fig.~\ref{fig:schematics}(a)], which constitutes a paradigmatic model for resonance fluorescence, coherent control, cavity- and waveguide-QED, quantum information processing, and superconducting and semiconductor quantum devices~\cite{Mollow_Power_1969,Leggett_Dynamics_1987,Cohen-Tannoudji_Atom-Photon_1998,You_Superconducting_2005,Astafiev_Resonance_2010,Gu_Quantum_2017}. The second 
example is a harmonically driven-dissipative coupled-TLS system [Fig.~\ref{fig:schematics}(b)] with 
dipole--dipole (position--position) interaction, relevant to interacting quantum emitters, Rydberg-atom platforms, superconducting qubit architectures, collective radiative systems, and hybrid light--matter devices~\cite{Ficek_Entangled_2002,Saffman_Quantum_2010,Fink_Climbing_2009,Plankensteiner_Selective_2015,Macovei_Dense_2024}. Each system is weakly coupled to Markovian baths.

We primarily compare two dressed-basis dissipative descriptions:
(i) the time-independent generalized master equation (TI-GME), in
which the system--bath coupling is resolved with respect to the
interaction-dressed eigenstates of the static Hamiltonian, and
(ii) the nonsecular Floquet generalized master equation (F-GME)
developed here, in which these dressed transition sectors are
subsequently promoted to the Floquet representation. When relevant,
we also compare with the corresponding bare-transition constructions:
the conventional time-independent master equation (TI-ME) and
Floquet master equation (F-ME). All four approaches employ the same
coherent time-dependent Hamiltonian in our comparisons; their
distinction therefore lies not in a unitary change of representation
but in how the system--bath coupling operator is spectrally resolved.
The TI-ME and F-ME use bare/local transition sectors, with only the
latter incorporating external Floquet dressing, whereas the TI-GME
and F-GME resolve the bath coupling after interaction-induced
dressing, with only the latter additionally incorporating the
drive-induced Floquet structure. Consequently, the approaches may
predict similar resonance positions while differing substantially in
dissipative weights and steady-state observables.

Using the Floquet extended-space formalism, in which the periodic drive is incorporated through an infinite-dimensional quasienergy basis, we explicitly identify the dominant dissipative channels and show that they are governed by transition frequencies,
\begin{equation*}
\Delta_{\alpha\beta l}
=
\varepsilon_{\beta}
-
\varepsilon_{\alpha}
+
l\omega_d,
\end{equation*}
where $\varepsilon_{\alpha}$ are Floquet quasienergies and $l\in\mathbb{Z}$ labels drive-assisted sideband processes. Within the present framework, not only the coherent dynamics but also the dissipative bath sampling itself becomes quasienergy resolved, so that dissipation proceeds through drive-assisted Floquet channels rather than through static transitions alone. This quasienergy-resolved dissipation can produce modified steady states, asymmetric multiphoton sidebands, and channel-competition effects that are not generally captured when either the interaction-induced or drive-induced dressing is omitted.



The breakdown of incomplete dissipative descriptions emerges once multiple Floquet channels contribute on comparable dissipative scales, so that quasienergy-resolved decay pathways and nonsecular couplings can no longer be neglected. In driven TLS systems, this regime is often accompanied by resolved Floquet sidebands in observables such as the Mollow-triplet spectrum, although the precise crossover depends on the interplay between the system energy scales, drive amplitude, drive frequency, and dissipation strength. In coupled-TLS systems, the same issue is sharpened by the coexistence of internal and external dressing: omitting either dressing mechanism can yield incorrect long-time observables.

This work makes two key contributions. First, we formulate a unified nonsecular F-GME framework in which interaction-induced (internal) and drive-induced (external) dressing are treated simultaneously and on an equal footing within a driven open quantum system. Second, using a Floquet--Liouville (FL) modal decomposition, we show that observable spectral resonances are governed not simply by isolated quasienergy transitions, but by the eigenmodes of the full FL generator itself. This enables us to distinguish between spectral features associated with nearly pure quasienergy channels and those arising from hybridized dissipative Floquet channels merged into FL modes generated through either the strong drive hybridization and mixing or the nonsecular channel mixing.

By resolving quasienergy channels explicitly, we identify when reduced descriptions constitute controlled approximations: the TI-GME is reliable when drive-induced Floquet channels are effectively reducible to static dressed transitions, while the conventional F-ME is reliable only when interaction-induced dressing is negligible or perturbative. However, the full F-GME is required when both interaction-induced and drive-induced dressing are non-negligible and when the relevant quasienergy gaps are not well separated compared with dissipative couplings. While aspects of this double-dressing physics have previously been explored for closed ultrastrongly coupled light--matter systems~\cite{Akbari_Floquet_2025}, its consequences for open driven-dissipative systems and observable FL spectroscopy remain largely unexplored.

Beyond the specific models considered here, our results establish a practical framework for diagnosing when Floquet-consistent generalized dissipative treatments become necessary, and clarify the physical origin of discrepancies between commonly used master-equation approaches for driven open quantum systems.

The rest of our paper is organized as follows. Section~\ref{sec:GME} presents the GME formalism for the open (hybrid--strongly interacting) quantum system. Section~\ref{sec:Observables_FL} introduces the long-time observables considered throughout this work and establishes the FL framework used to analyze the spectral structure of the observables. In Sec.~\ref{sec:TLS}, we apply the formalism to a single driven-dissipative TLS and compare populations and spectra obtained from the different dissipative approaches. In Sec.~\ref{sec:CoupledTLSs}, we investigate the coupled-TLS system with dipole--dipole interaction and analyze the consequences of internal and external dressing on the long-time observables. Finally, Sec.~\ref{sec:Conclusions} concludes the paper.

In addition to the main text, we also provide extended details on the Floquet theory of strongly driven strongly interacting closed quantum systems in App.~\ref{secS:FloquetTheory_ClosedSystem}, as well as the open quantum system version of those to derive the F-GME in App.~\ref{secS:FloquetTheory_OpenSystem}. Moreover, we give the full details of the FL formalism in App.~\ref{secS:FL}. Finally, we discuss some additional results in App.~\ref{secS:AdditionalResults} to supplement the results and discussion shown in the main text.

\section{Open driven system: Competing Dissipative Descriptions}
\label{sec:GME}

We consider a generic quantum system composed of (arbitrarily) interacting subsystems, labeled by $\Lambda$. The internal couplings between subsystems are assumed to be comparable in magnitude to the bare subsystem energies, so that hybridization can substantially modify the dynamics of the full system. In addition, one or more subsystems may be subjected to an external periodic drive whose amplitude can also be sufficiently strong to induce significant dynamical dressing.

Under periodic driving, the Hamiltonian satisfies $H(t+T)=H(t)$, with period $T=2\pi/\omega_d$, and can thus be expanded in a Fourier series, 
\begin{equation}
    H(t)
    =
    \sum_{m\in\mathbb{Z}}
    \ee^{-\ii m\omega_dt}\,H_m
    =
    H_0+H_d(t),
    \label{eq:Ht}
\end{equation}
where 
\begin{equation}
    H_d(t)
    \equiv
    \sum_{m\neq0}
    H_m\,\ee^{-\ii m\omega_dt}
\end{equation}
denotes the explicitly time-dependent driving contribution.

We allow the system to be  strongly interacting internally, such that the eigenstates of the bare subsystems no longer provide an adequate basis for describing the dynamics. It is therefore necessary to work in the dressed eigenbasis of the time-independent Hamiltonian $H_0$~\cite{Carmichael_Master_1973,Settineri_Gauge_2021}. After diagonalizing $H_0$, we denote its eigenstates and eigenenergies by $\ket{j}$ and $E_j$, respectively, and formulate the theory in this interaction-induced dressed picture.

Consistency of the formalism further requires that the driving Hamiltonian also be expressed in this dressed basis~\cite{Settineri_Dissipation_2018,Salmon_Gauge-independent_2022}. We will consider coherent (optical) driving of a chosen system operator,
\begin{equation}
     H_d(t)
     =
     \Omega_d\,S_d\,\sin(\omega_dt+\phi_d),
     \label{eq:Ht_d}
 \end{equation}
where $\Omega_d$ and $\phi_d$ denote the drive amplitude and phase, respectively. The dressed transition operator to be driven is defined via $S_d=\sum_\omega S_d(\omega)$ with 
\begin{equation}
S_{d}(\omega)
=
\langle j|S_{d}^{\rm bare}|k\rangle
\,|j\rangle\langle k|,
\label{eq:Sp_d}
\end{equation}
and $\omega=\omega_{jk}\equiv E_k-E_j$, where $\{E_j\}$ are the eigenenergies corresponding to the eigenstate $\{\ket{j}\}$ of the time-independent Hamiltonian $H_0$.

When coupled to an environment, each subsystem may additionally interact with one or more dissipative bath channels.
To describe the dynamics of such systems, we employ the generalized master equation (GME) framework for open quantum systems, formulated in the dressed-state basis and capable of incorporating structured environments~\cite{Settineri_Dissipation_2018,Salmon_Gauge-independent_2022}. The reduced density matrix $\rho(t)$ evolves according to
\begin{equation}
\frac{\mathrm{d}\rho(t)}{\mathrm{d}t}
=
\mathcal{L}(t)\rho(t),
\label{eq:ME}
\end{equation}
where $\mathcal{L}(t)$ is the total Liouvillian superoperator.

For driven systems, $\mathcal{L}(t)$ is generally time dependent and can be decomposed as
\begin{equation}
\mathcal{L}(t)=\mathcal{L}_{\mathrm{S}}(t)+\mathcal{L}_{\mathrm{diss}}(t),
\label{eq:TotalLiouvillian}
\end{equation}
where
\begin{equation}
\mathcal{L}_{\rm S}(t)\rho(t)=-\ii[H(t),\rho(t)],
\label{eq:SystemLiouvillian}
\end{equation}
governs the coherent driven dynamics, while
$\mathcal{L}_{\rm diss}(t)$ accounts for environmental relaxation and decoherence.
Throughout our paper, we  use natural 
units, with 
$\hbar=1$. 

We will compare two dissipative descriptions of periodically driven systems. In the first case, dissipation is constructed in the eigenbasis of the time-independent Hamiltonian, leading to a {\it time-independent} generalized master equation (TI-GME). In the second case, dissipation is formulated in the Floquet quasienergy basis, yielding a Floquet generalized master equation (F-GME), in which the dissipator inherits the periodicity of the drive. The contrast between these two approaches forms the central focus of this paper. Both master-equation formalisms are derived within the Born--Markov approximation~\cite{Breuer_Theory_2002}; see 
App.~\ref{secS:FloquetTheory_OpenSystem} for a detailed derivation.

\subsection{Time-independent generalized master equation}

In the TI-GME approach~\cite{Settineri_Dissipation_2018,Salmon_Gauge-independent_2022}, the coherent system Liouvillian (i.e., the part not related to dissipation) retains the explicit time dependence of the drive, whereas the dissipative contribution is constructed {\it only} from the eigenbasis of the time-independent Hamiltonian. Accordingly, the total Liouvillian is written as $\mathcal{L}(t)=\mathcal{L}_{\rm S}(t)+\mathcal{L}_{\rm diss}$, where the dissipator is time independent and given by (see App.~\ref{secS:FloquetTheory_OpenSystem} for details):
\begin{widetext}
\begin{equation}
    \begin{split}
      \mathcal{L}_{\rm diss}\rho(t)&=\sum_{\Lambda}\sum_{\omega,\omega'}
      \lcb\Gamma^{\Lambda}(\omega)\lsb{S}_{\rm B}^{\Lambda}(\omega) {\rho}(t){S}_{\rm B}^{\Lambda\dagger}(\omega')- {S}_{\rm B}^{\Lambda\dagger}(\omega'){S}_{\rm B}^{\Lambda}(\omega){\rho}(t)\rsb\right.
        \\
        &\hspace{3cm}\left.+
\Gamma^{\Lambda*}(\omega')\lsb {S}_{\rm B}^{\Lambda}(\omega){\rho}(t){S}_{\rm B}^{\Lambda\dagger}(\omega')-{\rho}(t) {S}_{\rm B}^{\Lambda\dagger}(\omega') {S}_{\rm B}^{\Lambda}(\omega) \rsb \rcb,
    \end{split}
    \label{eq:Ldiss_TIGME}
\end{equation}
\end{widetext}
where
\begin{equation}
S_{\rm B}^{\Lambda}(\omega)
=
\langle j|S_{\rm B}^{\Lambda,\rm bare}|k\rangle\,|j\rangle\langle k|,
\label{eq:S_B}
\end{equation}
with again $\omega=\omega_{jk}\equiv E_k-E_j$,
is the system-bath coupling operator in the time-independent dressed basis, and without loss of generality, it can be assumed to be Hermitian.

The operator $S_{\rm B}^{\Lambda}$ denotes the system observable coupled to bath channel $\Lambda$, expressed in the dressed basis. The problem-dependent dissipative transition frequencies are $\omega_{jk}$, while $\Gamma^\Lambda(\omega)$ is the Fourier transform of the corresponding bath correlation function;
this form is readily generalized to multiple independent dissipation channels labeled by $\Lambda$, such that $\mathcal{L}_{\rm diss}=\sum_\Lambda \mathcal{L}_{\rm diss}^{\Lambda}$. The same framework has also been successfully extended to more complex hybrid systems, including the generalized Dicke model~\cite{Akbari_Generalized_2023}.

Equation~\eqref{eq:Ldiss_TIGME} is the full nonsecular
Born--Markov dissipator considered here and therefore retains
interference between distinct transition channels $\omega$ and
$\omega'$. A full secular approximation neglects unequal-frequency
cross terms when $|\omega-\omega'|$ is large compared with the
relevant dissipative linewidths, while partial-secular filtering
retains only sufficiently close transition pairs. Independently, a
further rotating-wave-type approximation
(dissipator only) may remove rapidly
oscillating dissipative terms at frequencies
$\pm(\omega+\omega')$ and, when zero-frequency transition operators
are present, at $\pm\omega$ and $\pm\omega'$. These terms average out
when their oscillation frequencies are large compared with the
relevant dissipative rates
\cite{Settineri_Dissipation_2018,Salmon_Gauge-independent_2022};
the nonoscillating $S_{\rm B}(\omega=0),S_{\rm B}(\omega'=0)$ terms instead
describe pure dephasing. For more details on these optional approximations, see Ref.~\onlinecite{Settineri_Dissipation_2018} and also App.~\ref{secS:FloquetTheory_OpenSystem}.

The TI-GME designation instead refers to resolving the system--bath
operator with respect to transitions of the full interacting static
Hamiltonian $H_0$. In the absence of internal hybridization this
resolution reduces to the corresponding bare/local transition
structure, and hence to the TI-ME when the same dissipative
approximations are made. For strongly hybridized systems the dressed
transition resolution is essential, while structured reservoirs
enter through the frequency dependence of
$\Gamma^\Lambda(\omega)$
\cite{Settineri_Dissipation_2018,Salmon_Gauge-independent_2022}.

Drive effects may be incorporated perturbatively within a different framework, for example, through Van Vleck or related high-frequency expansions~\cite{VanVleck_Sigma-Type_1929}. However, when the drive is strong, far detuned, or induces multiple competing resonances, a perturbative treatment is generally insufficient. In that regime, one must retain the full time-dependence of the driven system and treat dissipation consistently beyond static transition frequencies. This motivates the Floquet-based formulation introduced next.

\subsection{Floquet generalized master equation}

We now extend the nonsecular GME framework to periodically driven systems by constructing the dissipator in the Floquet basis.

 The explicit time dependence of $H(t)$ gives rise to an infinite ladder of drive-assisted channels and, in general, many near-degeneracies within a chosen Brillouin zone. In such situations, secularization in the quasienergy basis can fail, and nonsecular couplings between nearby channels may become important~\cite{Hone_Statistical_2009,NafariQaleh_Enhancing_2022,Farina_Open-quantum-system_2019}.

The same evolution equation in Eq.~\eqref{eq:ME} applies along with Eqs.~\eqref{eq:TotalLiouvillian} and \eqref{eq:SystemLiouvillian}, but the dissipator is now constructed in the Floquet basis and becomes explicitly time periodic.
The reduced density matrix then evolves according to the F-GME,
where the total Liouvillian is
$\mathcal{L}(t)=\mathcal{L}_{\rm S}(t)+\mathcal{L}_{\rm diss}(t)$,
and the dissipator now inherits the periodicity of the drive~\cite{Grifoni_Driven_1998,Restrepo_Driven_2019,Mori_Floquet_2023,LeBoite_Theoretical_2020}. Its nonsecular form is (see App.~\ref{secS:FloquetTheory_OpenSystem} for the derivation)
\begin{widetext}
\begin{equation}
    \begin{split}
      {\mathcal{L}}_\mathrm{diss}(t){\rho}(t)          &=\sum_{\Lambda}\sum_{\Delta,\Delta'}\lcb\mathrm{e}^{-\ii l\omega_{d}t}\Gamma^{\Lambda}(\Delta)\lsb{S}_{\rm B}^{\Lambda}(\Delta;t) {\rho}(t){S}_{\rm B}^{\Lambda\dagger}(\Delta')- {S}_{\rm B}^{\Lambda\dagger}(\Delta'){S}_{\rm B}^{\Lambda}(\Delta;t){\rho}(t)\rsb\right.
        \\
        &\hspace{3cm}\left.+\mathrm{e}^{\ii l'\omega_{d}t}
\Gamma^{\Lambda*}(\Delta')\lsb {S}_{\rm B}^{\Lambda}(\Delta){\rho}(t){S}_{\rm B}^{\Lambda\dagger}(\Delta';t)-{\rho}(t) {S}_{\rm B}^{\Lambda\dagger}(\Delta';t) {S}_{\rm B}^{\Lambda}(\Delta) \rsb \rcb,
    \end{split}
    \label{eq:Ldiss_FGME_OriginalBasis}
\end{equation}
\end{widetext}
where the relevant transition frequencies are the quasienergy-resolved channels:
\begin{equation}
\Delta \equiv \Delta_{\alpha\beta l}
=
\varepsilon_\beta-\varepsilon_\alpha+l\omega_d,
\label{eq:Delta_abl}
\end{equation}
where $\varepsilon_\alpha$ are Floquet quasienergies (in the primary BZ) and $l\in\mathbb{Z}$ labels drive-assisted sideband processes. Equivalently, one may write $\Delta_{\alpha\beta l}=\varepsilon_{l'\beta}-\varepsilon_{l''\alpha}$ (where $\varepsilon_{l\alpha}$ are Floquet quasienergies in the $l$th BZ) with $l=l'-l''$.

The time-independent and time-dependent transition operators are defined, respectively,  as
\begin{equation}
S^\Lambda_{\rm B}(\Delta)
=
S^\Lambda_{{\rm B},\alpha\beta l}\,|\alpha\rangle\langle\beta|,
\end{equation}
and
\begin{equation}
S^\Lambda_{\rm B}(\Delta;t)
=
S^\Lambda_{{\rm B},\alpha\beta l}\,|\alpha(t)\rangle\langle\beta(t)|,
\end{equation}
with matrix elements
\begin{equation}
S^\Lambda_{{\rm B},\alpha\beta l}
=
\frac{1}{T}
\int_0^T dt\,
e^{-\ii l\omega_d t}
\langle\alpha(t)|S^\Lambda_{\rm B}|\beta(t)\rangle,
\label{eq:FelementS_ex}
\end{equation}
where $\ket{\alpha(t)}$ are the Floquet modes with $\ket{\alpha}\equiv\ket{\alpha(0)}$.

Equation~\eqref{eq:Ldiss_FGME_OriginalBasis} is the {\it full nonsecular}
Floquet--Born--Markov dissipator considered here and therefore retains
interference between distinct quasienergy-assisted transition channels
$\Delta$ and $\Delta'$. 
In a similar fashion to that of the TI-GME in Eq.~\eqref{eq:Ldiss_TIGME}, the full or partial secularization or dissipator-RWA can be obtained by the notational change of $\omega\to\Delta$ (see also App.~\ref{secS:FloquetTheory_OpenSystem}).


Unlike the TI-GME, the F-GME designation 
refers to resolving dissipation at the
full quasienergy-assisted transition frequencies
$\Delta_{\alpha\beta l}
=\varepsilon_\beta-\varepsilon_\alpha+l\omega_d$.
Since many Floquet sideband channels can become degenerate or nearly
degenerate, secular and filtering approximations can be substantially
more restrictive than in the corresponding static problem. In the
absence of external Floquet dressing, the quasienergy-resolved
transition structure reduces to the static dressed transitions and
the F-GME reduces to the TI-GME when the same dissipative
approximations are made. Conversely, in the absence of internal
hybridization, the static dressed transition sectors reduce to the
corresponding bare/local ones and the F-GME reduces to the
bare-transition-resolved F-ME. Structured reservoirs enter through
the frequency dependence of $\Gamma^\Lambda(\Delta)$, by sampling the quasienergy transitions.

Relative to the TI-GME, the key formal change in
the dissipator is the replacement
\begin{equation*}
\mathcal{L}_{\rm diss}\rightarrow \mathcal{L}_{\rm diss}(t).
\end{equation*}
Since the Floquet modes satisfy $|\alpha(t+T)\rangle=|\alpha(t)\rangle$, they can be Fourier expanded in terms of the Floquet sidebands via 
\begin{equation}
    \ket{\alpha(t)}=\sum_l\ee^{-\ii l\omega_d t}\,\ket{\alpha_l}.
    \label{eq:sideband_expansion}
\end{equation}
Thus, the dissipator itself is periodic and may be expanded as
\begin{equation}
\mathcal{L}_{\rm diss}(t)
=
\sum_n e^{-\ii n\omega_d t}\mathcal{L}_{{\rm diss},n},
\end{equation}
with $\mathcal{L}_{{\rm diss},n}$ given explicitly in App.~\ref{secS:FloquetTheory_OpenSystem}.
In contrast, the TI-GME retains only a static dissipative component. 

However, both theories use a time-harmonic system Liouvillian, $\mathcal{L}_{\rm S}=\sum_n\mathcal{L}_{{\rm S},n}\,\ee^{-\ii n\omega_dt}$ with $\mathcal{L}_{{\rm S},n}=-\ii[H_n\,,\,\cdot\,]$, since the Hamiltonian is time-harmonic.
Thus, the full Liouvillian obeys the same periodicity, $\mathcal{L}(t+T)=\mathcal{L}(t)$, so that
\begin{equation}
\mathcal{L}(t)=\sum_n\ee^{-\ii n\omega_dt}\mathcal{L}_n,
\end{equation}
where $\mathcal{L}_n=\mathcal{L}_{{\rm S},n}+\mathcal{L}_{{\rm diss},n}$.

We emphasize that both the TI-GME and F-GME originate from the same microscopic Born--Markov--Redfield construction. Their essential difference is not the bath approximation itself, but the basis in which system transitions are resolved: static dressed-state gaps in the TI-GME versus quasienergy channels in the Floquet treatment~\cite{Breuer_Theory_2002,Restrepo_Driven_2019}.
For all parameter regimes considered below, the nonsecular generators used here yield stable dynamics and positive density matrices in numerical simulations.
If, on the other hand, the Floquet formalism applied to the bare basis of the Hamiltonian, one recovers the conventional Floquet master equations (F-MEs).
Hence, the F-GME is the only formalism that incorporates the internal and external dressing on equal footing. If the external dressing is ignored, it reduces to the TI-GME, whereas if the internal dressing is not included, it reduces to F-ME.

\subsection{Key conceptual differences between models}

The distinction between the TI-GME and the F-GME is not in the coherent driven dynamics. In both approaches, the system evolves under the same time-dependent Hamiltonian $H(t)$ and therefore contains the same coherent Floquet sideband structure. 
Namely, the transitions are among the Floquet sidebands $\ket{\alpha_k}\leftrightarrow\ket{\beta_{k'}}$ assisted with $l=k-k'$ external quanta, or equivalently, $\Fket{k\alpha}\leftrightarrow\Fket{k'\beta}$ in the Floquet extended space.
The difference lies in how the system--bath coupling operator is spectrally decomposed when deriving the dissipator.

In the TI-GME, the bath couples to transitions between eigenstates of the {\it time-independent} Hamiltonian. Dissipation is therefore resolved at the static dressed-state  frequencies
\begin{equation*}
\omega_{jk}=E_k-E_j.
\end{equation*}

The drive enters only through the coherent (system) Liouvillian $\mathcal{L}_{\rm S}(t)$, while the dissipator remains tied to the static transition structure.
The dissipator therefore contains a hierarchy of drive-assisted channels labeled by $(j,k)$, and can be influenced by the important static dressed eigenenergy-resolved factors:
\begin{equation*}
\Gamma^\Lambda(\omega_{jk}),
\qquad
S^\Lambda_{\mathrm{B},jk}.
\end{equation*}

In the secular, or well-isolated-channel limit, the leading amplitude terms in the TI-GME dissipator scale as
\begin{equation*}
\Gamma^\Lambda(\omega_{jk})
\left|S^{\Lambda}_{{\rm B},jk}\right|^2,
\end{equation*}
whereas in the nonsecular regime considered here, different static channels can couple and hybridize through the full Liouvillian generator.
This quantity provides only a useful diagonal estimate in the static dressed basis representation; the actual decay rates and spectral weights are determined by the eigenmodes of the nonsecular Floquet--Liouville generator.

The secular approximation requires that the timescale of the system-bath interaction is much slower than the energy-level spacings. For the TI-GME this means
\begin{equation}
    \lvert\omega_{jk}-\omega_{j'k'}\rvert\gg\Gamma.
\end{equation}
Thus, the secular approximation fails when the system has degenerate or nearly degenerate static energy levels, or in very large systems where energy levels become closely spaced.
Therefore, TI-GME retains coherences between \emph{static dressed transitions}.

In contrast, in the F-GME, the bath couples to transitions between Floquet states (created by dynamical excitation from the drive). Dissipation is resolved at the {\it quasienergy-assisted} transition frequencies:
\begin{equation*}
\Delta_{\alpha\beta l}
=
\varepsilon_\beta-\varepsilon_\alpha+l\omega_d,
\end{equation*}
where $l$ labels the exchange of drive quanta with the periodic field. 
The dissipator now contains a hierarchy of drive-assisted channels labeled by $(\alpha,\beta,l)$, and can be influenced by the important static dressed eigenenrgy-resolved factors:
\begin{equation*}
\Gamma^\Lambda(\Delta_{\alpha\beta l}),
\qquad
S^\Lambda_{{\rm B},\alpha\beta l}.
\end{equation*}

\begin{table*}[t]
\caption{\textbf{Comparison of dressing mechanisms incorporated by different open-system approaches.} The conventional standard master equation (TI-ME) is valid for weakly-driven and weakly-interacting quantum systems, formulated in the bare state of the static Hamiltonian. The generalized master equation (specifically, the usual TI-GME) is valid for weakly-driven and strongly-interacting quantum systems, formulated in the dressed state of the static Hamiltonian. The conventional Floquet master equation (F-ME) is valid for strongly-driven and weakly-interacting quantum systems, formulated in the Floquet picture on the bare state basis of the static Hamiltonian. Finally, the Floquet generalized master equation (F-GME) is valid for strongly-driven and strongly-interacting quantum systems, formulated in the Floquet picture on the dressed state basis (of the static Hamiltonian).}
\begin{ruledtabular}
\begin{tabular}{lccc}
Methods & Internal dressing & External dressing & Double dressing \\
\hline
TI-ME  & $\times$     & $\times$     & $\times$ \\
TI-GME & $\checkmark$ & $\times$     & $\times$ \\
F-ME   & $\times$     & $\checkmark$ & $\times$ \\
F-GME  & $\checkmark$ & $\checkmark$ & $\checkmark$ \\
\end{tabular}
\end{ruledtabular}
\label{tab:dressing}
\end{table*}

In the secular limit,
the leading amplitude terms in the F-GME dissipator scale as
\begin{equation*}
\Gamma^\Lambda(\Delta_{\alpha\beta l})
\left|S^{\Lambda}_{{\rm B},\alpha\beta l}\right|^2,
\end{equation*}
whereas in the nonsecular regime, considered here, different Floquet sideband channels can couple and hybridize through the full Liouvillian generator.
This quantity provides only a useful diagonal estimate in the Floquet quasienergy state basis representation; the actual decay rates and spectral weights are determined by the eigenmodes of the nonsecular Floquet--Liouville generator.
The secular approximation, here then, requires 
\begin{equation}
    \lvert\Delta_{\alpha\beta l}-\Delta_{\alpha'\beta' l'}\rvert\gg\Gamma.
\end{equation}

Thus, the secular approximation generally fails when the system has degenerate or nearly degenerate quasienergy levels, or in very large systems where energy levels become closely spaced. Since the quasienergies are confined in a BZ energy scale of the drive frequency, in the Floquet theory, it is more vulnerable to form 
near-degeneracies or a large number of states in a finite energy range in the quasienergy space. Hence, the {\it nonsecularity in the F-GME is more crucial}.
Therefore, F-GME retains coherences between \emph{Floquet sidebands}.

When multiple channels contribute comparably or become nearly degenerate, the nonsecular Floquet dissipator can hybridize them into collective Floquet--Liouville modes, which we will discuss in more details, in the next section. In such a regime, the correct dissipative dynamics cannot, in general, be inferred from the static eigenfrequencies alone.

Consequently, even for a spectrally flat bath, where $\Gamma^\Lambda(\omega)\equiv\Gamma^\Lambda$ is frequency-independent, the two approaches are generally not equivalent. A flat bath removes spectral selectivity from the environment, but it does not remove the Floquet redistribution of system--bath matrix elements among sideband channels. The TI-GME collapses these channels into the static dressed-state transition structure, whereas the F-GME retains their quasienergy-resolved character.

The dynamics is therefore quasienergy-resolved ($\Delta_{\alpha\beta l}$-resolved) rather than static eigenenergy-resolved ($\omega_{jk}$-resolved), which manifests itself in the difference in the system-bath coupling operator content ($S_{{\rm B}, jk}\to S_{{\rm B}, \alpha\beta l}$), bath frequency sampling [$\Gamma(\omega_{jk})\to\Gamma(\Delta_{\alpha\beta l})$] and nonsecular dissipative channel hybridization [$(j,k)\to(\alpha\beta l)$ coherences], as discussed above. 
The TI-GME is therefore expected to provide a good approximation only when a single Floquet channel dominates the dynamics, or when the relevant quasienergy channels are well separated and weakly mixed.
The agreement between the TI-GME and F-GME is not governed by a monotonic threshold in drive strength 
$\eta_d$ or frequency $\omega_d$. Rather, the TI-GME is reliable when the dissipatively relevant Floquet channels can be effectively associated with the corresponding static transition families, with comparable rates and weak interchannel mixing. The F-GME becomes necessary when appreciable dissipative weight is transferred into Floquet channels that cannot be reduced to those static dressed transitions. In general, the resulting dynamics depends jointly on the channel matrix elements, bath rates, quasienergy separations, steady-state occupations, and nonsecular interference. As a heuristic indication, for a particular pair of channels,
\begin{equation}
    \lvert\Delta_{\alpha\beta l}-\omega_{jk}\rvert\lesssim\Gamma,
\end{equation}
implies that the Floquet and corresponding static transition frequencies are not dissipatively well resolved. If their associated matrix elements and bath rates are also comparable, the two channels may then contribute similarly to the dynamics, leading to an accidental local agreement between the F-GME and TI-GME.

For a summary of the dressing mechanisms incorporated in the different master-equation approaches, see Tab.~\ref{tab:dressing}. Our nomenclature distinguishes both the basis used to resolve the system--bath coupling and the treatment of the drive. A time-independent-basis master equation (TI-ME) denotes the conventional construction in the uncoupled bare subsystem basis, where the drive is treated perturbatively and the dissipative transition operators remain tied to local bare-state transitions. In the corresponding generalized master equation (TI-GME), the full interacting static Hamiltonian is first diagonalized and the system--bath coupling operator is resolved into transitions between its interaction-dressed eigenstates; in the present work, this construction is retained without secularization. 

The same distinction carries over to the Floquet framework, except that the drive is treated nonperturbatively, promoting the static basis and dissipator to time-periodic Floquet counterparts. Thus, the difference between the conventional Floquet-Markov or Floquet-Redfield description (F-ME) and the Floquet generalized master equation (F-GME) is not merely a change of matrix representation; it lies in whether the system--bath operator is spectrally resolved before the Floquet construction in the bare or interaction-dressed static basis. Because interaction dressing and transition-frequency resolution do not generally commute, the two procedures generate physically distinct quasienergy-resolved dissipative channels, even when they employ the same coherent time-dependent Hamiltonian.

\section{Observables and Floquet--Liouville Analysis}
\label{sec:Observables_FL}

In this section, we introduce the steady-state observables used throughout this work and present the Floquet--Liouville (FL) framework employed later to interpret the spectra and dissipative dynamics. The main text focuses on the key physical ingredients required for the discussion of the results, while the complete derivation of the FL formalism and its numerical implementation is presented in App.~\ref{secS:FL}.

\subsection{System Populations}

For a subsystem $\Lambda$, the instantaneous excitation population (i.e., time-dependent) is defined as
\begin{equation}
N_\Lambda(t)
=
\langle
s_{\rm B}^{\Lambda-}(t)
s_{\rm B}^{\Lambda+}(t)
\rangle,
\label{eq:N_t}
\end{equation}
where $s_{\rm B}^{\Lambda\pm}=\sum_{\omega\gtrless0}S^\Lambda_{\rm B}(\omega)$, with $S^\Lambda_{\rm B}(\omega)$ in Eq.~\eqref{eq:S_B}, denote the system dressed ladder operators. For noninteracting systems, these operators reduce to the familiar bare operators, but with a 
potential rotation.

In the present work, $N_\Lambda(t)$ is obtained directly from the density matrix evolved under the corresponding master equations appropriate for the bath model under consideration.

The quantity of primary interest is the long-time periodic average after the system reaches its periodic steady state, so that we can obtain periodic averages,
\begin{equation}
\overline{N}_\Lambda
=
\frac{1}{T'}
\int_0^{T'}dt\,
\lim_{t\rightarrow\infty}
N_\Lambda(t),
\label{eq:NLambda_avg}
\end{equation}
where $t\rightarrow\infty$ denotes times exceeding the relaxation time $t_{\rm ss}$ associated with the periodic steady state, and $T'\geq T$ is the period of the asymptotic observable. In the situations considered here, first-order observables remain $T$-periodic, while higher-order quantities may contain higher harmonics corresponding to shorter effective periods such as $T/2$. Accordingly, choosing $T'=T$ is sufficient in all calculations below.

\paragraph*{Phenomenological channel picture.}
A simplified analytical description can be obtained within the phenomenological Floquet approach introduced in Ref.~\onlinecite{Akbari_Floquet_2025}, 
yielding
\begin{equation}
N_\Lambda(t)
=
\sum_{\alpha\beta}
c_\alpha^*c_\beta
e^{i(\varepsilon_\alpha-\varepsilon_\beta)t
-\Gamma t(1-\delta_{\alpha\beta})}
\langle
\alpha(t)
|
s^{\Lambda-}_{\rm B}
s^{\Lambda+}_{\rm B}
|
\beta(t)
\rangle.
\label{N}
\end{equation}
Here, the assumption is that the system is prepared at $t_0$ ($=0$ here) in a pure state $|\psi_0\rangle$
(the ground state of the undriven, dressed Hamiltonian), which is
decomposed in the Floquet basis at the switch-on time,
\begin{equation}
|\psi_0\rangle
=
\sum_\alpha c_\alpha\,|\alpha(t_0)\rangle,
\quad
c_\alpha=\langle\alpha(t_0)|\psi_0\rangle
=\sum_l \mathrm{e}^{\ii l\omega_d t_0}\,\langle\alpha_l|\psi_0\rangle .
\label{eq:calpha_def}
\end{equation}

Setting $t_0=0$, in the long-time limit, Eq.~\eqref{N} reduces to
\begin{equation}
N_\Lambda^{\rm ss}(t)
=
\sum_{\alpha}
|c_\alpha|^2
\langle
\alpha(t)
|
s_{\rm B}^{\Lambda-}
s_{\rm B}^{\Lambda+}
|
\alpha(t)
\rangle,
\label{eq:Nss_t_phen}
\end{equation}
and thus the long-time averaging yields
\begin{equation}
\overline{N}^{\rm(phase-locked)}_\Lambda
=
\sum_{\alpha}
|c_\alpha|^2
\sum_l\langle
\alpha_l
|
s_{\rm B}^{\Lambda-}
s_{\rm B}^{\Lambda+}
|
\alpha_l
\rangle.
\label{eq:Nbar_phen}
\end{equation}

In the phenomenological 
approach~\cite{Akbari_Floquet_2025}, the coupling to a
bath is modeled by exponential damping of the Floquet
\emph{coherences} at a rate $\Gamma$, while the Floquet
\emph{populations} $|c_\alpha|^2$ are left untouched. 
For $t\gg\Gamma^{-1}$ the off-diagonal terms are extinguished and
the signal becomes strictly $T$-periodic,
whose one-period average follows from
Eq.~(\ref{eq:sideband_expansion}): the $l\neq l'$ cross terms carry
$e^{i(l-l')\omega_{\rm d}t}$ and vanish under the average, leaving Eq.~\eqref{eq:Nbar_phen}.
Thus, it is better to refer to
Eq.~\eqref{eq:Nbar_phen} as the \emph{phase-locked} (pl) phenomenological
result; while the Floquet coherences $\alpha\neq\beta$ have been
dephased, the weights
$|c_\alpha|^2=\big|\sum_l\langle\alpha_l|\psi_0\rangle\big|^2$
retain the \emph{inter-sideband} coherences of the initial-state
projection---the relative phases between the harmonics $|\alpha_l\rangle$
at the switch-on time. Consequently,
Eq.~\eqref{eq:Nbar_phen} depends on the launch time $t_0$
(equivalently, on the drive phase $\phi_{d}$) through
$c_\alpha=\langle\alpha(t_0)|\psi_0\rangle$; it yields  the long-time cycle-averaged occupation for a system
switched on at this particular phase of the drive.

Throughout this work, the 
{\it (phase-locked) phenomenological} Floquet approach is used as one of the reference models and compared with the results obtained from the master equations.

\paragraph*{Phenomenological channel picture with the initial-phase average.}
A closely related and experimentally motivated, initial-phase-independent quantity, is obtained by
additionally averaging Eq.~\eqref{eq:Nbar_phen} over the switch-on
time within one drive period. Using
$c_\alpha(t_0)=\sum_l e^{il\omega_{d}t_0}\langle\alpha_l|\psi_0\rangle$, then
\begin{equation}
\frac{1}{T}\int_0^T\!dt_0\,
|c_\alpha(t_0)|^2
=
\sum_l
\big|\langle\alpha_l|\psi_0\rangle\big|^2
\equiv p_\alpha,
\label{eq:pav}
\end{equation}
i.e., the phase average eliminates precisely the inter-sideband
cross terms $l\neq l'$ of the weights, yielding the
\emph{phase-averaged} counterpart of
Eq.~\eqref{eq:Nbar_phen}:
\begin{equation}
\overline{N}_{\Lambda}^{\rm (phase-averaged)}
=
\sum_{\alpha}
p_\alpha
\sum_l
\langle\alpha_l|s^{\Lambda-}_{\rm B}s^{\Lambda+}_{\rm B}|\alpha_l\rangle.
\label{eq:Nbar_pa}
\end{equation}

Equation~(\ref{eq:Nbar_pa}) is exactly Shirley's expression for
the long-time-averaged occupation
(see Ref.~\onlinecite{Shirley_Solution_1965}, in which it was derived for the time- and
phase-averaged transition probability), as given in Eqs.~\eqref{eq:P_t0_avg}, \eqref{eq:P_long_avg_def} and \eqref{eq:Pbar_llprime} of App.~\ref{secS:FloquetTheory_ClosedSystem}. Note, the Shirley limit is obtained by choosing the measurement operator as
$O=|f\rangle\langle f|$ instead of $N_\Lambda=s^{\Lambda-}_{\rm B}s^{\Lambda+}_{\rm B}$ and the initial state $|\psi_0\rangle=|i\rangle$ in Eq.~\eqref{eq:Nbar_pa} to reproduce
\begin{equation}
\overline{P}_{f\leftarrow i}
=
\sum_{\alpha}
\sum_{l l'}
\big|\langle f|\alpha_l\rangle\big|^2
\big|\langle\alpha_{l'}|i\rangle\big|^2 ,
\label{eq:ShirleyLimit}
\end{equation}
which is the standard (unitary) Floquet transition probability formula. 

The
phase-locked (\emph{sensitive} to the initial time or the drive phase) and phase-averaged (\emph{insensitive} to the initial time or the drive phase) weights obey the same sum rule,
\begin{equation}
\sum_\alpha |c_\alpha|^2=\sum_\alpha p_\alpha
=\langle\psi_0|\psi_0\rangle=1,
\label{eq:sumrule}
\end{equation}
by completeness of the Floquet basis at fixed time, so the two
formulations differ only in how the unit total weight is
distributed among the modes:
\begin{equation}
|c_\alpha|^2-p_\alpha
=
\sum_{l\neq l'}
e^{i(l-l')\omega_{d}t_0}
\langle\psi_0|\alpha_l\rangle
\langle\alpha_{l'}|\psi_0\rangle ,
\label{eq:weight_difference}
\end{equation}
where the right-hand side vanishes whenever $|\psi_0\rangle$ overlaps
essentially a single sideband of each mode. This is the situation in
the weak-drive/rotating-wave regime, where the two formulations
coincide; they branch precisely when counter-rotating processes
populate multiple sidebands of the same mode, so their difference is
a direct, basis-independent measure of the counter-rotating-induced
sideband coherence imprinted by the switch-on [cf. see Fig.~\ref{figS:TLSRWA_etad}(c) of App.~\ref{secS:AdditionalResults}].


In contrast, the GMEs perform the phase average dynamically. Every $c_\alpha$
with $\alpha \neq 0$ decays, so the steady result is independent of $\psi_0$
and of $t_0$; shifting the switch-on time just time-translates the limit cycle, and the one-period average is translation-invariant. Thus, $\overline{N}^{\rm (GME)}$ is automatically initial-phase-invariant, it lives on the same footing as $\overline{N}^{\rm (phase-averaged)}$, and comparing it to $\overline{N}^{\rm (phase-locked)}$ mixes in a coherence the GME cannot retain by construction. This justifies the earlier claim that $\overline{N}^{\rm (phase-averaged)}$ is the correct closed-system benchmark for GMEs.
These mechanisms are summarized in Tab.~\ref{tab:weights}.
\begin{table*}[!htbp]
\centering
\caption{The three formulations share the same Floquet channel
occupations $n_\alpha=\sum_l\langle\alpha_l|s^-s^+|\alpha_l\rangle$
and differ only in the statistical weight assigned to each mode,
$\overline{N}=\sum_\alpha w_\alpha\, n_\alpha$ (for the GME, at
leading order in $\gamma$).}
\label{tab:weights}
\begin{ruledtabular}
\begin{tabular}{lll}
Methods & Weight $w_\alpha$ of mode $\alpha$ & Origin \\[1mm]
\colrule\vspace*{3mm}
$\overline{N}^{({\rm phase-loc.})}$ &
$\lvert c_\alpha\rvert^2
=\big\lvert\sum_l\langle\alpha_l\vert\psi_0\rangle\big\rvert^2$ &
initial projection, fixed phase \\[2pt]
$\overline{N}^{({\rm phase-ave.})}$ &
$p_\alpha=\sum_l\big\lvert\langle\alpha_l\vert\psi_0\rangle\big\rvert^2$ &
initial projection, phase-averaged (Shirley) \\[2pt]
$\overline{N}^{({\rm GME})}$ &
$\bar p_\alpha$ from rate balance,
$\sum_\beta\big(\gamma_{\beta\to\alpha}\bar p_\beta
-\gamma_{\alpha\to\beta}\bar p_\alpha\big)=0$ &
bath golden-rule rates; $\psi_0$-independent \\
\end{tabular}
\end{ruledtabular}
\end{table*}

\subsection{Emission Spectra}

\paragraph*{First-order correlation function.}

To characterize the 
emission spectrum, we consider the first-order correlation function:
\begin{equation}
G_\Lambda^{(1)}(t,t')
=
\langle
\delta s_{\rm B}^{\Lambda-}(t)
\,
\delta s_{\rm B}^{\Lambda+}(t')
\rangle,
\label{eq:G1_def}
\end{equation}
where the fluctuation operators are defined by
\begin{equation}
\delta s_{\rm B}^{\Lambda\pm}(t)
=
s_{\rm B}^{\Lambda\pm}(t)
-
\langle
s_{\rm B}^{\Lambda\pm}(t)
\rangle.
\end{equation}

Subtracting the coherent expectation value isolates the incoherent contribution to the emitted signal, which is the important part.
Thus, the incoherent correlation function can be expressed as 
\begin{equation}
G_\Lambda^{(1)}(t,t')=\langle
s_{\rm B}^{\Lambda-}(t)
\,
s_{\rm B}^{\Lambda+}(t')
\rangle-G_{\Lambda,{\rm coh}}^{(1)}(t,t').
\end{equation}
with $G_{\Lambda,{\rm coh}}(t,t')=\langle
s_{\rm B}^{\Lambda-}(t)
\rangle\langle
s_{\rm B}^{\Lambda+}(t')
\rangle$. 
For continuous-wave (CW) driving, the coherent part generates a delta-function contribution in frequency space that is not relevant for the present analysis. In finite-time numerical calculations, this coherent contribution appears as a narrow sinc-like feature and may be efficiently removed 
numerically~\cite{Salmon_Gauge-independent_2022}.

The long-time averaged correlation function is defined as
\begin{equation}
\overline{G}^{(1)}_\Lambda(\tau)
=
\frac{1}{T'}
\int_0^{T'}dt\,
\lim_{t\rightarrow\infty}
G_\Lambda^{(1)}(t,t+\tau),
\label{eq:G1_avg}
\end{equation}
which allows us to also obtain the 
 associated incoherent emission spectrum,
 from
\begin{equation}
\mathsf{S}_\Lambda(\omega)
\propto
\mathrm{Re}
\int_0^\infty
d\tau\,
e^{i\omega\tau}
\,
\overline{G}^{(1)}_\Lambda(\tau).
\label{eq:Spectrum_def}
\end{equation}
Numerically, $\mathsf{S}_\Lambda(\omega)$ is evaluated using the quantum regression theorem over one period of the late-time periodic steady state, following Ref.~\onlinecite{Salmon_Gauge-independent_2022}.

In it interesting to note that the connection between the incoherent spectrum and the average number of excitation is not trivial. From Eq.~\eqref{eq:Spectrum_def}, one obtains
\begin{equation}
\int d\omega\,\mathsf{S}_\Lambda(\omega)
\propto
\overline{N}-\overline{\lvert\langle s^{\Lambda-}\rangle_{\rm ss}\rvert^2}.
\label{eq:Spectrum_Population_connection}
\end{equation}
This means the area under the spectrum is proportional to the average number of excitation when the coherent contribution is subtracted.

\paragraph*{Phenomenological channel picture.}

It is often convenient to interpret the spectrum as a sum of contributions arising from effective dissipative channels, using
\begin{equation}
\mathsf{S}(\omega)
=
\sum_\mu
\mathsf{S}_\mu(\omega),
\label{eq:S_decomp_main}
\end{equation}
where $\mu$ labels the relevant dynamical contributions. When a single channel dominates a given resonance, its spectral contribution may be approximated by a Lorentzian-like form,
\begin{equation}
\mathsf{S}_\mu(\omega)
\approx
L_\mu(\omega)
\propto
\frac{\gamma_\mu}
{(\omega-\Delta_\mu)^2+\gamma_\mu^2} \,,
\label{eq:Lorentzian_mu}
\end{equation}
with resonance frequency $\Delta_\mu$ and linewidth $\gamma_\mu$.

Within the Floquet picture, the relevant transition frequencies are the quasienergy-resolved channels
$\Delta_{\alpha\beta l}
=
\varepsilon_\beta-\varepsilon_\alpha+l\omega_d$.
If the corresponding channels remain spectrally well separated, one may identify
the following approximation,
\begin{equation}
\Delta_\mu
\simeq
\Delta_{\alpha\beta l},
\end{equation}
so that each spectral peak may be interpreted phenomenologically as originating from an individual Floquet transition. The effective spectral weight is then governed by both the bath spectral density evaluated at the transition frequency, $\Gamma(\Delta_{\alpha\beta l})$, and the corresponding Floquet matrix element:
\begin{equation}
\left|
(s_{\rm B}^{\Lambda+})_{\alpha\beta l}
\right|^2 .
\end{equation}

This picture is physically intuitive, but approximate. In the nonsecular regime, nearby quasienergy channels can couple through the dissipator, so that a measured spectral resonance need not correspond to a single isolated transition.

\paragraph*{Advantages of a Floquet--Liouville description.}

Direct numerical integration of the GMEs provides the density matrix and therefore the observables defined above, but  does not directly reveal how individual spectral features are connected to effective resonance frequencies, linewidths, and dissipative mode weights. A more systematic interpretation is obtained through a dynamical mode decomposition.

For periodically driven systems, however, the Liouvillian itself is explicitly time dependent, preventing a straightforward eigenmode analysis in ordinary Liouville space. The appropriate framework is therefore the FL extended (Sambe) space, in which the time-periodic Liouvillian is mapped onto a time-independent supermatrix. This construction permits the definition of complex eigenmodes, poles, and spectral residues in close analogy with stationary open quantum systems. The complete derivation is presented in App.~\ref{secS:FL}; below we summarize only the main ingredients needed for the discussion of the results.

\subsection{Floquet--Liouville Modal Decomposition}

\paragraph*{Liouville-space formalism.}

Analyzing the structure of the Liouvillian superoperator is 
useful for understanding the dynamics of open quantum systems. The Liouville space is constructed as
\(
\mathbb{H}_{\rm L}
=
\mathbb{H}_{\rm dressed}
\otimes
\mathbb{H}_{\rm dressed}
\),
and is spanned by the basis vectors
\(
\mathsf{B}_{\rm L}
=
\{
\Lket{jk}
\equiv
|j\rangle\langle k|
\}
\).
Its dimension is therefore
\(
{\rm Dim}(\mathbb{H}_{\rm L})
=
{\rm Dim}(\mathbb{H}_{\rm dressed})^2
\).

In this representation, operators acting in the physical (dressed) Hilbert space are vectorized and treated as vectors in Liouville space. For example,
\begin{equation}
\rho(t)
\;\rightarrow\;
\Lket{\rho(t)}
=
{\rm vec}[\rho(t)]
\in
\mathbb{C}^{{\rm Dim}(\mathbb{H}_{\rm L})}, 
\end{equation}
where $\mathbb{C}^{{\rm Dim}(\mathbb{H}_{\rm L})}$ is the vector space of complex continuous functions of dimension ${{\rm Dim}(\mathbb{H}_{\rm L})}$.

Diagonalization of the Liouvillian yields
\(
{\rm Dim}(\mathbb{H}_{\rm L})
\)
dynamical modes, so that the evolution of the open system is governed by the eigenmodes of the superoperator. For a time-independent open quantum system described by
\(
{\dd\rho(t)}/{\dd t}
=
\mathcal{L}\rho(t)
\),
the Liouvillian itself is time independent. Consequently, its modal decomposition provides a complete characterization of the dissipative dynamics and associated observables.

In periodically driven systems, however, the total Liouvillian becomes explicitly time dependent, making direct diagonalization inconvenient. Within the F-GME formulation, both the coherent and dissipative contributions are time periodic. In contrast, within the TI-GME formalism, only the coherent sector remains time dependent, while the dissipator itself is static. In that case, one may still diagonalize the dissipative contribution alone:
\begin{equation}
\mathscr{L}_{\rm diss}
\Lket{R_{{\rm diss},\mu}}
=
-\gamma_\mu
\Lket{R_{{\rm diss},\mu}},
\end{equation}
where $\mathscr{L}_{\rm diss}$ denotes the Liouville-space representation of $\mathcal{L}_{\rm diss}$ and $\gamma_\mu\ge0$ defines an effective dissipative decay rate. The dissipator spectrum naturally separates into population-relaxation modes with $\gamma_\mu=0$ and coherence-relaxation modes with $\gamma_\mu>0$. Nevertheless, this decomposition characterizes only the dissipative sector and does not fully capture the complete driven dynamics of the open system.

In the periodically driven case, the harmonic structure of the Liouvillian implies that the density matrix may be expanded as
\begin{equation}
\rho(t)
=
\sum_n
e^{-\ii n\omega_dt}
\rho_n(t).
\end{equation}

According to Floquet theory, the Liouville-space state can then be written as
\begin{equation}
\Lket{\rho(t)}
=
\sum_\mu
e^{\lambda_\mu t}
\Lket{R_\mu(t)},
\end{equation}
where
\begin{equation}
\Lket{R_\mu(t)}
=
\Lket{R_\mu(t+T)}
\end{equation}
are $T$-periodic Floquet--Liouville modes. These periodic modes admit the Fourier expansion
\begin{equation}
\Lket{R_\mu(t)}
=
\sum_n
e^{-\ii n\omega_dt}
\Lket{R_{\mu,n}}.
\label{eq:LiouvillianMode_FourierExpansion}
\end{equation}

The corresponding left and right Floquet--Liouville eigenvalue equations are
\begin{equation}
\begin{split}
\mathscr{L}^{\rm F}(t)
\Lket{R_\mu(t)}
&=
\lambda_\mu
\Lket{R_\mu(t)},
\\
\Lbra{L_\mu(t)}
\mathscr{L}^{\rm F}(t)
&=
\lambda_\mu
\Lbra{L_\mu(t)},
\end{split}
\label{eq:FL_EVP}
\end{equation}
where $\Lket{R_\mu(t)}$ and $\Lbra{L_\mu(t)}$ denote the right and left Floquet--Liouville modes, respectively, and
\(
\mathscr{L}^{\rm F}(t)
=
\mathscr{L}(t)-\partial_t
\).
The complex eigenvalues are written as
\begin{equation}
\lambda_\mu
=
-\gamma_\mu
-
\ii\Delta_\mu,
\end{equation}
where $\gamma_\mu\ge0$ defines the effective decay rate and $\Delta_\mu$ the oscillation frequency of mode $\mu$. Together, the eigenvalues and eigenmodes provide a complete dynamical characterization of the driven open quantum system.

In their original time-dependent form, however, these eigenvalue equations are generally difficult to solve directly. Owing to the harmonic time dependence of the Liouvillian, one may instead employ the Floquet--Liouville extended (Sambe) space formalism, in close analogy with the standard Floquet treatment of periodically driven Hamiltonians.

\paragraph*{Floquet--Liouville extended-space formalism.}

Thanks to the harmonic time-dependence of the total Liouvillian superoperator, the temporal dependence may be treated as an additional basis degree of freedom through the mapping
$e^{\ii l\omega_dt}
\rightarrow
|l)$.
This enlarges the Liouville space into the tensor-product FL extended (Sambe) space,
$\mathbb{H}^{\rm FL}_{\rm ex}
\equiv
\mathbb{H}_{\rm temp}
\otimes
\mathbb{H}_{\rm L}$,
where $\mathbb{H}_{\rm temp}$ denotes the temporal Fourier space~\cite{Sambe_Steady_1973}.

In this representation~\cite{Ho1986FLSM,Ho1983MMFT,Wang1987FLSM2}, the FL modes become time independent and may be expanded in the extended basis
\(
\FLket{l,jk}
\equiv
|l)\otimes\Lket{jk}
\)
as
\begin{equation}
\begin{split}
\FLket{R_\mu}
&=
\sum_{l\in\mathbb{Z}}
|l)\otimes\Lket{R_{\mu,l}}
=
\sum_{jkl}
R_\mu^{jkl}
\,
\FLket{l,jk},
\\
\FLbra{L_\mu}
&=
\sum_{l\in\mathbb{Z}}
(l|\otimes\Lbra{L_{\mu,l}}
=
\sum_{jkl}
L_\mu^{jkl}
\,
\FLbra{l,jk},
\end{split}
\label{eq:LiouvilleMode_ex}
\end{equation}
with
\(
|l)
\)
the Fourier basis of the temporal space $\mathbb{H}_{\rm temp}$.

Similarly, the Floquet Liouvillian may be represented in the extended basis as the time-independent supermatrix
\(
\mathscr{L}^{\rm F}(t)
\rightarrow
\mathscr{L}^{\rm F}_{\rm ex}
\)~\cite{Restrepo_Driven_2019,Restrepo_Quantum_2018}.
Substituting Eq.~\eqref{eq:LiouvilleMode_ex} into Eq.~\eqref{eq:FL_EVP} then yields the time-independent eigenvalue problem:
\begin{equation}
\begin{split}
\mathscr{L}^{\rm F}_{\rm ex}
\FLket{R_{l\mu}}
&=
\lambda_{l\mu}
\FLket{R_{l\mu}},
\\
\FLbra{L_{l\mu}}
\mathscr{L}^{\rm F}_{\rm ex}
&=
\lambda_{l\mu}
\FLbra{L_{l\mu}},
\end{split}
\label{eq:FL_EVP_ex}
\end{equation}
where $\mathscr{L}^{\rm F}_{\rm ex}$ is the Floquet Liouvillian in the extended space and
$\FLket{R_{l\mu}}$
($\FLbra{L_{l\mu}}$)
are its right (left) eigenvectors. Full derivational details are given in 
App.~\ref{secS:FL}.

Solving Eq.~\eqref{eq:FL_EVP_ex} gives the FL eigenvalues and their associated sideband structures, from which the physical FL modes may be reconstructed~\cite{Eckardt_High-frequency_2015,Restrepo_Quantum_2018,Restrepo_Driven_2019}. Since the extended space contains both temporal and Liouville degrees of freedom, its dimension is enlarged to
${\rm Dim}(\mathbb{H}^{\rm FL}_{\rm ex})
=
{\rm Dim}(\mathbb{H}_{\rm temp})
\times
{\rm Dim}(\mathbb{H}_{\rm L})$.
Accordingly, the extended-space eigenvalue problem produces a correspondingly larger set of eigenvalues and eigenvectors.

However, only
${\rm Dim}(\mathbb{H}_{\rm L})$
linearly independent Floquet modes are physically distinct. After solving the extended-space problem, one finds the replica structure:
\begin{equation}
\lambda_{l\mu}
=
\lambda_\mu
+
\ii l\omega_d,
\qquad
\lambda_\mu
\equiv
\lambda_{0\mu},
\end{equation}
together with the corresponding relation between FL modes,
\begin{equation}
\Lket{R_\mu^{[l]}(t)}
=
e^{-\ii l\omega_dt}
\Lket{R_\mu(t)},
\qquad
\Lket{R_\mu(t)}
\equiv
\Lket{R_\mu^{[0]}(t)}.
\end{equation}

This periodic repetition of the eigenvalues generates Brillouin-zone (BZ) replicas throughout the complex quasifrequency spectrum. In principle, it is therefore sufficient to retain only the set
$\{
\lambda_{l\mu},
\FLket{R_{l\mu}}
\}$
within a single BZ. Translation to another BZ yields a physically equivalent solution with a shifted imaginary part. Nevertheless, in practice it is often advantageous to work directly in the full extended space, both to avoid explicit BZ bookkeeping and to track high-order inter-zone structures and their associated modes in a more more transparent way. Accordingly, throughout this work, we relabel the composite index:
$l\mu\rightarrow\mu$,
such that
$\mu
=
0,\dots,
{\rm Dim}(\mathbb{H}^{\rm FL}_{\rm ex})-1$,
enumerates all modes across the entire FL extended space.

Within this representation, where a fully time-independent diagonalization of the total Liouvillian becomes possible, observables may be expressed directly in terms of FL modal expansions. Consequently, the FL analysis provides detailed dynamical information about the observables beyond what is directly accessible from the density matrix evolution alone.

The FL modal expansion for the time-dependent population in Eq.~\eqref{eq:N_t} involves only the population-sector modes (see App.~\ref{secS:FL} for details). After averaging over one period, only the zero-Fourier component survives,
\begin{equation}
\overline{N}_\Lambda
=
\mathcal{W}_{\mu_0}^{N_\Lambda},
\label{eq:AveN_FL}
\end{equation}
where $\mu_0$ denotes the neutral FL mode satisfying
\(
\lambda_{\mu_0}\simeq0
\),
and
\(
\mathcal{W}_{\mu_0}^{N_\Lambda}
\)
is determined by the corresponding left and right FL eigenvectors together with the projection of the number operator onto the periodic steady state and cycle average.

The  spectrum of interest may be written as
\begin{equation}
\mathsf{S}(\omega)
\propto
{\rm Re}
\left[
\sum_{\mu\notin\mathcal{P}}
\frac{\mathcal{W}_\mu}
{-\lambda_\mu-\ii\omega}
\right],
\label{eq:FL_spectrum_main}
\end{equation}
where $\lambda_\mu$ are the FL eigenvalues and $\mathcal{W}_\mu$ are the associated spectral residues determined by the left and right FL eigenvectors. The sum excludes the neutral steady-family modes associated with the coherent contribution to the emission.

Each term in Eq.~\eqref{eq:FL_spectrum_main} contributes a Lorentzian-like pole to the spectrum. The pole position and linewidth are determined by the complex FL eigenvalue according to
\begin{equation}
\Delta_\mu
=
-\mathrm{Im}[\lambda_\mu],
\qquad
\gamma_\mu
=
-\mathrm{Re}[\lambda_\mu].
\label{eq:omega_gamma_mu}
\end{equation}
Thus, $\Delta_\mu$ gives the resonance frequency associated with mode $\mu$, while $\gamma_\mu$ gives its effective decay rate.

In the closed-system limit (i.e., no loss), the natural Liouville-space basis is formed from coherences between Floquet states, whose oscillation frequencies are set by quasienergy differences. Once dissipation is included, these basis coherences acquire finite lifetimes and, in the nonsecular regime, may become coupled to nearby quasienergy channels (cf. Fig.~\ref{fig:schematics}). The exact FL right eigenmode can therefore be expanded as
\begin{equation}
\FLket{R_\mu}
=
\sum_{\alpha\beta l}
R_\mu^{\alpha\beta l}
\,
|l)
\otimes
|\alpha\rangle\langle\beta|,
\label{eq:Rmu_expand_main}
\end{equation}
where $R_\mu^{\alpha\beta l}$ is the complex amplitude of the quasienergy-resolved channel $(\alpha,\beta,l)$ in the exact mode $\mu$.

To measure the relative participation of each Floquet channel in a given FL mode, we define the normalized weight
\begin{equation}
r_\mu^{\alpha\beta l}
=
\frac{
|R_\mu^{\alpha\beta l}|^2
}{
\sum_{\alpha'\beta'l'}
|R_\mu^{\alpha'\beta'l'}|^2
},
\qquad
\sum_{\alpha\beta l}
r_\mu^{\alpha\beta l}
=
1.
\end{equation}
This quantity identifies whether a given Floquet--Liouville mode is dominated by a single quasienergy transition or instead contains appreciable contributions from several channels.

Accordingly, the index $\mu$ should not generally be interpreted as a single transition label.
Rather, it labels an eigenmode of the full dissipative FL generator, which may involve multiple coupled quasienergy channels. Only when dissipation is weak and the relevant channels are well separated does one recover the approximate identification
\begin{equation}
\mu
\longleftrightarrow
(\alpha,\beta,l),
\end{equation}
with
\begin{equation}
\lambda_\mu
\simeq
-\gamma_{\alpha\beta l}
-
\ii\Delta_{\alpha\beta l}.
\label{eq:lambda_mu_ab_l}
\end{equation}

In this isolated-channel limit, the pole expansion in Eq.~\eqref{eq:FL_spectrum_main} reduces to the phenomenological Lorentzian picture introduced above. More generally, however, the observable spectral resonances are determined by the eigenmodes of the full FL superoperator, rather than by bare quasienergy gaps alone.

For later use, we also define a practical peak-prominence measure, using
\begin{equation}
\mathcal{R}_\mu
\equiv
\frac{
{\rm Re}[\mathcal{W}_\mu]
}{
\gamma_\mu
}.
\end{equation}
Together with $\Delta_\mu$, $\gamma_\mu$, and the channel weights $r_\mu^{\alpha\beta l}$, this quantity allows us to identify the dominant spectral poles and determine whether a given peak is primarily a single-channel resonance or a mixed FL mode.
For strongly overlapping modes, an individual $\mathcal{R}_\mu$
 should be regarded as a modal-prominence diagnostic rather than an independent peak height, since the observed lineshape results from the coherent sum of the neighboring complex pole contributions (see App.~\ref{secS:FL} for details).

Finally, the same FL construction can be applied to the TI-GME. The difference is that the dissipative part of the TI-GME has only a zeroth Fourier component,
\begin{equation}
\mathcal{L}_0
=
\mathcal{L}_{\rm diss},
\qquad
\mathcal{L}_{n\neq0}=0,
\end{equation}
with $\mathcal{L}_{\rm diss}$ given in Eq.~\eqref{eq:Ldiss_TIGME}. The resulting poles and modal structures can therefore be compared directly with those obtained from the Floquet-consistent F-GME.

In general, in the fully dissipative analysis where we consider the master equation approaches, the master equation is solved via two methods (solvers): (1) the direct numerical integration where the partial differential equation is solved via a numerical approach such as RK4 (here, we use the QuTiP's $\mathsf{mesolve}$ solver~\cite{Johansson_Qutip_2012,johansson2013qutip,qutip5}), and (2) the FL super-matrix approach which utilizes the time-evolution propagator. Indeed, both solvers predict the same result as they must because they solve the same master equation.
For the unitary dynamics, the results of the phenomenological approach also match those of the QuTiP's Schr{\"o}dinger equation solver ($\mathsf{sesolve}$~\cite{qutip5}).

\section{Driven-dissipative Two Level System}
\label{sec:TLS}
The TLS constitutes one of the most fundamental building blocks of quantum optics and quantum information science, serving as a universal quantum interface between light and matter~\cite{Scully_Quantum_1999,Haroche_Cavity_1993}. Physically, TLSs can be realized in a wide variety of platforms, ranging from natural atomic systems to artificial quantum emitters such as semiconductor quantum dots, color centers, and superconducting qubits in circuit-QED architectures~\cite{Mabuchi_Cavity-QuantumElectrodynamics_2002,Wallraff_Strong_2004,Blais_Circuit_2021}. Coupled to optical or microwave resonators, they provide minimal yet remarkably powerful models for investigating cavity-QED phenomena, waveguide-QED, and nonequilibrium light--matter interactions.

Two level system platforms also underpin numerous quantum technologies, including quantum logic operations, quantum networking, and nonclassical light generation such as deterministic single-photon sources~\cite{Kimble_Quantum_2008,Lodahl_Chiral_2017}. 
In real systems, however, microscopic TLS 
inevitably couple to dissipation channels,
which limits coherence times~\cite{Martinis_Decoherence_2005,Muller_Towards_2019}.

The dissipative dynamics of periodically driven quantum
systems has been investigated extensively over the past
several decades. Early reviews established the foundations
of Floquet theory and Floquet transition rates in driven
open systems~\cite{Grifoni_Driven_1998}, while subsequent
works developed Floquet descriptions of transport
phenomena~\cite{Kohler_Driven_2005}, dissipative
Floquet--Van Vleck theory~\cite{Hausinger_Dissipative_2010},
Floquet--Gibbs states and thermalization~\cite{Shirai_Condition_2015},
and Floquet master equations formulated in extended
space~\cite{Restrepo_Driven_2019,Restrepo_Quantum_2018}.

Despite these important developments, a systematic
understanding of when time-independent-basis generalized
master equations and fully quasienergy-resolved Floquet
master equations yield equivalent predictions remains
largely absent. In particular, previous studies have not
explicitly identified the failure modes of time-independent-basis
dissipators, nor clarified whether discrepancies originate
from secularization, bath structure, internal dressing, or
the neglect of Floquet-resolved dissipative channels.
Moreover, the consequences of these differences for
steady-state populations have received considerably less
attention than their effects on spectra and transport
observables.
Here, we addresses this gap through a controlled
comparison between TI-GME and F-GME approaches in
minimal driven quantum systems, so that these discrepancies manifest directly in
steady-state populations rather than only in transient dynamics.

For a single TLS, 
the time-independent Hamiltonian in Eq.~\eqref{eq:Ht} is simply
 \begin{equation}
     H_0= \frac{\omega_a}{2}\sigma_z,
 \end{equation}
with $\omega_a$ being the TLS's transition frequency and $\sigma_i$ ($i=x,y,z$) represent the Pauli matrices, for the single TLS case.
We also choose $S_d^{\rm bare}=S_{\rm B}^{\rm bare}=\sigma_x$ so that $S_d=S_{\rm B}=s^++s^- \equiv \sigma^- + \sigma^+ \equiv \sigma_x$.

In this TLS case, the bare-state basis and the dressed-state basis coincide, i.e., $E_j \in \{E_g,E_e\}$ and $\ket{j} \in \{\ket{g},\ket{e}\}$ see Fig.~\ref{fig:schematics}(a). In contrast, in hybrid quantum systems with finite inter-subsystem coupling---such as coupled TLSs or cavity-QED systems in the ultrastrong-coupling regime---the interaction induces hybridization of the bare states, causing the dressed-state basis to differ from the bare-state basis [compare the static basis set to form the Floquet representation in Fig.~\ref{fig:schematics}(a) with that of Fig.~\ref{fig:schematics}(b)]. Consequently, the system dynamics and dissipation must be formulated in the dressed-state basis.

In this section, we investigate the dynamics of a driven-dissipative TLS and focus on steady-state observables, particularly the long-time averaged excitation number and the 
emission spectrum. This driven-TLS example is well studied in the literature~\cite{Mollow_Power_1969,Zhou_Resonance_1998,Yan_Resonance_2016,Alicki_PeriodicallyDriven_2013,Chen_FloquetOpen_2024,Mori_FloquetOpenSystems_2023,Restrepo_Driven_2014,Restrepo_Quantum_2018,Sambe_Steady_1973,Liu_Auxiliary_2021,Ruan_ResonanceFluorescence_2024,Moelbjerg_BeyondMollow_2012}, and can help to establish the reliability of the predictions. 

Popular and widely used forms of the bath function include the flat bath, Ohmic family, and Lorentzian models~\cite{Hughes_Phonon-mediated_2013,Wang_Dynamical_2019,Salmon_Gauge-independent_2022,Akbari_Generalized_2023}. 
For the TLS, for our purposes, we first examine the flat bath model with 
\begin{equation}
    \begin{split}
\Gamma_{\rm flat}(\omega)&=
\frac{\gamma}{2}.
    \end{split}
    \label{eq:BathModel_flat}
\end{equation}

Additionally, we investigate the effect of an structured bath, to make our study more complete. For this purpose, the Lorentzian bath model is a good choice~\cite{Picatoste_Dynamically_2024}. However, to avoid numerical instabilities in the super low or high frequency range, we add a background rate 
which is also there is realistic experiments, 
for example in the case of a cavity resonance in a 
 photonic band gap structures~\cite{Kristensen_Decay_2011,Devashish_Three-dimensional_2019}:
\begin{equation}
    \begin{split}
\Gamma_{\rm Lor}(\omega)=\Gamma_{\rm max}\frac{(\gamma/2)^2}{(\omega-\omega_0)^2+(\gamma/2)^2}+\frac{\gamma_{b}}{2},
    \end{split}
    \label{eq:BathModel_Lorentzian}
\end{equation}
where $\Gamma_{\rm max}=\gamma/2$ to have the HWHM of $\gamma/2$ (FWHM of $\gamma$) at $\omega=\omega_0\pm\gamma/2$, i.e., $\Gamma_{\rm Lor}(\omega_0\pm\gamma/2)=\gamma/2$, for the pure Lorentzian part excluding the background. Indeed, the choice of $\gamma_b$ is problem-dependent, and setting $\gamma_b=0$ gives a pure Lorentzian (and there is no numerical issue). In practical cases, often, $\gamma_b\neq0$, for example, $\gamma_{b}=0.1\gamma$ in the realistic photonic band gap reservoirs. 
For the sake of better comparison with the flat bath case and see how individual Floquet dissipative pathways can influence the dynamics, we use $\gamma_b=\gamma$ and conveniently set the Lorentzian peak on a Floquet sideband. Thus, except for that chosen sideband, every other dissipative channel feels almost the same decay rate as in the flat bath situation.

\begin{figure*}[!htpb]
\centering
\includegraphics[width=.75\linewidth]
{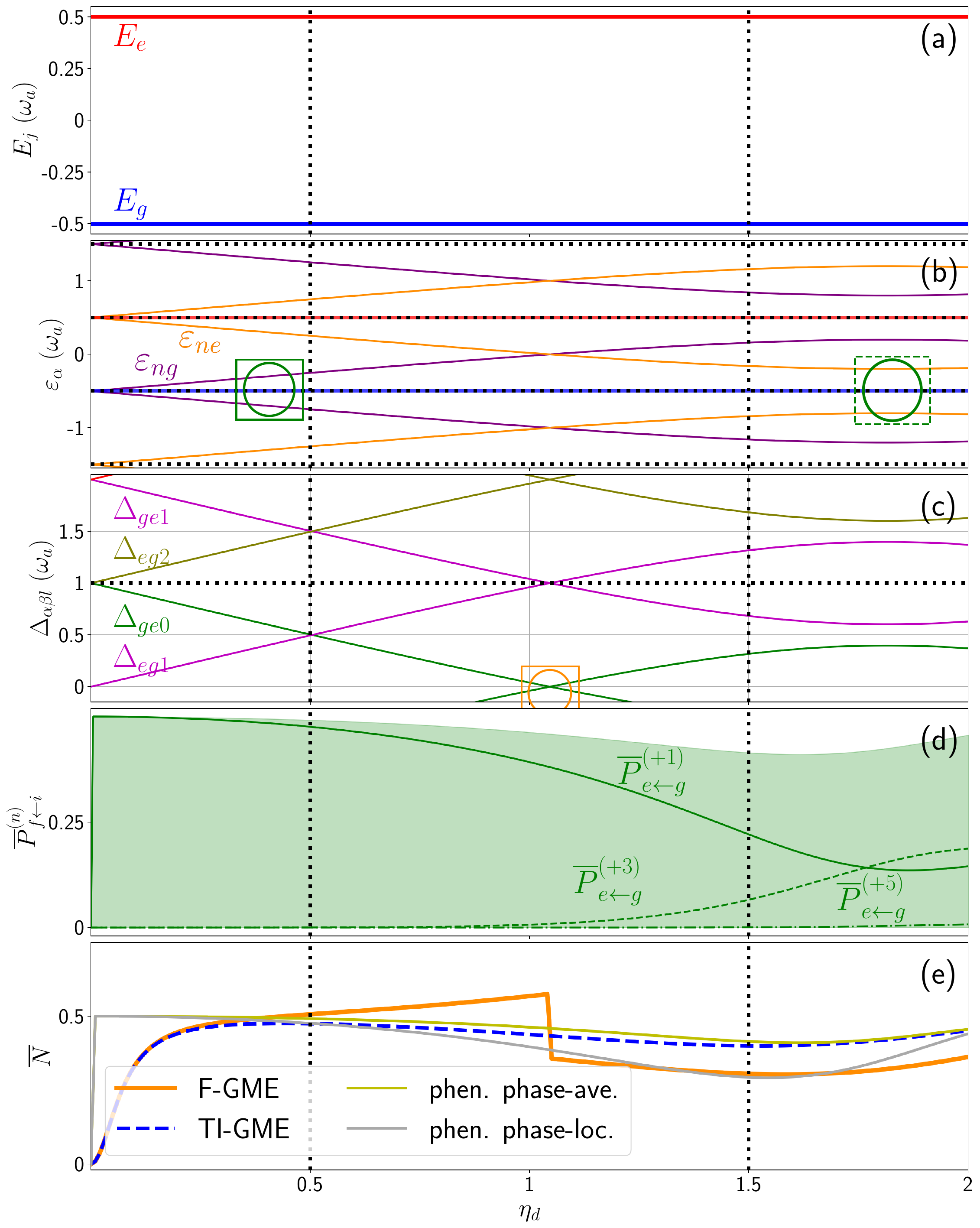} 
\caption[]{\textbf{Driven-dissipative TLS: drive-amplitude dependence of the quantum processes.}
(a) Static energy levels $E_g$ and $E_e$ of the undriven TLS. 
(b) Floquet quasienergies $\varepsilon_\alpha$ in the primary BZ. 
(c) Quasienergy-resolved transition frequencies $\Delta_{\alpha\beta l}=\varepsilon_\beta-\varepsilon_\alpha+l\omega_d$. 
(d) Long-time averaged closed-system transition probabilities $\overline{P}^{(n)}_{e\leftarrow g}$ for odd drive-assisted processes. 
(e) Long-time averaged excitation number $\overline{N}$ for $\gamma=0.1\omega_a$, comparing the F-GME, TI-GME, and phenomenological Floquet treatment. 
The horizontal black dotted lines in panels (a)--(c) indicate the BZ energy scale, while the vertical black dotted lines mark representative drive amplitudes analyzed later. In all panels, $\omega_d=\omega_a$, and we use nonsecular full form of dissipators in (e). 
}
\label{fig:TLS_etad}
\end{figure*}

\subsection{Numerical results with different drive amplitudes}

Figure~\ref{fig:TLS_etad} shows the drive-amplitude dependence of a resonantly driven dissipative TLS, with $\omega_d=\omega_a$, and compares the TI-GME (which, for a single TLS, is the same formalism as the TI-ME) and F-GME (which, for a single TLS, is the same formalism as the F-ME) predictions as the dimensionless drive amplitude $\eta_d$ is varied.

The basic structure is already visible at the closed-system level. In the absence of coherent driving, the relevant energies are simply the static TLS levels $E_g$ and $E_e$, and the  transition frequency remains fixed [see Fig.~\ref{fig:TLS_etad}(a)], since no drive is included. Under periodic driving, however, the appropriate energies are the Floquet quasienergies $\varepsilon_\alpha$, defined within a Brillouin zone [see Fig.~\ref{fig:TLS_etad}(b)]. The corresponding transition frequencies are no longer fixed resonance gaps $\omega_{jk}$, but quasienergy-resolved channels $\Delta_{\alpha\beta l}$
where $l$ labels the exchange of drive quanta. These channels reorganize strongly with increasing $\eta_d$, as shown in Fig.~\ref{fig:TLS_etad}(c).

A particularly important regime occurs near $\eta_d\simeq 1.05$, where several relevant quasienergy channels cross or become nearly degenerate. Around this field value, the Floquet channels $\Delta_{eg1}$ and $\Delta_{ge1}$ approach the same frequency scale as the static transition used in the TI-GME. This produces a {\it crossing} in the long-time excitation predicted by the two approaches [Fig.~\ref{fig:TLS_etad}(e)]. However, at this point, the quasienergy curves may also become nonpositive
 [orange indicator/circles in panel (c)], so that away from this crossing, the F-GME exhibits a sharp cusp-like feature, reflecting a rapid switching of the dominant quasienergy-resolved dissipative pathway through a discontinuity. This behavior is absent in the TI-GME because its dissipator remains tied to the static system transition.

The closed-system transition probabilities in Fig.~\ref{fig:TLS_etad}(d) provide complementary insight. At weak drive values, the dynamics is dominated by the fundamental one-photon process, denoted by $\overline{P}^{(+1)}_{e\leftarrow g}$, connected to the one-photon-assisted anticrossing [solid green indicator/cicles in panel (b)]. As $\eta_d$ increases, higher-order odd processes, especially $\overline{P}^{(+3)}_{e\leftarrow g}$ connected to the three-photon-assisted anticrossing [dashed green indicator in panel (b)] and eventually $\overline{P}^{(+5)}_{e\leftarrow g}$, acquire visible weight. This shows that, even before dissipation is included, the driven TLS is no longer governed by a single transition channel. Instead, the excitation probability is redistributed among several Floquet-assisted pathways.

The dissipative influence of this redistribution is shown in Fig.~\ref{fig:TLS_etad}(e), where we plot the long-time averaged excitation number $\overline{N}$ obtained from the phenomenological Floquet treatment 
(phase-locked: solid gray, phase-averaged: solid olive), the TI-GME (dashed blue), and the F-GME (solid orange). At small $\eta_d$, where one Floquet channel dominates, the TI-GME and F-GME agree well and follow the same general trend as the phenomenological Floquet result. As the drive amplitude increases, several quasienergy channels contribute simultaneously, and the two master equations separate. The origin of this discrepancy is structural: the TI-GME resolves dissipation using the static dressed-state transition frequency, whereas the F-GME resolves dissipation through the full set of Floquet channels $\Delta_{\alpha\beta l}$.

Note the phenomenological phase-locked results are highly sensitive to the initial phase. 
However, the phenomenological phase-averaged 
and GMEs' results are initial-condition-invariant; hence these results are expected to follow the same general trend, expect for the effects of the dissipator. 
For example, if the drive is more non-adiabatic, or if we simply add a $\phi$-phase to the drive, the phenomenological phase-locked results can acquire a three-photon-assisted peak for $\eta_d\gtrsim1.5$, while the results of other methods remain unchanged [cf. Fig.~\ref{figS:TLSRWA_etad} in App.~\ref{secS:AdditionalResults}].

Two features of the F-GME result in Fig.~\ref{fig:TLS_etad}(e)
merit further commentary: (i) the cycle-averaged excitation exceeding
$0.5$ and (ii) the sharp cusp near $\eta_d\simeq1.05$.
First, $\overline{N}>0.5$ is not intrinsically unphysical, but
corresponds to population inversion in the bare-energy basis,
\begin{equation*}
    \overline{N}
    =\frac{1}{T}\int_0^T dt\,
    \mathrm{Tr}\!\left[\rho^{\rm ss}(t)\ket{e}\bra{e}\right]
    >\frac{1}{2},
\end{equation*}
with the physical bound $0\leq\overline{N}\leq1$.
Population inversion in coherently driven dissipative two-level
systems has been demonstrated previously through bath-assisted
mechanisms (drive-bath interplay)~\cite{PhysRevLett.95.106801,Stace_DynamicalSteadyStates_2013}.
The familiar saturation value $\rho_{ee}\rightarrow1/2$ instead
applies to the standard resonantly driven optical-Bloch model with
the drive-RWA and ordinary bare-frequency spontaneous emission.

The F-GME considered here differs because both the
counter-rotating drive processes and the quasienergy-resolved
dissipative channels are retained. A physical bath transition occurs
at $\Delta_{\alpha\beta l}
    =\varepsilon_\beta-\varepsilon_\alpha+l\omega_d$,
so that $\Delta_{\alpha\beta l}>0$ need not correspond to a downward
transition according to either the primary-BZ quasienergy ordering
or the bare-state ordering; energy exchanged with the periodic drive
through $l\omega_d$ also contributes to the net bath-transition
energy. Consequently, zero-temperature relaxation between Floquet
states need not impose the equilibrium detailed-balance relation on
the bare-state populations. The drive supplies the energy while the
bath removes energy in the periodic steady state, allowing the
dissipation-selected Floquet distribution to yield
$\overline{N}>1/2$. Importantly, bare-state inversion does not require
an inverted population distribution among the Floquet states
themselves.

In other words, for the present
model and parameter regime, this is a genuine beyond-drive-RWA
feature rather than a numerical artifact. The full drive retains the
counter-rotating terms, which activate higher-order multiphoton
excitation pathways absent in the drive-RWA treatment. Within the
latter, the steady-state excitation approaches the usual saturation
value of $0.5$ [Eq.~\eqref{eq:rho_ee_RWA}], whereas the full
non-RWA dynamics permits a cycle-averaged bare-state inversion,
$\overline N>0.5$. This occurs in the deep-strong-drive regime,
$\Omega_d\gtrsim\omega_d$, where higher-order processes become
appreciable, consistent with the activation of the
$P^{(3,5)}_{e\leftarrow g}$ pathways identified in
Fig.~\ref{fig:TLS_etad}(d).

As a numerical consistency check, we have 
also verified throughout the
drive period that the periodic steady-state density matrix remains
positive semidefinite within numerical precision.

The feature seen at $\eta_d\approx1.05$ for the orange curve of the F-GME population in panel (e) of Fig.~\ref{fig:TLS_etad} points out the existence of a possible discontinuity in the Floquet dissipator. All the ingredients in the F-GME dissipator (the density matrix, system-bath coupling operator matrix elements and the bath function) are, in principle, continuous (analytical) functions of their argument, except at $T=0$ and where $\Delta$ passes zero as we let $\Gamma(\Delta)=0$ for $\Delta\leq0$. Near this particular point, due to the nonanalytical sudden jump in the bath function, one observes a sharp cusp-like discontinuity. The lower and higher neighborhood of this point is also a clear implication of the Floquet sidebands having a competing effect in the dissipator.
At this exact point of sudden reduction, the quasienergy lines cross, yielding an overlap between the Floquet channels $\Delta_{\alpha\beta l}$ with the same order $l$. The two most dominant channels here are $\Delta_{eg0}$ and $\Delta_{ge0}$ that compete, and the dissipative relaxation switch between these two pathways dominantly, then $\Delta_{eg1}$ and $\Delta_{ge1}$. 

In other words, the dominant Floquet relaxation pathway changes when a relevant channel crosses zero net transfer frequency. This pathway reorganization is physical, whereas the sharpness of the cusp reflects the nonanalytic zero-temperature prescription $\Gamma(\Delta\leq0)=0$.
We have verified that replacing the flat reservoir by an Ohmic spectral density, $\Gamma(\omega)=(\gamma/2)(\omega/\omega_a)\Theta(\omega)$, removes the sharp cusp while preserving the underlying Floquet-channel exchange.

As shown in Fig.~\ref{fig:TLS_etad}(e), the F-GME prediction can
exceed both phenomenological Floquet results. This is not
counterintuitive once the different steady-state constructions are
recognized. The phenomenological treatments describe the isolated
driven system and retain Floquet-state weights fixed by the initial
condition (or by its phase average). For example, the phase-locked
result is
\begin{equation*}
\overline{N}^{(\mathrm{phase\text{-}locked})}
=
\sum_{\alpha}
\left|
\langle\alpha(0)|\psi_0\rangle
\right|^2
\overline{n}_{\alpha},
\end{equation*}
with an analogous expression for the phase-averaged treatment.

In contrast, the F-GME determines the periodic steady state
dynamically through bath-induced transitions between Floquet
channels. Writing
$\rho^{\rm ss}(t)
=
\sum_{\alpha\beta}
\rho_{\alpha\beta}^{\rm ss}(t)
|\alpha(t)\rangle\langle\beta(t)|$,
the exact cycle-averaged excitation is
\begin{equation}
\overline{N}^{(\mathrm{F-GME})}
=
\frac{1}{T}\int_0^T dt
\sum_{\alpha}
\rho_{\alpha\alpha}^{\rm ss}(t)
N_{\alpha\alpha}(t)
+
\overline{N}_{\rm coherences},
\label{eq:FGME_population_decomposition}
\end{equation}
where
\begin{equation*}
\overline{N}_{\rm coherences}
=
\frac{1}{T}\int_0^T dt
\sum_{\alpha\neq\beta}
\rho_{\alpha\beta}^{\rm ss}(t)
\langle\beta(t)|N|\alpha(t)\rangle.
\end{equation*}

The diagonal steady-state populations are therefore not fixed by the
initial Floquet decomposition, but are selected by dissipative
transitions whose rates depend on
\begin{equation*}
\Gamma(\Delta_{\alpha\beta l}),
\qquad
\left|S_{B,\alpha\beta l}\right|^2,
\qquad
\Delta_{\alpha\beta l}
=
\varepsilon_\beta-\varepsilon_\alpha+l\omega_d .
\end{equation*}
Since different Floquet states carry different bare-excitation
content, this bath-induced redistribution can stabilize a periodic
steady state with a larger cycle-averaged excitation than either
isolated-system phenomenological prediction. Nonsecular steady-state
coherences may provide an additional contribution through
$\overline{N}_{\rm coherences}$.

At zero bath temperature this enhancement does not imply that the
bath supplies energy. The periodic drive remains the energy source,
while the bath redistributes the Floquet-state occupations and removes
energy from the driven system. In the periodic steady state,
\begin{equation*}
\overline{P}_{\rm drive}
+
\overline{J}_{\rm bath}
=0,
\qquad
\overline{P}_{\rm drive}>0,
\qquad
\overline{J}_{\rm bath}<0,
\label{eq:steady_energy_balance}
\end{equation*}
i.e., cycle-averaged power supplied by the external modulation cancels that of from the bath into the system.
Thus, the F-GME exceeding the phenomenological predictions reflects
drive-powered, dissipation-assisted redistribution among Floquet
channels rather than excitation supplied by the bath.

The distinction between the TI-GME and F-GME results remains important even for a flat bath. Although a flat bath removes frequency selectivity from $\Gamma(\omega)$, it does not remove the drive-dependent redistribution of system--bath matrix elements among Floquet sidebands. Thus, the F-GME can predict a different steady-state balance of excitation and relaxation from the TI-GME, even when the bath spectral density is structureless and the system--bath coupling is weak. This is a fundamental consequence when the external dressing is not properly integrated on an equal footing with the dissipation. 
Thus, it is predictable that while the secular approximation has no consequence here for the flat bath of the TI-GME results, it is detrimental for the F-GME and can lead to unphysical results as the competing dissipative channels emerge rather than a sole dissipative channel in the TI-GME theory 
(cf.~Fig.~\ref{figS:TLSRWA_etad} in App.~\ref{secS:AdditionalResults}).

The physical reason for the discrepancy in the TI-GME and F-GME results in Fig.~\ref{fig:TLS_etad}(e) is therefore more specific than the statement that Floquet sidebands exist. Both the TI-GME and F-GME include the same time-dependent Hamiltonian and hence the same coherent Floquet sideband structure. Their difference lies in the dissipator. The TI-GME collapses dissipative transitions onto static eigenfrequencies, whereas the F-GME retains the quasienergy-resolved dissipative channels. Consequently, the TI-GME is reliable only when a single Floquet channel dominates or when the relevant channels are well separated and weakly mixed. Once multiple channels become comparable or nearly degenerate, the steady state becomes intrinsically quasienergy-resolved, and the Floquet-consistent dissipator is required.

Figure~\ref{fig:TLS_etad} thus demonstrates that nonsecularity alone is not sufficient since both GMEs are nonsecular: the dissipator must also be decomposed at the correct transition frequencies.
In the TI-GME, nonsecular terms between static dressed-state 
transitions---governed by $S^{\pm}_{\rm B}(\omega)=S^{\pm}_{{\rm B},jk}\ket{j}\bra{k}$---are retained, but the dissipative channels remain tied to the undriven spectrum. In the F-GME, the same Born--Markov logic is applied after resolving the system into Floquet quasienergy channels so that the nonsecularity is promoted to the sideband interactions governed by $S^{\pm}_{\rm B}(\Delta)=S^{\pm}_{{\rm B},\alpha\beta l}\ket{\alpha}\bra{\beta}$. This difference is enough, even with the same equal decay rate $\gamma$, to produce qualitatively different steady states in a minimal driven TLS, even
within a flat-bath coupling regime.

The dependence on the dissipation rate should also be interpreted carefully. Although one might expect damping primarily to control the transient relaxation toward the steady state, in a driven open system the steady state itself is determined by the balance of excitation and relaxation pathways. Thus, even at zero temperature, the combination of coherent driving and dissipation produces a nonequilibrium steady state whose populations depend on how the dissipative channels are resolved. In the present parameter regime, the F-GME steady-state population is comparatively weakly dependent on the overall scale of $\gamma$, while the TI-GME shows a stronger dependence, reflecting its 
{\it incorrect redistribution} of dissipative weight.

These results expose the failure of TI-based dissipation directly at the level of a steady-state observable. The agreement between the TI-GME and F-GME appears to be valid only  when $\Omega_d/\omega_d$ is sufficiently small,  where the drive-RWA is valid; namely,  when $\eta_d\gtrsim0.1\omega_d/\omega_a$, Floquet theory is needed.
It seems that the failure of the conventional time-independent basis master equations is directly related to the emergence of higher-order quantum processes assisted by the strong drive where the Floquet sidebands obtain significant weight in the dissipation pathways. However, these sidebands may also be washed away by the effect of the higher dissipation linewidths. 

Consequently, one must also consider the role of $\gamma$ dissipation in the distinction 
between the F-GME and TI-GME results, which we investigate more in the spectral analysis. 
Earlier Floquet master-equation studies focus on spectral features or employ quasienergy-secular approximations~\cite{Grifoni_Driven_1998,Kohler_Driven_2005,Hausinger_Dissipative_2010,Shirai_Condition_2015,Restrepo_Quantum_2018}. However,
the minimal driven TLS considered here already shows that a time-independent dissipator can yield {\it qualitatively incorrect long-time populations} because it does not respect quasienergy-resolved dissipation, i.e., the dissipative pathways must also be considered to be externally dressed.

To resolve the population dynamics and the underlying dissipative channels more precisely, it is instructive to examine the de-excitation (emission) spectrum. The spectral structure provides a more detailed microscopic view of the discrepancy between the long-time averaged excitation predicted by the TI-GME and F-GME approaches, and clarifies why the TI-GME can in different parameter regimes either overestimate or underestimate the Floquet-consistent result.

When operating away from resonance, the relevant dissipative channels are generally centered on the drive-dressed transition frequencies rather than solely on the bare Bohr frequency, further restricting the validity of a static dissipator.

\subsection{Spectrally resolved results}
\begin{figure*}[!htpb]
\centering
\includegraphics[width=.8\linewidth]
{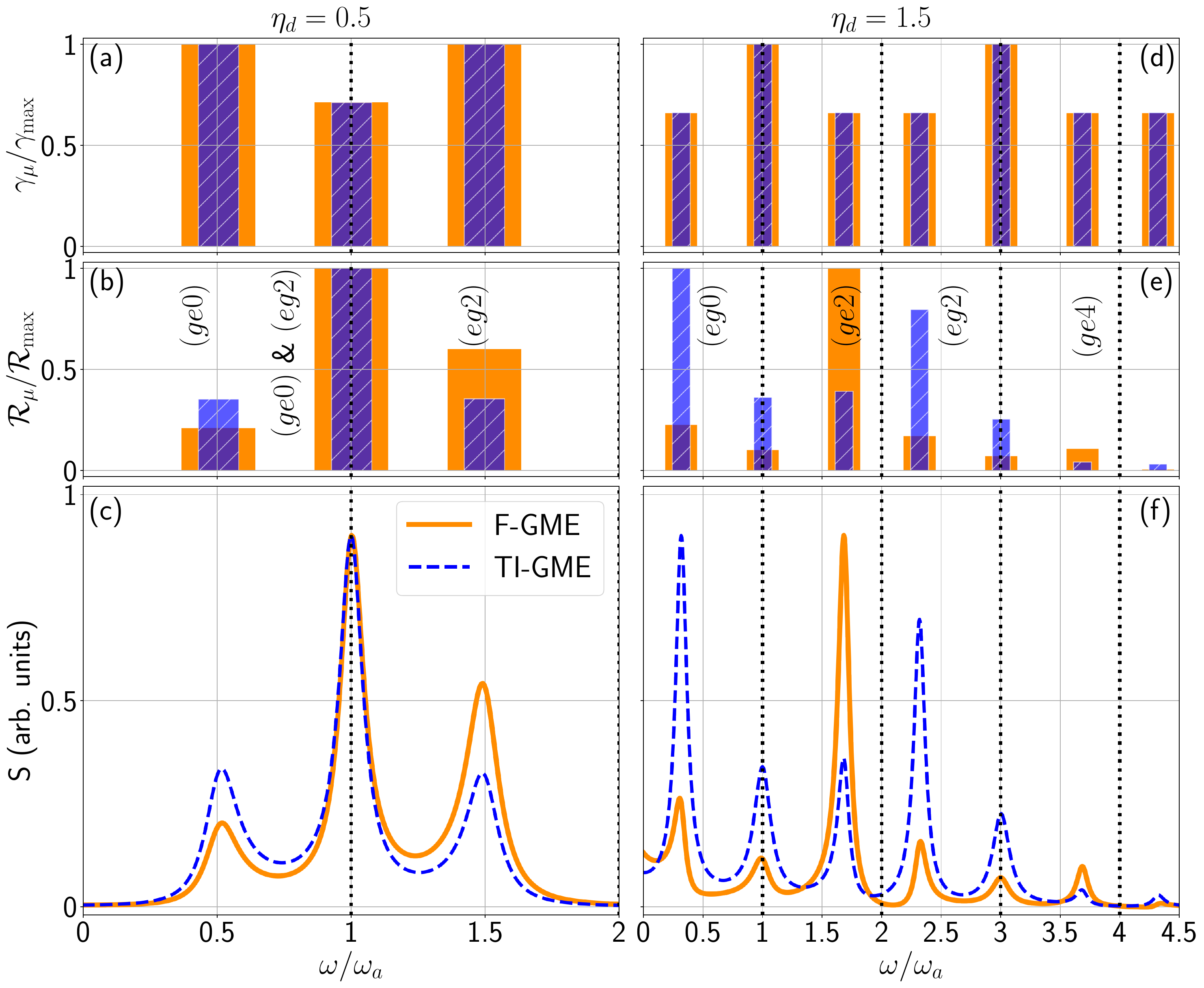}
\caption[]{\textbf{Driven-dissipative TLS: incoherent spectra.}
Comparison of the incoherent emission spectra obtained from the F-GME (solid orange) and TI-GME (dashed blue) for a resonantly driven TLS, $\omega_d=\omega_a$, coupled to a flat bath with $\gamma=0.1\omega_a$. The left column corresponds to $\eta_d=0.5$, and the right column to $\eta_d=1.5$. Panels (c) and (f) show the spectra. Panels (a),(d) show the normalized effective linewidths $\gamma_\mu/\gamma_{\max}$ of the dominant FL modes, while panels (b),(e) show the corresponding normalized peak-strength measure $\mathcal{R}_\mu/\mathcal{R}_{\max}$, with $\mathcal{R}_\mu=\mathrm{Re}(\mathcal{W}_\mu)/\gamma_\mu$ (see text). The orange and blue bars use the same convention as the spectra: thick solid bars for the F-GME and thinner dashed bars for the TI-GME. The labels indicate the dominant FL modes, and the vertical black dotted lines mark integer multiples of the drive frequency. The nonsecular full forms of dissipators are considered across the calculation.
}
\label{fig:TLSSpectra_wd1_Flat}
\end{figure*}
Figure~\ref{fig:TLSSpectra_wd1_Flat} compares the incoherent spectra of the resonantly driven TLS for two representative drive amplitudes, $\eta_d=0.5$ (moderately strong) and $\eta_d=1.5$ (deep strong).
These values correspond to the regimes marked by the vertical dotted black lines in Fig.~\ref{fig:TLS_etad}: a moderate large-drive regime where the TI-GME and F-GME give similar steady-state populations; and a more extreme deep-strong-drive regime, where the
two model predictions differ substantially.
At the bottom panels of Fig.~\ref{fig:TLSSpectra_wd1_Flat}, we show the spectra, whereas in the middle and top inset panel, we show their corresponding spectral individual peak amplitude by the bar plots of the normalized mode prominence $\mathcal{R}_\mu$ and normalized mode linewidth $\gamma_\mu$, respectively.

At $\eta_d=0.5$, both approaches reproduce the familiar Mollow-triplet-like spectral structure. This is not surprising, as the coherent part of the dynamics is governed by the same driven Hamiltonian in both theories, so the dominant spectral locations are largely determined by the same Floquet quasienergy structure. The key difference is instead in the dissipator. The TI-GME treats dissipation through static dressed-state transitions, whereas the F-GME resolves the bath coupling through quasienergy-assisted channels. As a result, spectra that look qualitatively similar can still correspond to different linewidths, residues, and steady-state populations, yielding noticeably differences such as 
spectral asymmetries.

To be more specific,
for the drive amplitude of $\eta_d=0.5$ [Figs.~\ref{fig:TLSSpectra_wd1_Flat}(a)--(c)], the spectrum is dominated by a Mollow-triplet structure at 
\begin{equation*}
\omega/\omega_a=\Delta_\mu/\omega_a\simeq 0.505,\qquad 1,\qquad 1.495,
\end{equation*}
with corresponding effective linewidths
\begin{equation*}
\gamma_\mu/\omega_a\simeq 0.073,\qquad 0.050,\qquad 0.073,
\end{equation*}
respectively. The modal decomposition shown in the bar plots reveals that the two side peaks near $0.505\,\omega_a$ and $1.495\,\omega_a$ are essentially dominated by single Floquet sideband channels (pure mode resonances). In the F-GME description, these resonances are primarily associated with the quasienergy-resolved transitions $(g,e,0)$ and $(e,g,2)$, respectively, each carrying nearly unit channel weight.
The normalized mode prominence factor $\mathcal{R}_\mu/\mathcal{R}_{max}$ in panel (b) clarifies the asymmetry/symmetry of the Mollow triplet in the F-GME/TI-GME prediction.

The central peak at $\omega=\omega_d$ exhibits a more intricate structure. Its dominant FL mode has 
the eigenvalue\footnote{Recall $ \lambda_\mu =  -\gamma_\mu - \ii \Delta_\mu $, as we define the total
Liouvillian from
$\dot{\rho}= \mathcal{L}\rho$.} 
\begin{equation*}
\lambda_\mu/\omega_a\simeq -0.050-\ii,
\end{equation*}
corresponding to a linewidth $\gamma_\mu\simeq0.050\,\omega_a$ and resonance frequency $\Delta_\mu=\omega_a$. The associated right eigenmode $\FLket{R_\mu}$ is not aligned with a single Floquet transition template, but is instead composed almost equally of the two channels $(g,e,0)$ and $(e,g,2)$, with normalized weights:
\begin{equation*}
r_\mu^{(g,e,0)}=
r_\mu^{(e,g,2)}\simeq0.499.
\end{equation*}

The corresponding Hamiltonian--Floquet transition frequencies are
\begin{equation*}
\Delta_{ge0}\simeq0.5041\,\omega_a,
\qquad
\Delta_{eg2}\simeq1.4959\,\omega_a,
\end{equation*}
whereas the observed Liouvillian resonance occurs at the intermediate value $\Delta_\mu=\omega_a$.

These results should not be interpreted as implying that the central Mollow line originates exclusively from nonsecular coupling, since a central peak at the drive frequency also emerges in conventional secular treatments of resonance fluorescence. Rather, the present FL analysis shows that, within the full nonsecular dissipative generator, the physical central resonance is carried by a collective Liouvillian eigenmode whose weight coming from the hybridization of the modalweights is distributed over multiple quasienergy transition channels. Nonsecular coupling therefore governs the detailed modal composition, linewidth renormalization, and spectral weight of the peak.

This interpretation also explains why the central resonance is simultaneously the narrowest and brightest feature of the triplet. Its reduced decay rate together with its large residue-to-linewidth ratio,
$\mathcal{R}_\mu=\mathrm{Re}(\mathcal{W}_\mu)/\gamma_\mu$,
makes it the dominant spectral contribution. More generally, the measured spectrum is determined by the eigenmodes of the full Floquet--Liouville generator rather than by isolated quasienergy transitions alone.

This result connects with the familiar resonance-fluorescence
spectrum of a coherently driven TLS derived by
Mollow~\cite{Mollow_Power_1969}, while the present FL/F-GME
formulation extends that picture by retaining the full periodic drive
and a nonsecular, Floquet-resolved dissipative description. The
drive-RWA and the leading Bloch--Siegert (BS) expansion are controlled by
the small parameter
\begin{equation}
\frac{\Omega_d}{\omega_a+\omega_d}
\ll 1.
\end{equation}
Accordingly, $\eta_d\lesssim0.1$ may be regarded as a conservative
perturbative regime rather than a sharp validity boundary. At the
moderate driving strength $\eta_d=0.5$ considered below, the
lowest-order BS correction provides only a qualitative
estimate, whereas our exact Floquet quasienergies and FL poles provide
the quantitatively reliable resonance positions.

In the drive-RWA limit, the periodically driven TLS Hamiltonian
\begin{equation}
H(t)=\frac{\omega_a}{2}\sigma_z
+\Omega_d\sigma_x\sin(\omega_dt+\phi_d)
\end{equation}
is approximated by
\begin{equation}
H_{\rm RWA}(t)
=
\frac{\omega_a}{2}\sigma_z
+
\frac{\Omega_d}{2\ii}
\left(
\sigma^-\ee^{\ii(\omega_dt+\phi_d)}
-
\sigma^+\ee^{-\ii(\omega_dt+\phi_d)}
\right).
\end{equation}
This Hamiltonian can be transformed into the drive's rotating frame via
\begin{equation}
U(t)=\ee^{-\ii(\omega_dt+\phi_d)\sigma_z/2},
\end{equation}
to yield the time-independent model
\begin{equation}
H_{\rm rot}
=
U^\dagger H_{\rm RWA}U
-\ii U^\dagger\dot U
=
\frac{\delta_a}{2}\sigma_z
-
\frac{\Omega_d}{2}\sigma_y,
\label{eq:HTLSrot_RWA}
\end{equation}
where $\delta_a=\omega_a-\omega_d$ is the detuning.
This relation is the $\pi/2$ phase-shifted version of the mostly seen rotating-frame Hamiltonian $({\delta_a}/2)\sigma_z+({\Omega_d}/{2})\sigma_x$ with the cosine drive.

Including spontaneous emission with rate $\gamma$, the dynamics are governed by the optical Bloch master equation,
\begin{equation}
\dot\rho
=
-\ii[H_{\rm rot},\rho]
+
\frac{\gamma}{2}\,\mathcal{D}[\sigma^-]\rho .
\end{equation}
where the dissipator is $\mathcal{D}[\sigma^-]=2\sigma^-\rho\sigma^+-\sigma^+\sigma^-\rho-\rho\sigma^+\sigma^-$.

The steady-state excited-state population and coherence are
\begin{equation}
\rho_{ee}^{\rm ss}
=
\frac{\Omega_d^2}
{4\delta_a^2+2\Omega_d^2+\gamma^2}.
\label{eq:rho_ee_RWA}
\end{equation}
and
\begin{equation}
\langle\sigma^-\rangle_{\rm ss}
=
\frac{
\Omega_d\left(\gamma-2\ii\delta_a\right)
}{
4\delta_a^2+2\Omega_d^2+\gamma^2
}.
\label{eq:sigma_minus_ss_RWA}
\end{equation}

Using the quantum regression theorem, at exact (nominal) resonance, $\delta_a=0$, the incoherent first-order correlation function becomes
\begin{equation}
\begin{split}
G^{(1)}_{\rm rot}(\tau)
&=
\langle\sigma^+(\tau)\sigma^-(0)\rangle_{\rm ss}
-
|\langle\sigma^-\rangle_{\rm ss}|^2
\\
&=
C_0\ee^{-\gamma\tau/2}
+
C_+\ee^{-3\gamma\tau/4+\ii\Omega_{\rm eff}\tau}
+
C_-\ee^{-3\gamma\tau/4-\ii\Omega_{\rm eff}\tau},
\end{split}
\label{eq:TLS_Correlator_RWA}
\end{equation}
where
\begin{equation}
\Omega_{\rm eff}
=
\sqrt{
\Omega_d^2-\left(\frac{\gamma}{4}\right)^2
},
\label{eq:Mollow_effective_splitting}
\end{equation}
and 
\begin{equation}
\begin{split}
    &C_0=\frac{\Omega_d^2}{2(2\Omega_d^2+\gamma^2)},
    \\
    &C_\pm=\frac{\Omega_d^2(2\Omega_d^2-\gamma^2)}{4(2\Omega_d^2+\gamma^2)}\mp\ii\frac{\Omega_d^2\gamma(10\Omega_d^2-\gamma^2)}{16\Omega_{\rm eff}(2\Omega_d^2+\gamma^2)^2},
\end{split}
\end{equation}
where $C_-=C_+^*$. In the Mollow (strong driving) limit~\cite{Mollow_Power_1969} of $\Omega_d\gg\gamma$, then $C_0\to1/4$ and $C_\pm\to1/8\mp\ii(5\gamma)/(32\Omega_d)$. Indeed, to leading order in $\gamma/\Omega_d$,  they reduce to $C_0=2C_-=2C_+=1/8$ so that $\lvert C_\pm\rvert\approx0.5\lvert C_0\rvert$.
From this correlation function,
we can obtain the incoherent resonance-fluorescence spectrum from
\begin{equation}
\mathsf{S}(\omega)
\propto
\mathrm{Re}
\int_0^\infty
d\tau\,
\ee^{\ii(\omega-\omega_d)\tau}
G^{(1)}_{\rm rot}(\tau),
\end{equation}
and the total
fluorescence spectrum also contains an elastic coherent contribution
\begin{equation}
2\pi
|\langle\sigma^-\rangle_{\rm ss}|^2
\delta(\omega-\omega_d).
\end{equation}

Thus,
the incoherent spectrum reduces to the familiar Mollow triplet: a central peak located at $\omega_{\rm center}=\omega_d$ 
and two symmetric sidebands located at
$\omega_\pm
=
\omega_d\pm\Omega_{\rm eff}$,
with the effective linewidths
\begin{equation}
\Gamma_0^{\rm HWHM}
=
\frac{\gamma}{2},
\qquad
\Gamma_\pm^{\rm HWHM}
=
\frac{3\gamma}{4},
\end{equation}
respectively.
For the present parameters $\eta_d=\Omega_d/\omega_a=0.5$ and $\gamma=0.1\omega_a$, amd one obtains
$\Omega_{\rm eff}
\simeq
0.4994\,\omega_a$,
yielding spectral resonances at
\begin{equation*}
\omega/\omega_a
\simeq
0.5006,\qquad
1.0000,\qquad
1.4994,
\end{equation*}
with corresponding linewidths,
\begin{equation*}
\Gamma/\omega_a
\simeq
0.075,\qquad
0.050,\qquad
0.075,
\end{equation*}
respectively. These values are close to the dominant FL resonances extracted from both the TI-GME and F-GME calculations.

To further improve this beyond the drive-RWA approximation, one may include the BS correction~\cite{Bloch_Magnetic_1940} (and higher order corrections as well) due to the counter-rotating drive terms. For
the drive-amplitude convention adopted here, its leading-order value is
\begin{equation}
\delta_{\rm BS}
=
\frac{\Omega_d^2}
{2(\omega_a+\omega_d)}.
\label{eq:Bloch_Siegert_shift}
\end{equation}
Near resonance, $\omega_d\simeq\omega_a$, this becomes
$\delta_{\rm BS}\simeq\Omega_d^2/(4\omega_a)$. The corresponding
effective detuning is, then,
\begin{equation}
\delta_{\rm eff}
=
\delta_a+\delta_{\rm BS}.
\label{eq:BS_effective_detuning}
\end{equation}
At the nominal resonance, one therefore has
$\delta_{\rm eff}\simeq\delta_{\rm BS}$ rather than exact effective
resonance. The associated optical-Bloch poles satisfy
\begin{equation}
\left(\lambda+\frac{\gamma}{2}\right)
\left[
\left(\lambda+\frac{\gamma}{2}\right)
\left(\lambda+\gamma\right)
+
\Omega_d^2
\right]
+
\delta_{\rm eff}^{\,2}
\left(\lambda+\gamma\right)
=
0.
\label{eq:BS_Bloch_poles}
\end{equation}
For $\eta_d=0.5$ and $\gamma=0.1\omega_a$, the leading-order estimate
$\delta_{\rm BS}=0.0625\omega_a$ yields
\begin{equation*}
\lambda_0\simeq-0.05077\omega_a,
\qquad
\lambda_\pm
\simeq
-0.07462\omega_a
\pm\ii\,0.50325\omega_a.
\end{equation*}

The corresponding spectral poles occur approximately at
$\omega/\omega_a\simeq0.49675$, $1$, and $1.50325$. Thus, the
BS correction produces only a modest change in the
sideband separation relative to the resonant RWA result. Nevertheless,
$\delta_{\rm BS}$ is comparable to the dissipative linewidths and can
affect the sideband weights and spectral asymmetry. Since
$\eta_d=0.5$ lies outside the strictly controlled perturbative regime,
these values should be regarded as qualitative benchmarks; the exact
Floquet quasienergies and FL poles provide the quantitatively reliable
non-RWA resonance positions. This qualification is consistent with
strong-drive experiments in which corrections beyond the
lowest-order BS approximation are required~\cite{PhysRevLett.105.257003}.

The RWA and leading-order BS treatments reproduce the spectral poles reasonably well because the corresponding eigenvalues are only modestly renormalized. Nevertheless, they {\it fail to reproduce the sideband asymmetry} because the asymmetry, which is governed primarily by the FL residues rather than by the pole locations. 
As seen in the left panels of Fig.~\ref{fig:TLSSpectra_wd1_Flat}, clearly the mode prominences, $\mathcal{R}_{\mu:\text{right lobe}}>\mathcal{R}_{\mu:\text{left lobe}}$ characterizing the height of the peaks are asymmetric while the positions of the peaks and their linewidths agree. 
Note the $\gamma$'s are equal for the lobes so that the asymmetry appears as the discrepancy in the modal residues. Here, $\mathrm{Re}[\mathcal{W}_{\mu:\text{right lobe}}]/\mathrm{Re}[\mathcal{W}_{\mu:\text{left lobe}}]\approx2.86$.

These residues depend on the full non-RWA Floquet eigenvectors, the quasienergy-resolved dissipative steady state, and the biorthogonal excitation and detection overlaps (they are mathematically hidden in the modal residues $\mathcal{W}_\mu$), none of which is contained in a simple BS shift.
In the FL formulation the a peak's height is not simply an oscillator strength, but it is related to the residue $\mathcal{W}_\mu$ that is
\begin{equation}
    \mathcal{W}_\mu=\frac{1}{T}\int_T dt\,\Lbraket{s^-_{\rm B}}{R_\mu(t)}\,\Lbra{L_\mu(t)}s^+_{\rm B}\Lket{\rho^{\rm ss}(t)}.
\end{equation}
where the derivation and explicit expression in the extended-space is given in Eq.~\eqref{eq:Wmu_extended} of App.~\ref{secS:FL}.
For the fluorescence correlator of a TLS from the quantum regression theorem, $G^{(1)}(\tau)=\mathrm{Tr}[\sigma^+\mathrm{e}^{\mathcal{L}\tau}(\sigma^-\rho^{\rm ss})]$, one thus has
\begin{equation}
    \mathcal{D}_\mu=\Lbraket{\sigma^+}{R_\mu(t)},
\end{equation}
as the detection/emission overlap, and
\begin{equation}
    \mathcal{Q}_\mu=\Lbra{L_\mu(t)}\sigma^-\Lket{\rho^{\rm ss}(t)},
\end{equation}
as the preparation/source overlap.
Altogether, then, $\mathcal{W}_\mu$ contains two conceptually different pieces: the first factor gives information on how strongly the FL mode radiates into the detected dipole channel, while the second term informs how strongly that mode is excited by applying the correlation-function source operator to the periodic steady state.

First, the counterrotating drive terms change the Floquet wavefunctions, not just the quasienergies. They generate additional harmonic components in $\ket{\alpha(t)}$, so that the Fourier-resolved transition amplitudes associated with the two conjugate sideband processes are no longer related by the simple resonant-RWA symmetry. Thus, the radiative matrix elements entering $\mathcal{D}_\mu$ are different.
On the other hand, the same non-RWA Floquet structure enters the F-GME dissipator, via $\Gamma(\Delta_{\alpha\beta l})$ and $S_{{\rm B},\alpha\beta l}$, which determines the periodic steady state. Even for a flat positive-frequency bath, $\Gamma(\omega)=\Gamma$ being constant does not force the two total Floquet transition rates to be equal because the Fourier matrix elements and accessible sideband channels differ. Consequently, the steady-state Floquet populations and coherences entering $\mathcal{Q}_\mu$ are also asymmetric.
The BS calculation mainly improves the
estimation of the eigenvalues, hence $\Delta_\mu$ and, to some extent, $\gamma_\mu$; but the sideband asymmetry lives primarily in the eigenvectors and their biorthogonal residues $\mathcal{W}_\mu$. A scalar BS energy shift does not reconstruct the exact non-RWA Floquet micromotion, the Floquet-resolved dissipative redistribution, or the left/right FL overlaps entering $\mathcal{W}_\mu$.

The FL decomposition further shows that, in the full nonsecular F-GME, the central resonance is carried by a collective mode with substantial 
contributions from several Floquet coherence channels. This should not be interpreted as implying that the central Mollow line originates from nonsecular coupling: the central resonance is already present in the secular dressed-state description. Rather, nonsecular coupling reorganizes its detailed Floquet-channel composition and modifies its residue and linewidth.

These results demonstrate that the TI-GME can already fail in this minimal setting, namely for a weakly dissipative driven TLS coupled to a flat bath.
This may seem like
a surprising result, as
it is well known that 
a flat-bath system with radiative decay produces symmetric emission spectra within a drive-RWA;
however, asymmetric Mollow spectra are 
possible outside this regime, even with a flat bath.
For a resonantly and CW-driven TLS, within the drive-RWA,
the secular approximation in the Rabi-dressed transition basis, and
radiative population-decay coupling only, the incoherent
resonance-fluorescence spectrum is symmetric about the drive
frequency $\omega_d$. In the usual strong-driving Mollow limit, the
red and blue sidebands at $\omega_d\pm\Omega_d$ have equal heights
and equal integrated weights. The integrated areas of the red
sideband, central peak, and blue sideband occur in the ratio
$1:2:1$, while their half-widths at half maximum are
$3\gamma/4$, $\gamma/2$, and $3\gamma/4$, respectively
\cite{Mollow_Power_1969,Cohen-Tannoudji_Atom-Photon_1998}.

Mollow-sideband symmetry can be broken through several physically distinct mechanisms. First, reservoir-induced processes, including pure dephasing, phonon-assisted scattering, and frequency-dependent photonic coupling, which can modify the dressed-state relaxation and coherence rates and thereby produce unequal sideband weights or linewidths \cite{Ulhaq:13,McCutcheonNazir2013,RoyHughes2011}. In semiconductor quantum dots, these effects have been observed experimentally and described microscopically using phonon-based master-equation approaches \cite{Ulhaq:13}.

Second, and most relevant to the present results, sufficiently strong monochromatic driving invalidates the drive-RWA. Counter-rotating terms modify the Floquet quasienergies and transition amplitudes and can produce asymmetric Mollow sidebands even for a broadband radiative reservoir and in the absence of pure dephasing \cite{Yan2013CounterRotatingFluorescence,Yan_Resonance_2016}. Yan \emph{et al.} traced this driving-induced asymmetry to the violation of the balance between the relevant Floquet radiative pathways through the combined modification of Floquet transition quantities and stationary populations \cite{Yan_Resonance_2016}.

Third, more general temporal driving protocols, including bichromatic, multiharmonic, frequency-comb, and pulsed excitation, 
whose temporal structure breaks the generalized Floquet symmetry underlying a symmetric fluorescence spectrum~\cite{Yan2018Multiphoton,Yan2019GeneralizedParity}.
Thus, they can substantially modify the conventional Mollow structure and generate additional or strongly redistributed fluorescence sidebands \cite{Yan_Resonance_2016,Konthasinghe2014,He2015DoublyDressed,Moelbjerg2012}.

The F-GME framework can accommodate all three classes within a unified construction, provided the corresponding system--bath coupling operators, reservoir spectra, and temporal Fourier components of the drive are included. Reservoir-induced effects enter through the quasienergy-dependent bath rates and coupling operators, whereas strong-drive and multiharmonic effects enter through the Floquet quasienergies, modes, and transition matrix elements. The calculations presented here deliberately isolate the second mechanism: a monochromatically and continuously driven TLS coupled only to a radiative bath, with no pure dephasing. Our resulting spectral asymmetry therefore connects directly to the beyond-drive-RWA fluorescence physics,
extending previous  pictures through a fully nonsecular F-GME, a direct comparison with the TI-GME, and the FL modal decomposition.

Next we consider the drive strength of $\eta_d=1.5$ [see Figs.~\ref{fig:TLSSpectra_wd1_Flat}(d)--(f)], 
which is now significantly beyond a RWA regime; the spectrum is thus substantially reorganized relative to the moderate-drive regime, and we see multiple resonance peaks. The dominant resonances now occur near (extending also beyond a simple triplet)
\begin{equation*}
\omega/\omega_a
\simeq
0.317,\qquad
1.683,\qquad
2.317,\qquad
3.683,\ldots,
\end{equation*}
indicating that higher-order Floquet sideband processes have become the principal radiative channels. 

In contrast to the $\eta_d=0.5$ case, where the brightest feature was the central resonance at $\omega=\omega_d$, the strongest peaks at larger drive are displaced away from the integer harmonics and are predominantly associated with individual quasienergy-resolved sideband transitions. The odd integer-harmonic structures at $\omega/\omega_a=1,3,\ldots$ remain visible according to the selection rule since $\sigma_x$ is an odd operator, but they now appear as weaker collective resonances with substantially reduced spectral weight. Increasing the drive amplitude therefore reverses the modal hierarchy observed at moderate pumping: isolated sideband transitions become dominant, while drive-centered hybrid resonances become secondary.

This behavior is consistent with the well-known strong-driving breakdown of the simple RWA Mollow picture. In the conventional resonance-fluorescence treatment, the spectrum consists of a central peak at $\omega_d$ and sidebands at $\omega_d\pm\Omega_d$. At very strong driving, however, the counter-rotating drive terms can no longer be treated perturbatively, and the fluorescence spectrum develops higher-order multiphoton sidebands. In this regime, the spectral resonances are more accurately associated with the harmonics:
\begin{equation}
\omega
\simeq
n\omega_d
\pm
\Omega'_{\rm eff},
\end{equation}
where $\Omega_{\rm B}$ is a drive-renormalized quasienergy splitting. 

A simple analytical estimate based on strong-driving treatments beyond the drive-RWA yields~\cite{Lu_Effects_2012}
\begin{equation}
\Omega'_{\rm eff}
\approx
2\omega_a
J_1\!\left(
\frac{2\Omega_d}{\omega_d}
\right),
\label{eq:Omegaprim_eff}
\end{equation}
with $J_1$ the Bessel function of the first kind. For the present parameters $\Omega_d=1.5\,\omega_a$ and $\omega_d=\omega_a$,
then
\begin{equation*}
\Omega'_{\rm eff}
\simeq
0.678\,\omega_a.
\end{equation*}

This immediately predicts the dominant resonance families:
\begin{equation*}
\omega/\omega_a
\simeq
1\pm0.678,
\qquad
3\pm0.678,
\end{equation*}
yielding
\begin{equation*}
\omega/\omega_a
\simeq
0.322,\qquad
1.678,\qquad
2.322,\qquad
3.678,
\end{equation*}
in excellent agreement with the FL resonances extracted numerically from both the TI-GME and F-GME spectra.

The bar plots in panels (d) and (e)
of Fig.~\ref{fig:TLSSpectra_wd1_Flat} make clear, however, that agreement in resonance locations does not imply agreement in dissipative physics. Panel (d) shows that the effective linewidths $\gamma_\mu$ differ quantitatively between the F-GME and TI-GME descriptions, while panel (e) demonstrates that the spectral residues ($R_\mu$) are affected even more strongly. In particular, the TI-GME assigns excessively large spectral weights to several resonances, especially those near the lower-frequency sideband structures and integer harmonics of the drive. This produces the much larger dashed-blue peaks in Fig.~\ref{fig:TLSSpectra_wd1_Flat}(f), despite the fact that the corresponding steady-state populations differ only moderately between the two theories.

The discrepancy, and failure of the TI-GME,  therefore originates primarily from the modal residues rather than from the coherent quasienergy structure itself (similar to the $\eta_d=0.5$ example, but now with multiple resonances involved). The resonance frequencies are governed mainly by the driven Hamiltonian and can therefore still be reproduced semi-quantitatively even by the TI-GME. In contrast, the spectral amplitudes depend sensitively on how the dissipator distributes decay rates and residues across the active Floquet sideband channels. The TI-GME collapses these processes into static dressed-state transitions, whereas the F-GME preserves the full quasienergy-resolved frequency structure of the bath coupling. As a consequence, the TI-GME overestimates the contribution of several collective resonances once many Floquet sidebands become simultaneously active.

The FL modal decomposition further clarifies the microscopic origin of the dominant resonances. The mode located near $\omega/\omega_a\simeq1.683$, corresponding approximately to $\Delta_\mu\simeq\Delta_{ge2}$, is almost entirely dominated by the single Floquet channel $(g,e,2)$. Likewise, the resonances near
\begin{equation*}
\omega/\omega_a
\simeq
0.317,\qquad
2.317,\qquad
3.683,
\end{equation*}
correspond predominantly to the isolated quasienergy-resolved channels $(e,g,0)$, $(e,g,2)$, and $(g,e,4)$, respectively. These modes therefore behave approximately as pure sideband transitions.

In contrast, the resonances centered near the integer harmonics $\omega/\omega_a\simeq1$ and $\omega/\omega_a\simeq3$ are strongly hybridized collective modes. The mode near $\omega/\omega_a\simeq1$ receives dominant contributions from the channels:
\begin{equation*}
\Delta_{eg0}\approx0.318\,\omega_a,
\qquad
\Delta_{ge2}\approx1.68\,\omega_a,
\end{equation*}
with normalized overlaps
\begin{equation*}
r_\mu^{(e,g,0)}
\simeq
r_\mu^{(g,e,2)}
\simeq
0.33,
\end{equation*}
together with smaller contributions from
\begin{equation*}
\Delta_{ge0}\approx-3.18\,\omega_a,
\qquad
\Delta_{eg2}\approx2.318\,\omega_a,
\end{equation*}
with
\begin{equation*}
r_\mu^{(g,e,0)}
\simeq
r_\mu^{(e,g,2)}
\simeq
0.164.
\end{equation*}
Similarly, the mode near $\omega/\omega_a\simeq3$ is composed primarily of the channels $(e,g,2)$ and $(g,e,4)$ together with weaker admixtures of $(g,e,2)$ and $(e,g,4)$.

These results demonstrate that the computed spectrum is governed not by isolated quasienergy gaps alone, but by the poles of the full FL generator,
$\lambda_\mu
=
-\gamma_\mu
-\ii\Delta_\mu$.
In the weak-mixing regime, each pole can often be associated predominantly with a single Floquet transition channel. In the deep-strong-driving 
regime, however, many sideband channels become simultaneously active and strongly coupled through the dissipator, producing collective Liouvillian resonances whose linewidths and residues are highly nontrivial. The Floquet--Liouville decomposition therefore identifies not only the resonance frequencies, but also the microscopic channel composition and dissipative origin of each spectral line.

Accordingly, Fig.~\ref{fig:TLSSpectra_wd1_Flat} reinforces the central conclusion already suggested by Fig.~\ref{fig:TLS_etad}: the difference between the TI-GME and F-GME does not originate from the coherent Hamiltonian dynamics, which are common to both approaches, but from their fundamentally different dissipative frequency resolutions. Even for a flat spectral bath, the redistribution of system--bath coupling across Floquet sideband channels remains essential once multiple harmonics participate. The TI-GME collapses this structure into static dressed-state transitions, whereas the F-GME preserves the quasienergy-resolved dissipative pathways. This is why the two theories may exhibit qualitatively similar resonance positions while predicting substantially different spectral amplitudes, linewidths, and steady-state observables in the strong-driving regime.

The frequency-resolved results show that the emergence of Floquet pathways can distinguish the need for F-GME. 
That is, in general, when the Floquet channels are activated and depart from the static channels, $\Delta_{\alpha\beta l}\neq\omega_{jk}+k\omega_d$, one is required to use the correct channels in the dissipator. 
This can, for example, appear in the emergence of a sideband lobe in the Mollow-triplet structure, where the drive can overcome the effect of dissipation. Hence,  when $\Omega_d/\omega_d>\gamma/\omega_a$ (which is a similar condition to the conventional strong coupling regime of quantum light-matter interaction or cavity-QED), the failure of the time-independent master equation approaches are expected and the use of the Floquet master equation approaches is a  requirement
for accurate predictions. 

The lack of proper dressing of the transition pathways (time-independent not externally-dressed dissipator versus Floquet externally-dressed dissipator) is the key issue that can be the source of unphysical predictions when the counter-rotating wave effects are non-negligible. This includes the drive-RWA and the secular approximation failures shown in Fig.~\ref{figS:TLSRWA_etad} of 
App.~\ref{secS:AdditionalResults}.
It is also advisable to use the TI-GME model~\cite{Settineri_Dissipation_2018,Salmon_Gauge-independent_2022,Akbari_Generalized_2023,Mercurio_Regimes_2020} (e.g., with $\Omega_d\ll\omega_a$), or use the first-order Floquet perturbation theory~\cite{Macri_Spontaneous_2022} when the coherent drive amplitude is 
sufficiently low (so its effect on the dressed states for the GME is neglected).

The deep-strong-driving spectra predicted by the F-GME qualitatively reproduce several experimentally established signatures of beyond-RWA resonance fluorescence, including multiphoton sidebands, harmonic spectral replicas, and strong redistribution of spectral weight away from the conventional Mollow-triplet structure~\cite{FornDiaz_Observation_2010,Yoshihara_BlochSiegert_2017,Yan_Resonance_2015,Astafiev_Resonance_2010}. In particular, the dominant resonances observed near $\omega\simeq n\omega_d\pm\Omega'_{\rm eff}$ in Fig.~\ref{fig:TLSSpectra_wd1_Flat}(f) are reminiscent of experimentally observed higher-order Mollow and strong-field resonance-fluorescence structures, where sufficiently strong driving generates additional sidebands beyond the conventional Mollow triplet~\cite{Wang2021HighOrderMollow,Boos2024DynamicallyDressed,Liu2024DynamicRF}, with the breakdown of the drive-RWA approximation.

Importantly, while both the TI-GME and F-GME reproduce the coherent quasienergy structure semi-quantitatively, only the F-GME preserves the quasienergy-resolved dissipative pathways responsible for the experimentally relevant redistribution of spectral weights among the active Floquet sidebands.

The same FL-mode hybridization mechanism persists away from resonance, although the detailed channel composition depends on the drive frequency.

\subsection{The effect of structured baths}

\begin{figure*}[!htpb]
\centering
\includegraphics[width=\linewidth]{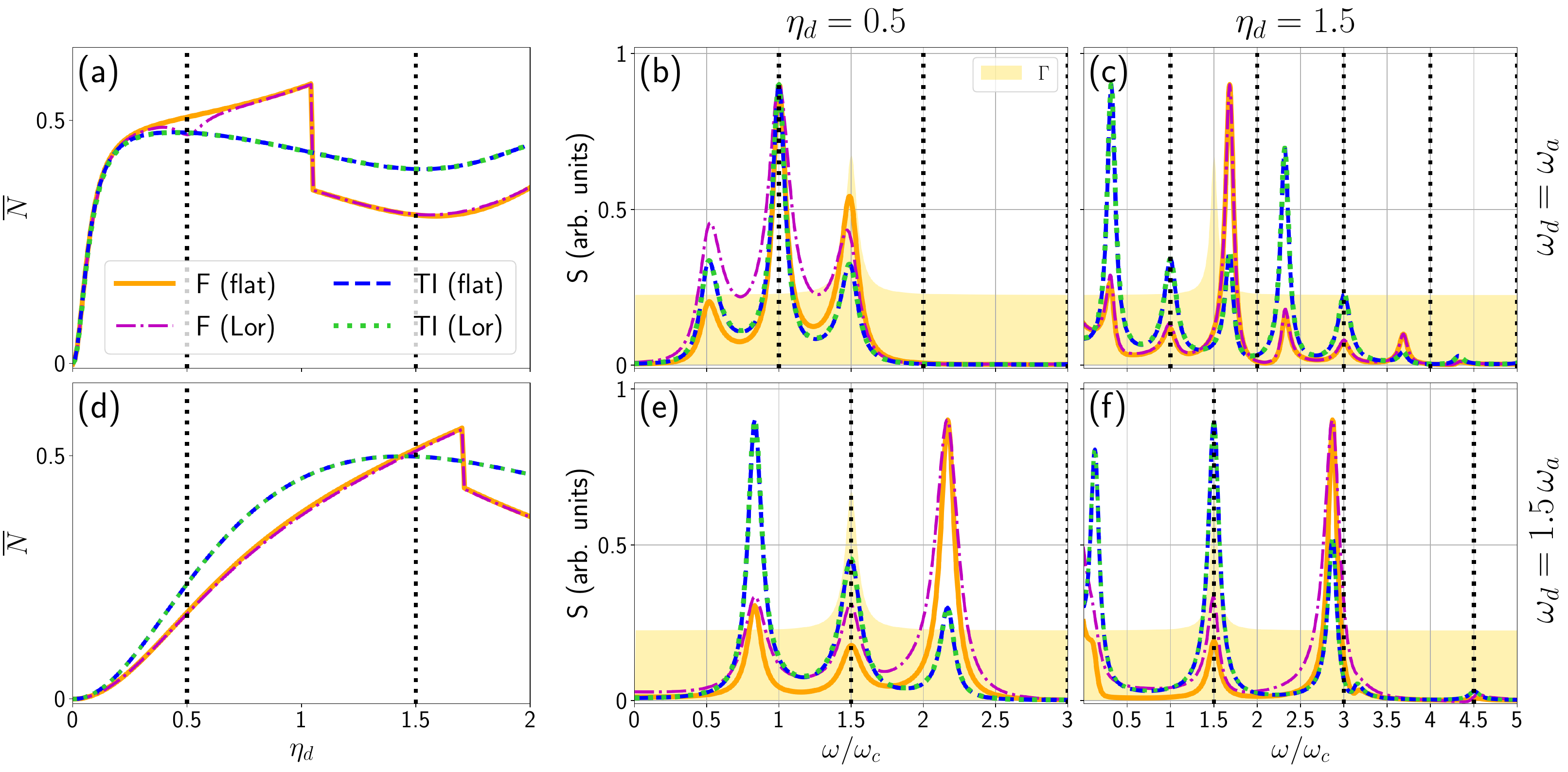}
\caption[]{\textbf{Driven-dissipative TLS: the effect of a structured bath.}
(a,d) Long-time averaged excitation population versus $\eta_d$, and (b,c,e,f) incoherent emission spectra of a driven-dissipative TLS. The results with a Lorentzian bath (red dash-dotted curves from the F-GME and dotted green curves for the TI-GME) centered at $\omega_0=1.5\omega_a$ with linewidth and amplitude parameter $\gamma=0.1\omega_a$ are compared with those of with the reference flat bath curves. Here, for the top panels $\omega_d=\omega_a$ and the bottom panels $\omega_d=1.5\omega_a$. The spectra are obtained for the strong drive of $\eta_d=0.5$ (b,e) and deep-strong drive of $\eta_d=1.5$ (c,f). The vertical dotted black lines represent the drive's specifics, i.e., in panels (a) and (d) they mark drive's representative amplitudes considered in the spectra calculation, while they mark the integer-multiples of the drive frequencies in panels (b,c) and (e,f).
The background damping rate for the structured bath is $\gamma_b=\gamma$.
}
\label{fig:TLS_Lor}
\end{figure*}

We next investigate the influence of a structured bath environment, with results shown in Fig.~\ref{fig:TLS_Lor}, when we replace the flat bath with a Lorentzian spectral density centered at $\omega_0=1.5\,\omega_a$ and the same decay rate of $\gamma=0.1\omega_a$ as for the flat bath (and, with $\gamma_b=\gamma$). 
Two detuning regimes of the drive frequency are considered:  the resonant case of $\omega_d=\omega_a$ (top row). as well as the detuned drive case of $\omega_d=1.5\,\omega_a$, where the drive is deliberately tuned with the bath resonance peak (bottom row).
The resulting long-time averaged excitation populations versus the full range of the drive amplitude is shown in the left panels of Fig.~\ref{fig:TLS_Lor}, whereas
the spectra are depicted for two distinct drive amplitude regimes of the strong driving case with $\eta_d=0.5$ (middle panels) and the deep-strong driving case with $\eta_d=1.5$ (right panels). 

The general rule here is that the dynamics depends on  how the bath peak (or, different values) align with the dissipative pathways in each description. 
In the TI-GME, dissipation samples the bath only at the {\it static} dressed state transition frequency $\omega_{ge}=\omega_a$, yielding an effective decay rate $\Gamma_{\rm Lor}(\omega_a)\approx0.051$, which remains close to the flat-bath value $\Gamma_{\rm flat}=\gamma/2=0.05$. Consequently, the structured bath only weakly modifies the TI-GME predictions. In contrast, the F-GME samples the bath at the full set of Floquet transition frequencies $\Delta_{\alpha\beta l}$. Since these sideband-resolved dissipative channels probe different regions of the Lorentzian reservoir, their relaxation rates become strongly redistributed. The structured bath therefore amplifies the distinction between the two theories, leading to substantially larger deviations between the TI-GME and F-GME populations than in the flat-bath case. To observe the system's response, below we analyze, in more details, the population and spectra in different regimes of the drive.

For the tuned resonant drive, $\omega_d=\omega_a$, comparing the flat- and Lorentzian-bath results within either theory (TI-GME and F-GME) shows that the
structured reservoir does not 
noticeably affect the population
[Fig.~\ref{fig:TLS_Lor}(a)] except generating a smooth cusp-like  dip near $\eta_d\approx0.5$ in the F-GME result (magenta curve). 
In this case, the principal drive-dressed sidebands are approximately
$\omega_d\pm\Omega_{\rm eff}\approx(1+\eta_d)\omega_a$.
With the Lorentzian bath centered at
$\omega_0=1.5\omega_a$, the upper sideband crosses the bath resonance approximately when
$1+\eta_d\approx1.5$ yielding $\eta_d\approx0.5$.
That is exactly where the Lorentzian-bath F-GME curves develop the dip, while the TI-GME curves do not. This is the expected signature of Floquet-sideband-resolved dissipation in a structured bath.

We next look at the spectra in panels (b) and (c) of Fig.~\ref{fig:TLS_Lor}.
In general, the population and the normalized emission spectrum need not respond
in the same way to the structured reservoir. The period-averaged
population is determined by the periodic steady-state modes,
$\overline N=\llangle N\mid R_{\rm ss,0}\rrangle$ [Eq.~\eqref{eq:AveN_FL}], whereas the
incoherent spectrum depends on the eigenvalues and residues of the
decaying FL coherence modes [Eq.~\eqref{eq:FL_spectrum_main}].
Although their integrated quantities are related through
Eq.~\eqref{eq:Spectrum_Population_connection}, equality or near
equality of $\overline N$ does not imply equality of the normalized
spectral line shape. 

However,  for the same reason that a single dissipative channel  is not affected by the bath peak, in the
incoherent emission spectra of Fig.~\ref{fig:TLS_Lor}(b,c),
the flat-bath TI-GME and the Lorentzian-bath TI-GME shapes overlap. 
In the F-GME, on the other hand, the Lorentzian reservoir
redistributes the rates among different Floquet channels
$\Gamma(\Delta_{\alpha\beta l})$. Their contributions to the
steady-state population may partially cancel, while the same
redistribution modifies the decay rates and residues of the
spectroscopic FL modes. Consequently, the flat- and
Lorentzian-bath F-GME populations can remain similar even when their
normalized spectra differ substantially.
The bath peak lands on the Mollow triplet's upper lobe in (b), therefore, suppress this individual peak and increase its linewidth, with respect to the other peaks in the normalized spectrum. 

As the drive increases, additional drive-induced processes become accessible and coherent
pumping progressively dominates, driving the TLS toward saturation.
The steady-state population is then governed by the balance between pumping and dissipation
rather than by the detailed frequency dependence of the reservoir. Consequently, within
each theory the flat- and Lorentzian-bath predictions converge at large $\eta_d$ (deep-strong driving regime $\eta_d\gtrsim1$), as seen in Fig.~\ref{fig:TLS_Lor}(c). For the
F-GME this occurs on the high-field branch beyond the channel-crossing cusp at
$\eta_d\approx1.05$.

We emphasize that this high-field convergence is a flat-versus-structured scenario
\emph{within} a given theory; the F-GME and TI-GME themselves do not converge to the same values, but remain separated by the quasienergy-resolved redistribution of dissipative weight. The residual
offset of the saturated values is consistent with the bath model used here
[Eq.~(\ref{eq:BathModel_Lorentzian})]; away from $\omega_0$ the Lorentzian reduces to the flat
background $\gamma_b$, so channels far from the peak relax at this background rate while
only channels near $\omega_0$ sample the enhanced rate.

Our analysis also naturally reduces to the three-resonance dressed-state dissipator model in Refs.~\onlinecite{Kowalewska-kudlaszyk_Generalized_2001,Ge_Accessing_2013}.
In the drive-RWA, secular and the lowest-order-Floquet limit, the F-GME reduces exactly to the
dressed-state dissipators of Kowalewska-Kud\l aszyk and Tana\'s~\cite{Kowalewska-kudlaszyk_Generalized_2001}, and of
Ge \emph{et al.}~\cite{Ge_Accessing_2013}. Both perform external dressing \emph{first} and then
couple to the bath, thus both align with the F-GME approach and they contrast against the TI-GME (but within a drive-RWA approximation).

For the resonantly driven TLS ($\omega_d=\omega_a$, $S_{\rm B}=\sigma_x$), the rotating frame similar to Eq.~\eqref{eq:HTLSrot_RWA}, but with a cosine drive, gives $H_{\rm rot}=({\Omega_d}/{2})\sigma_x$, with the
Rabi-dressed states $\ket{\pm}=(\ket{e}\pm\ket{g})/\sqrt2$ and splitting $\Omega_d$.
The bath couples through the \emph{lab-frame} $\sigma_x$, which in the rotating frame is
$\tilde S_B(t)=\sigma^+e^{i\omega_d t}+\sigma^- e^{-i\omega_d t}$; re-expressing $\sigma^\pm$
in the dressed basis yields exactly three positive-frequency channels:
\begin{equation*}
\Delta_{\alpha\beta l}=\omega_d-\Omega_{\rm eff},\ \omega_d,\ \omega_d+\Omega_{\rm eff},\qquad l=1.
\end{equation*}
These are precisely the three Markov contributions $\omega_d-\omega,\,\omega_d-\omega\pm\Omega_d$
identified in Ref.~\onlinecite{Kowalewska-kudlaszyk_Generalized_2001}. The TI-GME, lacking external dressing, retains only
$\Delta=\omega_d=\omega_a$ and so cannot reproduce this structure.

Moreover, the secular FL generator is then block-diagonal, giving three isolated poles
$\lambda_\mu=-\gamma_\mu-\ii\Delta_\mu$ with $\Delta_\mu=\omega_d,\omega_d\pm\Omega_d$ and
\begin{equation*}
\gamma_{\mu_{\rm center}}=\frac12\big[\Gamma(\omega_d-\Omega_d)+\Gamma(\omega_d+\Omega_d)\big],
\end{equation*}
for the Mollow triplet's main peak, and
\begin{equation*}
\gamma_{\mu_{\pm}}=\frac14\big[\Gamma(\omega_d-\Omega_d)+4\,\Gamma(\omega_d)+\Gamma(\omega_d+\Omega_d)\big],
\end{equation*}
for the Mollow triplet's side lobes.
When ignoring the potential impact of dephasing, these are \emph{identical} to the dressed population/coherence rates
$\Gamma_{\rm pop},\Gamma_{\rm coh}$ of Ref.~\onlinecite{Kowalewska-kudlaszyk_Generalized_2001} [their Eq.~(47)] and of Ref.~\onlinecite{Ge_Accessing_2013} [their Eqs.~(13)--(14)], under the identification
$\Gamma(\Delta)\;\longleftrightarrow\;
\pi J_{\rm ph}(\Delta)$,
so the generic bath function plays the role of the mode density 
in Ref.~\onlinecite{Kowalewska-kudlaszyk_Generalized_2001}
and proportional to the projected local density of states (LDOS) of Ref.~\onlinecite{Ge_Accessing_2013}; the three above channels
are their three dressed resonances.

However, this simple reduction holds only under drive-RWA, secularization, and lowest Floquet order.
Relaxing each yields the new effects reported here: 
$\omega\simeq n\omega_d\pm\Omega'_{\rm eff}$ with $\Omega'_{\rm eff}$ in Eq.~\eqref{eq:Omegaprim_eff}
(many $\Delta_{\alpha\beta l}$ rather than three); full nonsecular FL hybridization
($\Delta_\mu\neq\Delta_{\alpha\beta l}$; cf. Figs.~\ref{fig:TLSSpectra_wd1_Flat} and \ref{figS:Spectra_etad1_Flat}), going beyond nonsecular shift terms, and population-level discrepancies in $\overline N$.

For the higher drive frequency of $\omega_d=1.5\,\omega_a$ [Fig.~\ref{fig:TLS_Lor}(d)], the
population rises more gradually than in Fig.~\ref{fig:TLS_Lor}(a)
because the drive is detuned from the bare TLS transition, and a
larger amplitude is required to approach saturation. 
As in Fig.~\ref{fig:TLS_Lor}(a), the F-GME exhibits a channel-crossing cusp absent from the
TI-GME; here it is shifted to $\eta_d\approx1.7$, since the Floquet-channel crossing
condition $\Delta_{\alpha\beta l}=0$ [i.e., $\Gamma(0^-)\neq\Gamma(0^+)$] depends on $\omega_d$. Both
Floquet flat- and Lorentzian-bath cusp at the same drive, the crossing being set by the bath-independent
coherent Floquet structure.
Also, we again observe that, for the same reason explained for Fig.~\ref{fig:TLS_etad}(e), that is a genuine beyond-drive-RWA feature, for the present
model and parameter regime. 

Remarkably,
the F-GME populations obtained with the flat and Lorentzian baths
remain nearly identical over most of the drive range, even though
their emission spectra differ substantially
[Figs.~\ref{fig:TLS_Lor}(e),(f)].
This behavior should not be interpreted using an on-resonance
optical-Bloch steady-state formula, since the present drive is both
detuned and treated beyond the drive-RWA. Rather, the relevant
observation is that the Lorentzian reservoir is centered at the
drive frequency, $\omega_0=\omega_d=1.5\,\omega_a$, and is therefore
symmetric about $\omega_d$, with 
\begin{equation*}
\Gamma_{\rm Lor}(\omega_d-\delta)
=
\Gamma_{\rm Lor}(\omega_d+\delta).
\end{equation*}
Consequently, for the dominant Floquet-channel pairs lying approximately
symmetrically about the drive, the structured bath does not introduce
an additional imbalance between the corresponding forward and
backward dissipative pathways. Their absolute rates may change
relative to the flat bath, but their ratio---which primarily controls
the steady-state population balance---can remain nearly unchanged.
This explains why the zero-mode contribution determining
$\overline N$ is only weakly modified.

Importantly, the bath symmetry does not imply a symmetric emission
spectrum. In the full non-RWA F-GME, the Floquet transition matrix
elements, stationary populations, and nonsecular coupling between
channels are generally asymmetric, so that
$|S_{B,\alpha\beta l}|^2\Gamma_{\rm Lor}(\Delta_{\alpha\beta l})$
need not be equal for the red- and blue-sideband processes even when
the bath function itself takes equal values at symmetrically placed
frequencies. The structured reservoir can therefore leave
$\overline N$ nearly unchanged while substantially modifying the
linewidths and residues of the decaying FL (coherence) modes.

The TI-GME's flat- and Lorentzian-bath curves likewise nearly coincide, but for a different reason; the
time-independent dissipator samples only at the bare transition $\omega_{ge}=\omega_a$, which lies on the
Lorentzian background irrespective of $\omega_d$, so $\Gamma_{\rm Lor}(\omega_a)\approx
\gamma/2$ in both baths. The flat/Lorentzian agreement within the F-GME is thus physical
(symmetric sideband sampling), whereas within the TI-GME it reflects the pathological
sampling at the atom rather than at the drive. The F-GME and TI-GME nonetheless remain
distinct, separated by the quasienergy-resolved redistribution of dissipative weight.

For this off-resonant drive case, the spectra for the Lorentzian bath are also compared with their flat bath counterparts, for the strong and deep-strong drive amplitudes in panels (e) and (f) of Fig.~\ref{fig:TLS_Lor}, respectively.
Both panels (e) and (f) are slices of panel (d); (e) at $\eta_d=0.5$, (f) at $\eta_d=1.5$ [the two dotted vertical black lines in (d)]. At both panels, Floquet flat- and Lorentzian bath approximately coincide in the population (the symmetric-sideband cancellation), yet (e) and (f) both show the Floquet spectra differing---based on Eq.~\eqref{eq:Spectrum_Population_connection} dissociation rule. Thus, (e) and (f) together demonstrate that the population insensitivity at $\omega_d=1.5\omega_a$
 persists across drive strength while the spectral sensitivity not only carries on but grows with $\eta_d$.

The dominant structure is a Mollow-triplet-like structure centered at $\omega=\omega_d$. Approximately, the spectral structure can be rendered by the beyond-RWA calculation of the side lobes at $\omega_d\pm\Omega_{\rm eff}$ and  $\omega_d\pm\Omega'_{\rm eff}$, for the strong and deep-strong driving cases, respectively, shown in panels (e) and (f).

In Fig.~\ref{fig:TLS_Lor}(e) and (f), the bath frequency sampling is at the drive frequency, and as expected, the TI-GME fails to correctly predict  the spectral linewidths and weights, as also known in the drive-RWA case~\cite{Ge_Accessing_2013}.
This effect is also known in the context of the dynamical decoupling of ions and qubits~\cite{Morong_Engineering_2023,Ezzell_Dynamical_2023,Evert_Syncopated_2025} as well as dissipation reduction (e.g., in electron-phonon systems~\cite{PhysRevA.100.042112,Gustin_Efficient_2020,PRXQuantum.2.040354,3fd4-x4d1}, where the laser frequency takes on a interesting time-dependence to sweep around the bath density of states as population transfer is maximized.
In these panels, the main Mollow-triplet-like peak is majorly affected, while Floquet Lorentzian bath reshaping is more dramatic and more selective in (f) than (e).

This driven TLS plus bath example illustrates a key physical distinction between the two GME approaches. In the TI-GME, the structured bath is sampled only through a small set of static transitions.
In the F-GME, however, the bath acts directly on the hierarchy of quasienergy-resolved Floquet channels $\Delta_{\alpha\beta l}$. Structured environments, therefore, provide a particularly sensitive probe of quasienergy-resolved dissipation, making the limitations of time-independent dissipative descriptions substantially more pronounced than in flat-bath systems.
In the deep-strong drive range, however, the spectra (similar to the population) are not qualitatively affected by the structure of the bath as there is the dynamical decoupling from the bath---an effect discussed above, and also measured with electron-phonon coupling for coherently excited quantum dots~\cite{PRXQuantum.2.040354,nanolett.2c01783,qute.202300359}

\section{Driven-dissipative coupled-TLS}

\label{sec:CoupledTLSs}
Next we turn our attention to two dipole-coupled 
TLSs, which provides a minimal strongly interacting qubit--qubit platform in which internal dressing arises from coherent inter-qubit coupling, while external dressing is induced by the applied periodic drive, 
cf.~Fig.~\ref{fig:schematics}(b). 

Such models are directly relevant to various experimental settings. For example, in superconducting circuits, strong and tunable qubit--qubit interactions can be engineered through direct capacitive/inductive coupling or mediated by tunable couplers and microwave resonators, providing a controllable realization of interacting artificial atoms~\cite{You_Superconducting_2005,Blais_Circuit_2021,Bialczak_Fast_2011,Yan_Tunable_2018}. In atomic and solid-state emitter platforms, closely spaced dipoles or Rydberg atoms realize strong position-dependent dipole--dipole interactions, leading to collective level shifts, correlated decay, blockade physics, and modified resonance fluorescence~\cite{Ficek_Entangled_2002,Saffman_Quantum_2010,Jenkins_Collective_2016,Plankensteiner_Selective_2015}. Quantum optical experimental and theoretical studies also include the interacting quantum dots in various settings~\cite{Laucht_Mutual_2010,Carlson_Theory_2019,Vora_Strong_2019}.
The driven coupled-TLS system therefore represents the simplest nontrivial model in which the competition between interaction-induced dressing and drive-induced Floquet dressing can be tested in an open quantum system.

The general system of two dipole-coupled TLSs is modeled by the Hamiltonian
\begin{equation}
     H_0= \frac{\omega^{(1)}_a}{2}\sigma_z^{(1)}+\frac{\omega^{(2)}_a}{2}\sigma_z^{(2)}+g \,\sigma_x^{(1)}\sigma_x^{(2)},
     \label{eq:H0_CoupledTLS}
 \end{equation}
where $g$ is the (position-position) interaction coupling rate between the two TLSs for the coupled-TLS system.  
The coupled-TLS Hamiltonian can then be diagonalized straightforwardly as a set of spin eigenstates of $\vert j\rangle$ of energies $E_j$. 
 This can be done analytically since the system preserves the parity, and hence the Hamiltonian can be decomposed into two disjoint even- and odd-subspaces. 
 
 The even-subspace spans by $\{\ket{gg},\ket{ee}\}$ and the odd-subspace spans by $\{\ket{ge},\ket{eg}\}$, while the interacting term $g\,\sigma_x^{(1)}\sigma_x^{(2)}$ flips both states so that $\ket{gg}\leftrightarrow\ket{ee}$ and $\ket{ge}\leftrightarrow\ket{eg}$. The even and odd disjoint Hamiltonian read 
 \begin{equation}
 \begin{split}
     H^{(\rm e)}=\lsb\begin{array}{cc}
       \Sigma   & g \\
        g  & -\Sigma
     \end{array}\rsb,
\qquad
     H^{(\rm o)}=\lsb\begin{array}{cc}
       \delta   & g \\
        g  & -\delta
     \end{array}\rsb,
 \end{split}
\end{equation} 
respectively, with $\Sigma=(\omega_1+\omega_2)/2$ and $\delta=(\omega_1-\omega_2)/2$. 
Consequently, the eigenenergies, in the even- and odd-subspace, read 
\begin{equation}
 \begin{split}
     E^{(\rm e)}_\pm=\pm\sqrt{\Sigma^2+g^2},
\qquad
     E^{(\rm o)}_\pm=\pm\sqrt{\delta^2+g^2},
 \end{split}
\end{equation} 
respectively.

These {\it internally dressed} states have physical implications so that we rename them as such, in order: as the dressed ground state $\ket{G}$ with $E_G=E_-^{\rm (e)}$, the dressed lower-polariton state $\ket{L}$ with $E_L=E_-^{\rm (o)}$, the dressed 
upper-polariton state $\ket{U}$ with $E_U=E_+^{\rm (o)}$, and the dressed double exciton state $\ket{T}$ with $E_T=E_+^{\rm (e)}$. The dressed polariton states are often labeled
as superradiant and subradiant states, because of their modified decay rates.

Within a RWA, the Hamiltonian becomes number conserving. Thus, the Hilbert space splits into 
$N=0,1,2$ manifolds, with $N=0$ having the eigenstate $\ket{g_1g_2}$ and eigenenergy $E_{gg}=-\Sigma$; $N=1$ spanning $\{\ket{g_1e_2},\ket{e_1g_2}\}$ and eigenenergies $E_{\pm}=\pm\sqrt{\delta^2+g^2}$; and $N=2$ having the eigenstate $\ket{e_1e_2}$ and eigenenergy $E_{ee}=\Sigma$.

For our calculations, we will also consider applying an optical drive to coherently pump the first TLS, in the dressed basis, via $S_d^{\rm bare}=\sigma_x^{(1)}$.
Moreover, for the system-bath coupling operator $S^{\Lambda,\rm bare}_{\rm B}=\sigma_x^\Lambda$ for each subsystem,
which is also used for the ladder operators in the dressed picture.
For convenience, we also assume the TLSs are identical, so that $\omega_a^{(1)}=\omega_a^{(2)}\equiv\omega_a$.

\subsection{Results for different drive amplitudes}

\begin{figure*}[!htpb]
\centering
\includegraphics[width=.75\linewidth]
{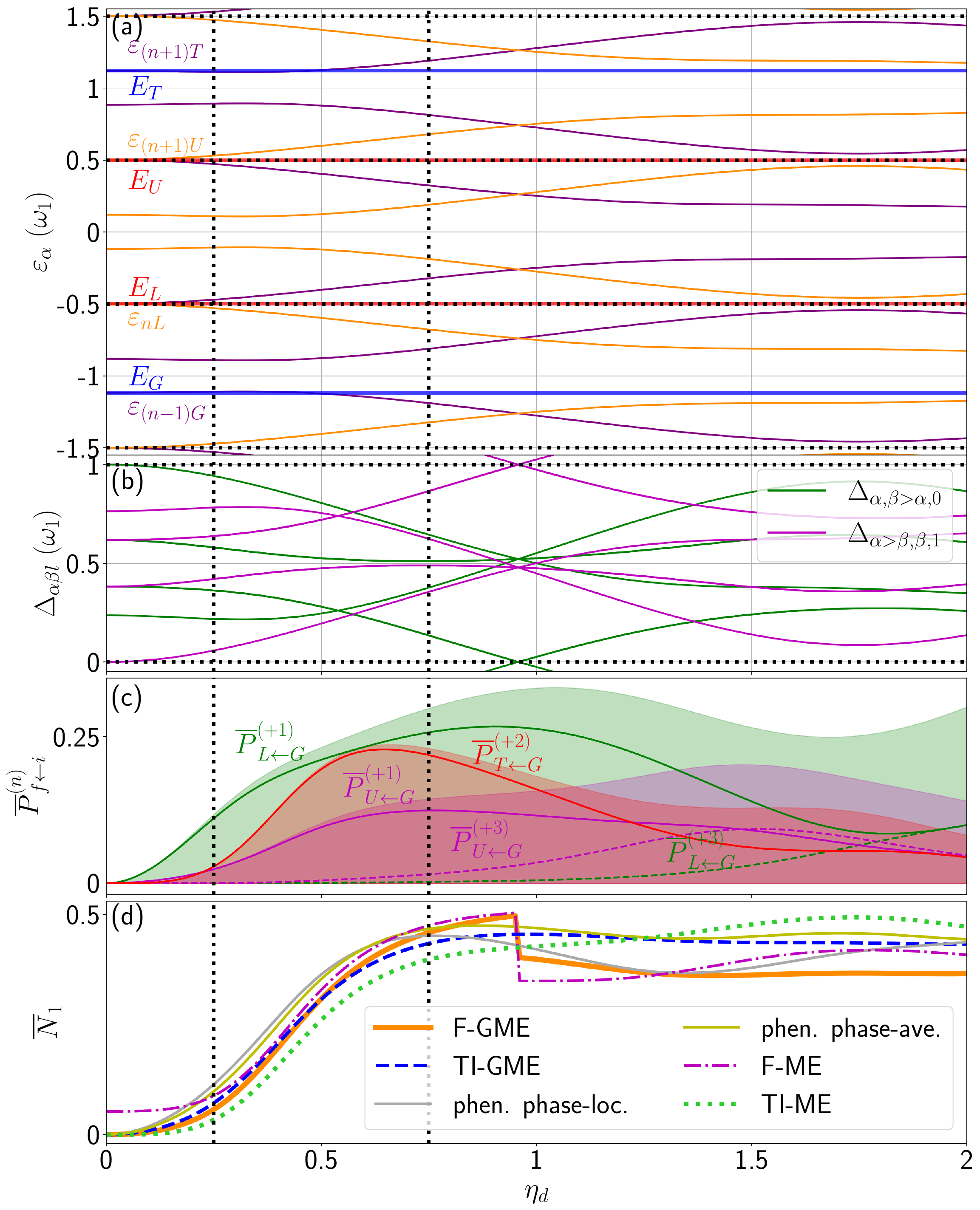} 
\caption[]{\textbf{Driven-dissipative coupled-TLS: drive amplitude dependence of the dynamics.} (a) Energy basis; shown are the static eigenenergies by the thick horizontal lines distinguished by their parities (blue: even, red:odd) and thin solid Floquet quasienergies lines distinguished by their generalized parities (purple:even, orange:odd), (b) extended space quasienergy gaps, (c) excitation transition probability when $\sigma_x^{(1)}$ is driven, and (d) long-time average number of excitation for $\gamma=0.1\omega_a$. For all panels, $\omega_d=\omega_a$ is the drive frequency which are shown by the vertical black dotted lines to mark BZs.  The sudden dip (discontinuous features) in (d) for the F-GME theory occurs when the same order quantum processes form crossings. Vertical dotted black line mark the two representative drive normalized amplitude of $\eta_d=(0.25,0.75)$ for the future furthur investigation.
}
\label{fig:CoupledTLS_etad}
\end{figure*}

Similar to the single-TLS case, we first compute  the steady-state time-averaged populations for the driven-dissipative coupled-TLS model, and aim to connect to the results from GMEs to the energy basis of the system, as shown in
Fig.~\ref{fig:CoupledTLS_etad}.

In panel (a) of Fig.~\ref{fig:CoupledTLS_etad},  we plot the energy basis eigenfrequencies of the system versus the drive normalized strength $\eta_d=\Omega_d/\omega_a^{(1)}$.  The horizontal thick solid lines show the static eigenenergies of the system which are the eigenvalues of the Hamiltonian in Eq.~\eqref{eq:H0_CoupledTLS} with their parities (blue: even, red: odd), and the thin solid lines represent the Floquet quasienergies in three primary BZs (distinguished by horizontal dotted black lines) with their generalized parities (purple: even, orange: odd). Note the quasienergies emerge from their parent eigenenergies (the nomenclature follows the same) where they also inherit their generalized parities from at the limit of $\eta_d\to0^+$, while keeping their parities throughout the entire increasing range of $\eta_d$.
Next, in panel (b), we show the quasienergy channels $\Delta_{\alpha\beta l}=\varepsilon_\beta-\varepsilon_\alpha+l\omega_d$.

In Fig.~\ref{fig:CoupledTLS_etad}(c), we show the unitary Floquet excitation transition probabilities.
As seen, the most probable excitation process is from the ground dressed state to the lower polariton state assisted by one external photon from the drive, $\overline{P}_{L\leftarrow G}^{(1)}$.
The long-time averaged excitation population is shown in Fig.~\ref{fig:CoupledTLS_etad}(d). The thin gray (phenomenological phase-locked) and olive (phenomenological phase-averaged) curves correspond to the unitary Floquet evolution and serves as a reference for the coherent driven dynamics in the absence of dissipation. The solid orange, dashed blue, dotted magenta and dotted light blue curves show the results obtained from the F-GME, TI-GME, F-ME and TI-ME approaches, respectively.
The TI-ME's prediction is indeed the weakest of all for such a strongly interacting system and can yield unphysical predictions as with $\eta\sim\omega_{1,2}$ the internal interaction and with $\eta_d\sim\eta$ the external interaction are considerable and cannot be neglected. 

Once again we find, in panel (d) of Fig.~\ref{fig:CoupledTLS_etad}, that the phenomenological results may roughly follow the same general trend as those of the GMEs', excluding any effect rooted in the dissipator such as the sharp cusp in the F-GME result; 
since  they do not include the dissipator, they are not able to capture the correct dissipative effects and thus they become different.
Also, the phase-averaged one may be slightly  closer to the trend of GMEs' results, especially to that of the F-GME---since it is not sensitive to the initial condition as the GMEs.

The comparison of the four master
equation
[Fig.~\ref{fig:CoupledTLS_etad}(d)] 
helps to illustrate the distinct roles of interaction-induced and drive-induced dressing. We highlight that the TI-GME incorporates the internal dressing generated by the dipole--dipole interaction but neglects the external Floquet dressing in the construction of the dissipative channels.
Thus, one can assume that it can approximate the correct predictions for the range of $\eta_d\lesssim\eta$, where  the internal dressing is more crucial than external dressing.
Conversely, the F-ME incorporates the 
external Floquet dressing but neglects the internal dressed-state structure of the interacting system. Only the F-GME consistently treats both mechanisms simultaneously. As a result, neither the TI-GME nor the F-ME reproduces the F-GME prediction over the full range of drive amplitudes, demonstrating that the dissipative dynamics of a strongly driven strongly interacting system are governed by the combined effect of both dressing mechanisms.

We note the F-ME does not fail because it is numerically represented in the bare basis,  as the unitary basis transformation does not change the physics. It fails because its dissipator is resolved using bare (lab) transition channels before the dipole-dipole interaction is incorporated into the dissipative eigenoperators. Consequently, the interaction can enter the coherent Floquet dynamics while remaining absent from the physical definition of the bath-induced transition channels. The F-GME instead resolves the bath coupling after diagonalizing the interacting Hamiltonian, so both the interaction-induced hybridization and the drive-induced Floquet dressing enter the dissipator.

A notable feature, in Fig.~\ref{fig:CoupledTLS_etad}(d), is the cusp-like discontinuity near $\eta_d\approx0.95$ in the F-GME result. This feature originates from a reorganization of the dominant quasienergy-resolved dissipative channels when they pass zero, as seen by the green lines in panel (c), for the same reason that explain before in Fig.~\ref{fig:TLS_etad}(e).

\subsection{Spectrally resolved results}

\begin{figure*}[!htpb]
\centering
\includegraphics[width=.85\linewidth]
{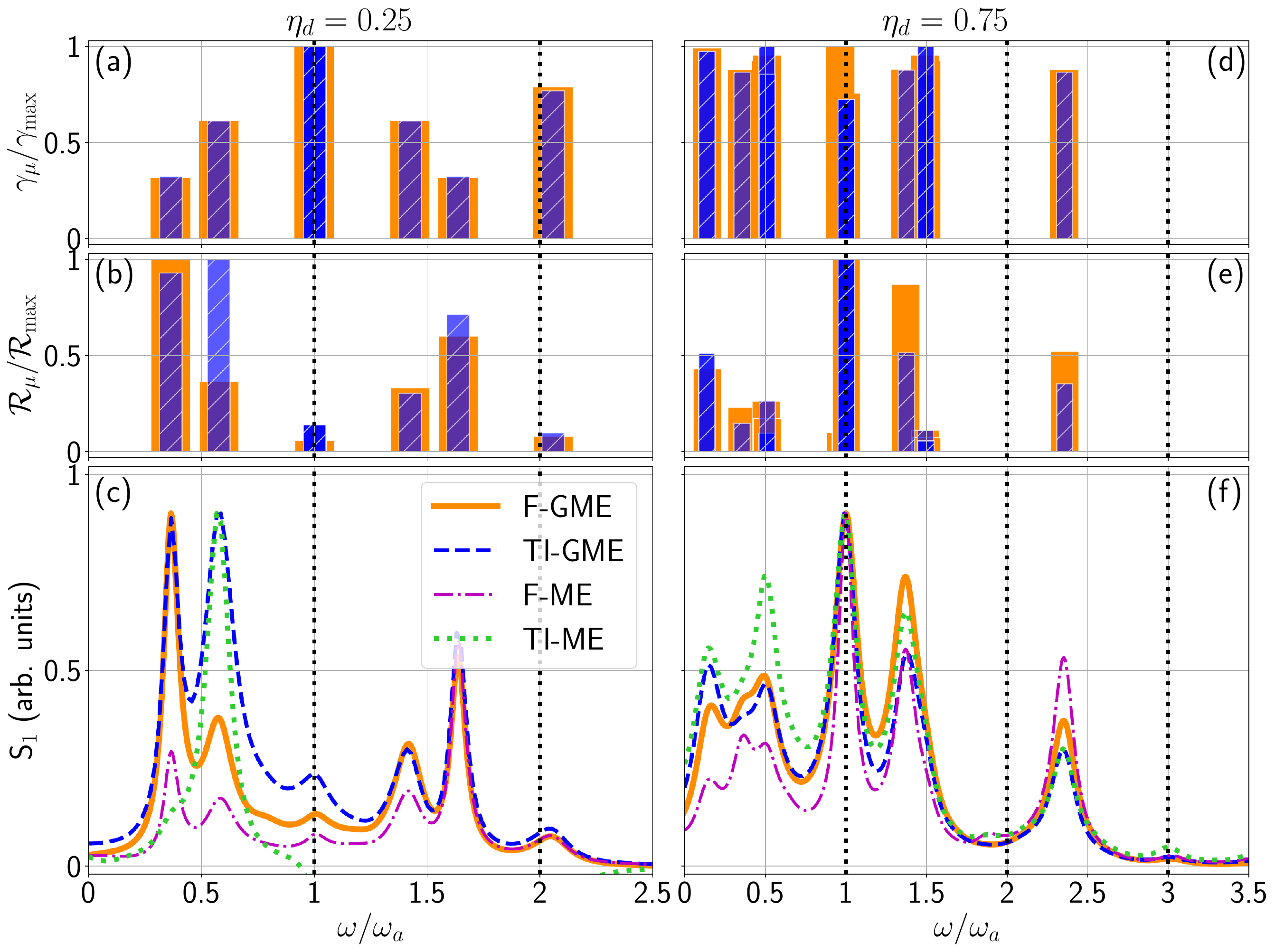}
\caption{
\textbf{Driven-dissipative coupled-TLS: spectra.}
Shown are the incoherent emission spectra, $\mathsf{S}_1(\omega)$ [Eq.~\eqref{eq:Spectrum_def}], and the corresponding FL pole analysis for two coupled two-level systems, where the optical drive acts coherently on TLS 1. The parameters are $\omega_d=\omega_a$, $\eta=0.5$, and $\gamma=0.1\,\omega_a$. The left column corresponds to $\eta_d=0.25$, and the right column to $\eta_d=0.75$. In each column, the top panel shows the normalized decay rates $\gamma_\mu/\gamma_{\max}$ of the dominant positive-frequency FL modes, the middle panel shows the corresponding normalized peak-strength factors $\mathcal{R}_\mu/\mathcal{R}_{\max}$, and the bottom panel shows the incoherent spectrum of the first TLS (the one that is also directly driven) obtained from the F-GME/FL approach (solid orange), the TI-GME approach (dashed blue) and the F-ME approach (dotted magenta). The vertical dotted lines mark the integer multiples of the drive frequency. The bar panels compare the dominant poles predicted by the two approaches and indicate how the observable resonances are organized in the driven-dissipative coupled system.
The TI-ME spectrum in (c) becomes unphysical because of the wrong resolution of the dissipative channels in the nonsecular dissipator (see text). 
}
\label{fig:CoupledTLSSpectra1_wd1_Flat}
\end{figure*}

Figure~\ref{fig:CoupledTLSSpectra1_wd1_Flat} shows the  spectrum of the first TLS (i.e., the one that is periodically pumped), $\mathsf{S}_1$ from Eq.~\eqref{eq:Spectrum_def}, in the coupled-TLS system;
the first TLS is driven coherently on resonance, $\omega_d=\omega_a$, for two drive amplitudes, $\eta_d=0.25$ (strong drive $0.1\lesssim\eta_d\lesssim\eta$) and $0.75$ (deep-strong drive $\eta_d\gtrsim\eta$). In contrast to the single-TLS problem, the coupled system possesses a richer quasienergy structure and therefore a larger set of allowed Floquet transition channels is now possible. Thus, the visible spectral resonances are best understood not directly in terms of individual Floquet transition frequencies $\Delta_{\alpha\beta l}$, but rather in terms of the exact FL poles
$\lambda_\mu=-\gamma_\mu-\ii \Delta_\mu $,
which determine the true resonance frequencies, linewidths, and modal strengths in the driven-dissipative problem. The top and middle panels display, respectively, the normalized linewidths $\gamma_\mu$ and peak-strength factors (mode prominence) $\mathcal{R}_\mu$ of the dominant modes, while the bottom panels show the resulting spectra.

To help appreciate the key role of the double dressing on 
an equal footing, the results of the F-GME (solid orange) are compared with those of the TI-GME (dashed blue)---which only include the internal dressing, and those of the F-ME (dotted magenta) and TI-ME (dotted light blue)---which F-ME includes external dressing only, whereas TI-ME includes neither.

For the strong drive strength, $\eta_d=0.25$ [left column, panels (a)--(c) of Fig.~\ref{fig:CoupledTLSSpectra1_wd1_Flat}], the dominant positive-frequency resonances occur at
\begin{equation*}
   \Delta_\mu/\omega_a\simeq 0.364,\;0.581,\;0.781,\;1,\;1.42,\;1.637,\;2.057, 
\end{equation*}
in the F-GME calculation, with the TI-GME prediction giving very similar values.
This close agreement in the pole positions explains why both approaches predict the same overall spectral pattern in panel (c). The resonance at $\omega/\omega_a\approx0.36$ is the strongest mode in both descriptions, and the two side structures around $\omega/\omega_a\approx 0.58$ and $1.64$ are also reproduced by both approaches at nearly the same frequencies. Likewise, the broader shoulders around $\omega/\omega_a\approx 0.36$ and $1.42$ and the weak feature near $\omega/\omega_a\approx 2.06$ are present in both model treatments.

However, the different master equation theories do not predict identical spectra
(the TI-ME  predicts totally unphysical spectrum as it is negative and indeed wrong). The difference is 
not in the resonance locations, but in the distribution of spectral weight among them. This is already evident from panel (b)
of Fig.~\ref{fig:CoupledTLSSpectra1_wd1_Flat}, where the relative values of $\mathcal{R}_\mu$ differ more noticeably than the linewidths shown in panel (a). In particular, the TI-GME assigns more weight to the lower side structure near $\omega/\omega_a\sim 0.58$, whereas the F-GME enhances the resonance near $\omega/\omega_a\sim 1.64$. This redistribution is reflected directly in panel (c): the dashed TI-GME curve exhibits a larger low-frequency side peak, while the solid F-GME curve produces a stronger higher-frequency side peak. Therefore, even though both theories agree well on the pole frequencies and linewidth hierarchy, the nonsecular Floquet treatment predicts a quantitatively different spectral-weight arrangement on the spectrum.

This behavior indicates that, already at $\eta_d=0.25$, the coupled-TLS spectrum is governed by exact dissipative modes rather than by a naive one-to-one assignment to isolated Floquet transitions. The visible peaks are controlled by $\Delta_\mu$, while the underlying Floquet channels $\Delta_{\alpha\beta l}$ merely reveal which coherent processes participate in each exact mode. In other words, the observed resonances are collective driven-dissipative modes of the periodic Liouvillian.

When the drive is increased to $\eta_d=0.75$ [right column, panels (d)--(f) of Fig.~\ref{fig:CoupledTLSSpectra1_wd1_Flat}], the spectral reorganization becomes more pronounced. The dominant F-GME resonances now appear at
\begin{align*}
    \Delta_\mu/\omega_a \simeq \  &0.137,\;0.354,\;0.504,\;0.513, \\ \nonumber 
    &1,\;1.372,\;1.496,\;2.354, \nonumber
\end{align*}
and roughly, the similar values for the TI-GME.
The strongest resonance remains the central peak at $\omega=\omega_d=\omega_a$ as the drive pushes the system dominantly in this case, but the low-frequency sector below $\omega_a$ and the sideband sector around $\omega/\omega_a\approx 1.4$ become substantially more structured. In panel (f), the TI-GME predicts a visibly stronger pedestal and stronger low-frequency side features near $\omega/\omega_a\approx 0.14$, $0.35$, and $0.5$, whereas the F-GME suppresses these low-frequency contributions and transfers more weight toward the peak near $\omega/\omega_a\approx 1.37$ and especially the feature near $\omega/\omega_a\approx 2.35$. This shift is consistent with the bar analysis in panel (e), where the strongest $\mathcal{R}_\mu$ values occur at different off-central frequencies in the two approaches.

A particularly important feature of the deep-strong-drive case is the emergence of closely spaced modes in the low-frequency sector and around $\omega/\omega_a\approx 0.5$. Although these modes are individually resolved at the FL level, they interfere in the final spectrum and generate broad shoulders or composite spectral features rather than completely isolated Lorentzian peaks. The exact resonance peaks at $\Delta_\mu\approx 0.5037$ and $0.5131$, for example, lie very close to one another and therefore contribute jointly to the structured sideband just below the central peak. Likewise, the modes near $\Delta_\mu\approx 1.37$ and $1.50$ build the higher-frequency sideband structure to the right of the central resonance. Thus, the spectrum is no longer well described as a sum of independent single-transition peaks; instead, several nearby dissipative modes combine to produce hybrid sidebands.

Crucially, Fig.~\ref{fig:CoupledTLSSpectra1_wd1_Flat} shows that the modal strengths $\mathcal{R}_\mu$, rather than the pole positions alone, provide the most sensitive distinction between the different dissipative descriptions. From Eq.~\eqref{eq:FL_spectrum_main}, while $\Delta_\mu$ determines where a resonance occurs and $\gamma_\mu$ determines its characteristic width, $\mathcal{R}_\mu$ quantifies how strongly that FL mode contributes to the detected spectrum. Similar to our arguments before, since the F-GME and TI-GME employ the same coherent driven Hamiltonian, their pole frequencies can remain close even when their dissipative dynamics differ substantially. In contrast, the mode residue $\mathcal{W}_\mu$, and hence mode prominence $\mathcal{R}_\mu$, depends on the left and right FL eigenmodes together with the periodic steady-state source and detection overlaps, and is therefore directly sensitive to how the dissipative channels are resolved and mixed.
These values are 
shown in the bar plots of Fig.~\ref{fig:CoupledTLSSpectra1_wd1_Flat}(b) and (e), for $\eta_d=0.25$ and $0.75$, respectively, where for the latter, the redistribution becomes substantially stronger as several Floquet channels become nearly resonant and hybridize through the nonsecular dissipator. Moreover, the discrepancy between the dashed blue and solid orange bars reveals that the resulting FL eigenvectors, and therefore their residues, differ appreciably between the two theories even though the pole locations remain similar. 

The comparison of the F-GME curves and the F-ME curves also points out that models that assume only external dressing are not sufficient to yield the correct prediction for these observables that we show. Assigning the transition pathways of $\Delta_{\alpha\beta l}$ on the basis of the bare representation $g_{1,2},e_{1,2}$ can also produce wrong results. Hence, the external dressing must be applied to the correct internal basis for arbitrary interacting systems, which is the internally (interaction-induced) dressed basis, $\ket{\alpha(t)}=\sum_{l,j}\mathrm{e}^{-\ii l\omega_dt}\,f_{\alpha l j}\ket{j}$. Finally, when the internal strong interaction, strong drive and dissipation coexist, one is required to include the 
{\it double dressing} simultaneously with Floquet theory. 

This stronger hybridization also clarifies the role of the F-GME. In the coupled-TLS system, periodic driving does not merely shift transition energies; rather, it also reshapes the dissipative couplings among the Floquet sectors. When different Floquet channels become nearly resonant, the nonsecular terms retained in the F-GME modify the mixing among them and thereby alter the residues of the exact FL poles. The result is that the F-GME and TI-GME may produce similar resonance frequencies but substantially different intensities and lineshapes, especially away from the main central peak. This is precisely what is observed in panel (f): the resonance positions remain close, while the spectral weights are redistributed in a nontrivial way.

Overall, Fig.~\ref{fig:CoupledTLSSpectra1_wd1_Flat} shows that the coupled-TLS spectrum is organized by the exact FL modes of the periodically driven open system. At strong drive fields ($0.1\lesssim\eta_d\lesssim\eta$), these modes already control the relative sideband intensities even though the spectral pattern remains fairly simple. At stronger drive or deep-strong drive, ($\eta_d\gtrsim\eta$), the mode mixing becomes appreciable, the sideband structure becomes hybridized, and the difference between the F-GME and TI-GME predictions is more evident. The FL analysis, therefore, provides the natural microscopic interpretation of the spectra; the frequencies $\Delta_\mu$ determine where the resonances occur, the linewidths $\gamma_\mu$ determine how broad they are, and the modal strengths $\mathcal{R}_\mu$ explain why the F-GME and TI-GME spectra differ even when their peak positions remain close.

\section{Conclusions and Discussions}
\label{sec:Conclusions}

We have formulated a Floquet--Markov generalized master equation (F-GME) for a general 
periodically-driven open quantum system that is weakly coupled to a Markovian environment,  and applied it to two minimal yet representative models: 
(i) a driven-dissipative TLS. and 
(ii) a driven-dissipative pair of coupled TLSs (which are effectively dipole coupled). This allowed us to investigate the microscopic origin of dissipation in periodically driven quantum systems without invoking either the rotating-wave approximation  or secularization.

The comparison was performed against a widely used class of generalized master equations formulated in a time-independent basis for strongly hybridized quantum systems~\cite{Settineri_Dissipation_2018,Salmon_Gauge-independent_2022}, referred to here as TI-GMEs. These approaches incorporate the effects of internal (interaction-induced) dressing through the diagonalization of the static interacting Hamiltonian, but treat dissipation using transition channels defined in a time-independent basis. Consequently, while they correctly capture the internal structure of strongly interacting systems, they do not fully resolve the additional Floquet sideband channels generated by external periodic driving. The present work thus provides a controlled framework for assessing when such time-independent-basis descriptions remain reliable and when a fully quasienergy-resolved Floquet treatment becomes essential.

By comparing the F-GME with its time-independent counterpart, we demonstrated a qualitative failure of a widely used modeling strategy for periodically driven open quantum systems (TI-GME). Remarkably, this failure already appears in the most elementary setting considered here, namely: a driven TLS coupled weakly to a flat Markovian bath. Moreover, the discrepancy manifests directly in steady-state observables rather than only in transient dynamics or subtle higher-order quantities. Our results, therefore, show that even in the system-bath weak-coupling regime, with flat baths, and a fully nonsecular approach, time-independent dissipative descriptions can systematically misweight Floquet transition channels and consequently predict qualitatively incorrect steady states, despite sometimes reproducing apparently correct emission spectra.

The dipole-dipole pair of TLSs provides an even more revealing test of our theory because the interacting Hamiltonian must already be treated in the dressed basis prior to the application of the drive. In this case, both TI-GME and F-GME consistently incorporate the effects of internal dressing, whereas only the F-GME accounts for the additional external (drive-induced) dressing and its associated Floquet sideband channels. Comparison with the (conventional) bare-basis Floquet master equation further demonstrates that neither internal nor external dressing alone is sufficient. Instead, a consistent description requires both forms of dressing to be treated on an equal footing, such that the dissipator resolves the quasienergy channels generated by the jointly dressed system. Thus, the discrepancy between TI-GME and F-GME originates not from the treatment of the internal structure of the interacting system, but from the treatment of the externally generated Floquet channels.

Although the driven TLS and coupled-TLS systems studied here are minimal models, they reveal several important and general insights into the dissipation mechanism of periodically driven quantum systems. Most notably, we find that quasienergy-resolved dissipation remains essential even for spectrally flat baths. This observation is both nontrivial and counterintuitive, since it is often implicitly assumed that a flat bath washes out or renders irrelevant the underlying Floquet structure. 

Our analysis demonstrates that this is not the case. Furthermore, the correct description of dissipation is shown to depend not merely on whether nonsecular terms are retained, but on how the dissipative channels are resolved in frequency space. By comparing nonsecular versions of both TI-GME and F-GME, we demonstrated that the discrepancy persists even when secularization is completely eliminated. The decisive ingredient is therefore the quasienergy-resolved frequency decomposition of the dissipative processes rather than secularization itself, thereby resolving a long-standing ambiguity regarding the origin of the mismatch between Floquet and time-independent dissipative descriptions.

A particularly important observation is that agreement at the level of emission spectra does not guarantee agreement at the level of steady-state populations, or vice versa. Several examples considered here exhibit nearly identical spectral peak positions and overall lineshapes while predicting substantially different populations. Consequently, spectral agreement alone cannot be regarded as sufficient validation of a dissipative model for a periodically driven open quantum system.

Beyond the comparison of master-equation approaches, the Floquet--Liouville (FL) analysis developed here provides a physically transparent description of the dissipative dynamics. The FL eigenmodes constitute the natural dynamical modes of the driven open quantum system, with each mode characterized by a complex eigenvalue whose real and imaginary parts determine the effective decay rate and oscillation frequency, respectively. The decomposition of these modes into quasienergy channels reveals how multiple Floquet transitions may hybridize into collective dissipative modes, thereby providing a direct connection between the microscopic Floquet structure and observable spectral features. In this way, the FL formalism extends the quasienergy-channel picture beyond the master equation itself and provides a natural framework for interpreting both populations and spectra.

The microscopic origin of the discrepancy between TI-GME and F-GME can be understood directly from the channel structure of the dissipator.
This is manifested in their difference in the system-bath coupling operator content, bath function, and the nonsecularity of channels. 
In the Floquet description, the total transition strength is distributed among multiple quasienergy channels $(\alpha,\beta,l)$ 
that satisfy:
\begin{equation*}
\sum_{jk}
\left|S^\pm_{B,jk}\right|^2
=
\sum_{\alpha\beta l}
\left|S^\pm_{B,\alpha\beta l}\right|^2,
\end{equation*}
whereas the TI-GME effectively collapses this distribution into a single primary channel with $l=0$. Consequently, the dissipative weights are redistributed incorrectly whenever multiple Floquet sidebands contribute appreciably. The resulting population dynamics may therefore differ qualitatively even though the total transition strength remains unchanged. From this perspective, the discrepancy originates from the redistribution of dissipative weight among quasienergy channels rather than from bath structure or secularization itself.

The present work also provides a channel-resolved realization of the conceptual framework developed by Hone, Ketzmerick, and Kohn~\cite{Hone_Statistical_2009}; while that work established the importance of quasienergy near-degeneracies and nonsecular effects in Floquet statistical mechanics, our analysis makes these effects explicit at the level of individual quasienergy-difference channels. By tracking the activation and reorganization of the transitions,
\[
\Delta_{\alpha\beta l}
=
\varepsilon_\beta-\varepsilon_\alpha+l\omega_d,
\]
we directly connect avoided crossings in the extended quasienergy spectrum to observable signatures in steady-state populations and emission spectra.

We also investigated the off-resonantly driven cases and the effect of structured baths. Note the bath frequency sampling can influence a wide range of Floquet sidebands, while the only static gaps are affected in the time-independent dissipators.
The combination of drive parameters and bath structure can be utilized in the engineering of the operating regime.
When off resonance, the correct bath frequency sampling should be at the laser, and hence, the static master equations and GMEs would fail dramatically.

Broadly, our results lead to three central conclusions: (i) periodic driving generates multiple Floquet sideband channels that contribute to the dissipative dynamics; (ii) time-independent dissipators effectively employ a single set of rates associated with static energy gaps and therefore cannot generally describe the redistribution of dissipation among Floquet channels; and (iii) quasienergy-resolved dissipators naturally assign channel-dependent rates and weights, leading to the correct steady-state populations and spectral response. Accordingly, the TI-GME remains reliable only when a single Floquet channel dominates the dynamics or when the relevant quasienergy channels are sufficiently well separated that their mutual mixing is negligible.

Although demonstrated here for minimal driven TLS and coupled-TLS systems, the underlying mechanism is completely general and applies to periodically driven open quantum systems ranging from cavity-QED and circuit-QED platforms to quantum simulators, quantum transport devices, and driven many-body systems. The results therefore suggest that quasienergy-resolved dissipation and FL modal analysis should form the natural framework for understanding relaxation and spectroscopy in driven open quantum matter.
Dissipation in periodically driven open quantum systems is intrinsically quasienergy resolved, and the correct dynamical degrees of freedom are Floquet channels and FL modes rather than static dressed-state transitions.

In the very weak-drive regime, the drive-induced populations become vanishingly small, so spectral observables become increasingly difficult to resolve numerically--most acutely near the $\eta_d\to0$
 splitting of the quasienergy lines in the resonant case. In this limit, the F-GME continuously reduces to the static generalized master equation, and for a single standalone system driven on resonance, such as the driven-dissipative TLS, either description suffices for practical purposes. This equivalence is not generic, however, and the time-independent and Floquet descriptions' disagreement might not necessarily be monotonic within the drive parameter space. Already for a single TLS driven off resonance, the bath must be sampled at the drive frequency rather than at the bare transition, so the time-independent dissipator fails even at weak drive~\cite{Kowalewska-kudlaszyk_Generalized_2001,Ge_Accessing_2013}. More broadly, in strongly interacting systems, the quasienergies are confined to a Brillouin zone of width $\omega_d$ and may become near-degenerate, activating nonsecular dissipative couplings even outside the high-field regime. A fully quasienergy-resolved treatment is therefore required whenever the relevant transition channels depart from the static gaps, $\Delta_{\alpha\beta l}\neq\omega_{jk}$
 and $\lvert\Delta_{\alpha\beta l}-\Delta_{\alpha'\beta'l'}\rvert\lesssim\Gamma$, independently of drive strength.

\acknowledgements
 This work was supported by the Natural Sciences and Engineering Research Council of Canada (NSERC),
 the Canadian Foundation for Innovation (CFI), and Queen's University, Canada.
 The authors acknowledge the 
 use of ChatGPT by OpenAI and Claude AI for some assistance with manuscript preparation and coding. 
 All AI-assisted output was directed, reviewed, corrected, and independently verified by the authors. The final scientific claims, equations, numerical implementations, figures, and conclusions are the sole responsibility of the authors.
 We thank Franco Nori and Alberto Mercurio for the useful discussion.

\appendix

\vspace{1cm}

\section{Floquet theory for the closed driven system in the interaction-induced dressed picture}
\label{secS:FloquetTheory_ClosedSystem}

 For driven quantum systems, such as those studied in the main text,  the full time-dependent Hamiltonian is periodic, ${H}(t)={H}(t+T)$ with $T=2\pi/\omega_d$, where $\omega_d$ is the drive frequency. 
 To study the dynamics of the system, we seek to solve the time-dependent Schrodinger equation
\begin{equation}
    \ii \frac{\dd\lvert\psi(t)\rangle}{\dd t}={H}(t)\lvert\psi(t)\rangle,
    \label{TDSE}
\end{equation} 
with a $T$-periodic time-dependent Hamiltonian, that
can, in general, be expanded as a Fourier series 
 given in Eq.~\eqref{eq:Ht} of the main text, $H(t)=\sum_mH_m\,\exp(-\ii m\omega_d t)$.

In a strongly hybridized quantum, working in the dressed basis is a must~\cite{Settineri_Dissipation_2018,DiStefano_Resolution_2019}.
The dressed picture starts with the diagonalization of the {\it time-independent} Hamiltonian, $H_0$, as a set of eigenstates basis $\mathsf{B}_{\rm dressed}=\lcb\ket{j}\rcb$ of energies $\{E_j\}$, to span the physical Hilbert space $\mathbb{H}_{\rm dressed}$. 
Accordingly, all operators must be redefined to accommodate correct physical transitions among the dressed basis rather than the bare basis before including any interaction. Specifically, the ladder operators in the bare picture $s^{\Lambda\pm}_{\rm bare}$ for a subsystem $\Lambda$ forming the system operator $S^{\Lambda,{\rm bare}}=s^{\Lambda+,\rm bare}+s^{\Lambda-,\rm bare}$, is defined in the dressed picture via
\begin{equation}
\begin{split}
s^{\Lambda+}
&=\sum_{j,k>j}\braket{j\vert S^{\Lambda,\rm bare}\vert k}\ket{j}\bra{k},
\end{split}
\label{eq:ladderOp_dressed}
\end{equation} 
with $s^{\Lambda-}= [s^{\Lambda+}]^\dagger$ and $S^\Lambda=s^{\Lambda+}+s^{\Lambda-}$.
Then a periodic CW field (optically) coherently pumping the system, can be added in the form of Eq.~\eqref{eq:Ht_d} of the main text.

After properly accounting for the internal (interaction-induced) dressing, we proceed to incorporate the external dressing arising from the time-periodic drive using Floquet theory. The resulting externally dressed system is no longer described by the eigenstates of the static interacting Hamiltonian, but rather by Floquet states characterized by quasienergies and their associated sidebands. 
We therefore seek a Floquet representation of the driven time-periodic Hamiltonian, in which the dynamics can be expressed in terms of these quasienergy states~\cite{Grifoni_Driven_1998,LeBoite_Theoretical_2020}, i.e., $\lvert\psi(t)\rangle=\sum_\alpha c_\alpha\lvert\psi_\alpha(t)\rangle$, where the Floquet states read 
\begin{equation}
\begin{split}
\lvert\psi_\alpha(t)\rangle
&=\mathrm{e}^{-\ii\varepsilon_\alpha t}\lvert\alpha(t)\rangle,
\end{split}
\label{FloquetState}
\end{equation} 
with $c_\alpha\equiv c(t_0)=\langle\alpha(t_0)\vert\psi(t_0)\rangle$  being the time-independent complex coefficients, and $\vert \alpha(t)\rangle$ the Floquet modes corresponding to the Floquet quasienergy $\varepsilon_\alpha$~\cite{Nikishov_Quantum_1964}, i.e.,
\begin{equation}
    \begin{split}
        {H}^\mathrm{F}(t)\vert \alpha(t)\rangle=\varepsilon_\alpha\vert \alpha(t)\rangle, 
    \end{split}
    \label{eq:FloquetTDSE}
\end{equation}
with the Floquet Hamiltonian 
\begin{equation}
    {H}^\mathrm{F}(t) \equiv{H}(t)-\ii\partial_t.
    \label{eq:HF_t}
\end{equation}

Since the Floquet modes are periodic, they can also be expanded in the Fourier series,
\begin{equation}
\begin{split}
\lvert \alpha(t)\rangle&=\sum_{l\in\mathbb{Z}}\mathrm{e}^{-\ii l\omega_d t}\vert {\alpha_l}\rangle,
\end{split}
\label{FloquetMode}
\end{equation}
where $\lvert \alpha_l\rangle$ are the Fourier coefficient states called the Floquet sidebands. 
Moreover, in a general quantum  
basis set $\{\lvert j\rangle\}$, obtained for the time-independent portion of the Hamiltonian, ${H}_0$,  one can always expand the Floquet sidebands as a superposition 
$\lvert \alpha_l\rangle=\sum_{j}f_{\alpha lj}\lvert j\rangle$, and hence expand the Floquet modes in terms of the time-independent dressed, to present the the double dressing.

\subsection{Floquet extended-space (Sambe-space) formalism for the closed system}
\label{}

A central advantage of Floquet theory is that an explicitly time-periodic Schr\"odinger problem can be mapped onto a time-independent eigenvalue problem in an enlarged Hilbert space known as the \emph{Floquet extended space} or \emph{Sambe space}~\cite{Sambe_Steady_1973}. This construction allows periodically driven quantum systems to be treated using methods closely analogous to those employed for static Hamiltonians.

The key observation is that all periodic quantities appearing in the Floquet formalism may be expanded in harmonics of the drive frequency. One may therefore regard the oscillatory functions $\mathrm{e}^{-\ii l\omega_d t}$ as basis states of an auxiliary temporal Hilbert space. Introducing the correspondence
\begin{equation}
\mathrm{e}^{-\ii l\omega_d t}
\longleftrightarrow
\tket{l},
\end{equation}
the physical Hilbert space can be enlarged to
\begin{equation}
\mathbb{H}_{\rm ex}
=
\mathbb{H}_{\rm temp}
\otimes
\mathbb{H}_{\rm dressed},
\end{equation}
where $\mathbb{H}_{\rm temp}\equiv L_2[0,T]$ is the space of square-integrable periodic functions and $\mathbb{H}_{\rm dressed}$ denotes the physical Hilbert space of the system. The resulting space $\mathbb{H}_{\rm ex}$ is known as the Floquet extended space or Sambe space~\cite{Sambe_Steady_1973}.

\subsubsection{Temporal Hilbert space and extended-space basis}

The temporal Hilbert space $\mathbb{H}_{\rm temp}$ 
has states that satisfy the inner product
\begin{equation}
(f,g)
=
\frac{1}{T}
\int_0^T dt\,
f^*(t)g(t).
\end{equation}
A convenient orthonormal basis is the Fourier basis, 
\begin{equation}
\mathsf{B}_{\rm temp}
=
\{\tket{l}\}_{l\in\mathbb Z},
\qquad
(t|l)
=
\ee^{-\ii l\omega_d t},
\end{equation}
which satisfies
\begin{equation}
(l|l')
=
\delta_{ll'}.
\label{eq:temporal_orthogonality}
\end{equation}

The extended space $\mathbb{H}_{\rm ex}$ is therefore
spanned with the basis
\begin{equation}
\mathsf{B}_{\rm ex}
=
\{
\Fket{l,j}
\equiv
\tket{l}\otimes\ket{j}
\},
\qquad
l\in\mathbb Z,
\end{equation}
where $\{\ket{j}\}$ denotes an arbitrary basis of the physical Hilbert space. Throughout this work, ordinary kets $\ket{\cdot}$ refer to vectors in the physical Hilbert space, whereas double kets $\Fket{\cdot}$ denote vectors in the extended space.

The natural scalar product in the extended space is the period-averaged inner product
\begin{equation}
\Fbraket{\phi}{\chi}
\equiv
\frac{1}{T}
\int_0^T dt\,
\braket{\phi(t)|\chi(t)}.
\label{eq:Sambe_inner}
\end{equation}

Any $T$-periodic operator $O(t)$ admits the Fourier expansion
\begin{equation}
O(t)
=
\sum_{m\in\mathbb Z}
\ee^{-\ii m\omega_d t}
O_m,
\end{equation}
with Fourier components
\begin{equation}
O_m
=
\frac{1}{T}
\int_0^T dt\,
\ee^{\ii m\omega_d t}
O(t).
\label{eq:O_fourier}
\end{equation}
Thus, its representation in the extended space is
\begin{equation}
O_{\rm ex}
=
\sum_{m\in\mathbb Z}
F_m\otimes O_m,
\label{eq:O_extended}
\end{equation}
where $F_m$ acts on the temporal Hilbert space as a (temporal) shift operator~\cite{Restrepo_Driven_2014},
\begin{equation}
F_m\tket{l}=
\tket{l+m}.
\label{eq:TemporalShift_Op}
\end{equation}

It is also useful to introduce the temporal number operator
\begin{equation}
N_{\rm temp}\tket{l}
=
l\tket{l},
\end{equation}
which generates translations between Floquet sidebands.

\subsubsection{Time-independent Floquet eigenvalue problem in the extended space}

The Floquet quasienergy operator (or, the Floquet Hamiltonian) is defined in Eq.~\eqref{eq:HF_t}.
Using the Fourier representation of the Hamiltonian,
the matrix elements of the Floquet operator in the extended basis become
\begin{equation}
\begin{aligned}
\Fbraket{l,j}{H^{\rm F}_{\rm ex}|l',j'}
&=
\frac{1}{T}
\int_0^T dt\,
\ee^{\ii l\omega_d t}
\bra{j}
\left[
H(t)-\ii\partial_t
\right]
\ket{j'}
\ee^{-\ii l'\omega_d t}
\\
&=
\bra{j}
H_{l-l'}
\ket{j'}
-
l\omega_d
\delta_{ll'}
\delta_{jj'}.
\end{aligned}
\label{eq:Hf_matrix_elements}
\end{equation}

Equivalently,
\begin{equation}
H^{\rm F}_{\rm ex}
=
\sum_{m\in\mathbb Z}
F_m\otimes H_m
-
\omega_d
N_{\rm temp}\otimes \mathbf{1}_{\rm dressed},
\label{eq:Hf_extended}
\end{equation}
where $\mathbf{1}_{\rm dressed}$ is the identity in the physical dressed Hilbert space.

The Floquet problem therefore becomes a time-independent eigenvalue problem in the extended space,
\begin{equation}
H^{\rm F}_{\rm ex}
\Fket{\alpha}
=
\varepsilon_\alpha
\Fket{\alpha},
\label{eq:Floquet_eigenproblem_ex}
\end{equation}
which is formally identical to an ordinary stationary Schr\"odinger equation.
The corresponding extended-space eigenvector may be expanded as
\begin{equation}
\Fket{\alpha}
=
\sum_{l\in\mathbb Z}
\tket{l}
\otimes
\ket{\alpha_l}
=
\sum_{l,j}
f_{\alpha lj}
\Fket{l,j},
\label{eq:Floquet_extended_vector}
\end{equation}
where $\ket{\alpha_l}$ are the Fourier components (Floquet sidebands) of the periodic Floquet mode.

The physical Floquet mode is reconstructed directly from the extended-space eigenvector:
\begin{equation}
\ket{\alpha(t)}
=
\sum_{l\in\mathbb Z}
\ee^{-\ii l\omega_d t}
\ket{\alpha_l},
\label{eq:Floquet_mode_fourier}
\end{equation}
or equivalently
\begin{equation}
\ket{\alpha(t)}
=
(t\Fket{\alpha}
=
\sum_{j,l}
\ee^{-\ii l\omega_d t}
\Fbraket{l,j}{\alpha}
\ket{j}.
\label{eq:alpha_t_from_extended}
\end{equation}

The Fourier coefficients are therefore identified directly as
\begin{equation}
\braket{j|\alpha_l}
=
\Fbraket{l,j}{\alpha},
\label{eq:Fourier_coeff_extended}
\end{equation}
and the eigenvectors of $H_{\rm ex}^{\rm F}$ may be chosen orthonormal and complete,
\begin{equation}
\Fbraket{\alpha}{\beta}
=
\delta_{\alpha\beta},
\qquad
\sum_\alpha
\Fket{\alpha}
\Fbra{\alpha}
=
\mathbf{1}_{\rm ex}.
\label{eq:orthonormality_extended}
\end{equation}

Thus, by solving the time-independent eigenvalue problem in Eq.~\eqref{eq:Floquet_eigenproblem_ex}, one simultaneously obtains the quasienergies, Floquet sidebands, Floquet modes, and ultimately the complete driven dynamics of the closed system~\cite{Eckardt_High-frequency_2015,Restrepo_Quantum_2018,Restrepo_Driven_2019}.

\subsubsection{Quasienergy Brillouin zones and sideband redundancy}

Since the Hamiltonian is periodic in time, quasienergies are defined only modulo the drive frequency. Indeed, if $\ket{\alpha(t)}$ is a Floquet mode with quasienergy $\varepsilon_\alpha$, then
\begin{equation}
\ket{\alpha^{[n]}(t)}
\equiv
\ee^{+\ii n\omega_d t}
\ket{\alpha(t)}
\label{eq:FloquetMode_BZShift}
\end{equation}
is also a valid Floquet mode with shifted quasienergy
\begin{equation}
\varepsilon_\alpha^{[n]}
=
\varepsilon_\alpha
+
n\omega_d,
\qquad
n\in\mathbb Z.
\label{eq:FloquetQuasienergy_BZShift}
\end{equation}

The corresponding Schr\"odinger state remains unchanged,
\begin{equation}
\ee^{-\ii \varepsilon_\alpha^{[n]}t}
\ket{\alpha^{[n]}(t)}
=
\ee^{-\ii\varepsilon_\alpha t}
\ket{\alpha(t)},
\end{equation}
showing that these solutions are physically equivalent.

Consequently, the extended-space eigenvalue problem possesses infinitely many replica solutions,
\begin{equation}
H^{\rm F}_{\rm ex}
\Fket{n\alpha}
=
\varepsilon_{n\alpha}
\Fket{n\alpha},
\label{eq:Floquet_eigenproblem_ex_modified}
\end{equation}
with
\begin{equation}
\varepsilon_{n\alpha}
=
\varepsilon_\alpha
+
n\omega_d.
\end{equation}
These replicas form quasienergy Brillouin zones throughout the spectrum. Since all replicas correspond to the same physical Floquet state, it is sufficient to retain one representative per equivalence class, for example by restricting the quasienergies to the first Brillouin zone,
\begin{equation}
-\frac{\omega_d}{2}
\le
\varepsilon_\alpha
<
\frac{\omega_d}{2}.
\label{eq:first_BZ}
\end{equation}

The associated sidebands satisfy a simple translation rule. Writing
\begin{equation}
\ket{\alpha^{[n]}(t)}
=
\sum_l
\ee^{-\ii l\omega_d t}
\ket{\alpha_l^{[n]}},
\end{equation}
one finds
\begin{equation}
\ket{\alpha_l^{[n]}}
=
\ket{\alpha_{l+ n}},
\label{eq:sideband_shift}
\end{equation}
which simply reflects the redundancy between neighboring Brillouin zones.
By convention, we therefore define
\begin{equation}
\varepsilon_\alpha
\equiv
\varepsilon_{0\alpha},
\qquad
\Fket{\alpha}
\equiv
\Fket{0\alpha},
\end{equation}
and perform all calculations using the representatives within a single Brillouin zone.

We emphasize that for strongly driven systems, the relevant symmetries, selection rules, and transition channels are determined by the Floquet quasienergy operator $H_{\rm ex}^{\rm F}$ in the extended space rather than by the instantaneous Hamiltonian $H(t)$ or the undriven static Hamiltonian. Consequently, Floquet quasienergies and their associated sidebands provide the natural basis for describing both coherent dynamics and, as shown in the following sections, dissipative processes in periodically driven quantum systems.

\subsection{Symmetry, parity and selection rules}
The transitions now occur among the Floquet quasienergy states in the extended space. Thus, it is not only instructive but also crucial to understand the selection rules for the transitions within the original Hilbert space and its extension to the extended space, a concept that is related to identifying the symmetry properties of the system. 
For example, for the original Hamiltonian in the bare Hilbert space, we define the parity operator as $\mathcal{P}_{\rm bare}=\mathrm{e}^{\ii\pi \mathcal{N}_\mathrm{bare}}$, where $\mathcal{N}_\mathrm{bare}=\sigma^+\sigma^-$ is the total number of excitations operator, $\mathcal{N}_\mathrm{bare}\lvert j\rangle=j\lvert j\rangle$, in the Coulomb gauge. To identify the parity of the eigenstates $j$, we must calculate the matrix element $\langle j\vert \mathcal{P}_{\rm bare}\vert j\rangle$. If this matrix element is negative/positive, then the parity of the state $j$ is odd/even. 
However, the external harmonic drive can modify the symmetry of the system. 

Thus, in the Floquet picture, since the transitions are among the Floquet sidebands, one must understand the parity of the sidebands in the extended space, identified by the generalized parity transformation:
\begin{equation}
    \begin{split}
        \mathcal{P}^{\rm F}&:\lcb\begin{array}{l}
            x\to-x, \\
            t\to t+T/2.
        \end{array}\right.
    \end{split}
\end{equation}

To do this, we need to
find the parity operator in the temporal Hilbert space via $\mathcal{P}_{\rm temp}=\mathrm{e}^{\ii\pi {N}_{\rm temp}}$, where ${N}_{\rm temp}$  is a number operator, ${N}_{\rm temp}\lvert l)=l\lvert l)$ with $\lvert l)$ being the temporal number (eigen)states, in the temporal Hilbert space $\mathbb{H}_{\rm temp}$, i.e., $\mathsf{B}_{\rm temp}=\{\lvert l)\}_{l=-l_\mathrm{max}}^{l_{\rm max}}$ and ${\rm Dim}(\mathbb{H}_{\rm temp})=2l_\mathrm{max}+1$.
One can now identify the generalized parity of the Floquet modes in the Floquet extended picture by $\Fbra{l\alpha} \mathcal{P}^{\rm F}_{\rm ex}\Fket{l\alpha }\gtrless0$ for even/odd parity, where  
\begin{equation}
    \mathcal{P}^{\rm F}\to\mathcal{P}^{\rm F}_{\rm ex}=\mathcal{P}_{\rm temp}\otimes\mathcal{P}_{\rm dressed}
\end{equation}
is the generalized parity operator in the Floquet extended space, with $\mathcal{P}_{\rm dressed}=\sum_{jk}\bra{j}\mathcal{P}_{\rm bare}\ket{k}\,\ket{j}\bra{k}$ being the parity operator in the dressed basis.
Proper understanding of the generalized parity helps one to identify the selection rules in the driven quantum system, and recognize how many externally assisting quanta from the external classical field is needed for a specific transition to occur.

\subsection{Time-evolution operator}

The unitary time-evolution operator, generated by ${H}(t)$, is
\begin{equation}
U(t,t_0)=\mathcal{T}\exp\!\left[-\ii\int_{t_0}^{t}dt'\,\mathcal{H}(t')\right],
\label{eq:U_def}
\end{equation}
where $\mathcal{T}$ denotes time ordering. Using the completeness of the Floquet solutions, one obtains the standard Floquet representation
\begin{equation}
U(t,t_0)=
\sum_\alpha
\ee^{-\ii \varepsilon_\alpha (t-t_0)}
\ket{\alpha(t)}\bra{\alpha(t_0)}.
\label{eq:U_Floquet}
\end{equation}
Equation~\eqref{eq:U_Floquet} is exact and will be used repeatedly below.

It is often more convenient to separate the dynamics into a stroboscopic part and a periodic micromotion. In Floquet theory, one may write~\cite{Eckardt_High-frequency_2015}
\begin{equation}
U(t,t_0)=U^{\rm F}(t)\,\ee^{-\ii H^{\rm F}_{\rm eff}(t-t_0)}\,U^{{\rm F}\dagger}(t_0),
\label{eq:U_micromotion}
\end{equation}
where 
\begin{equation}
U^{\rm F}(t)\ket{\alpha}=\ket{\alpha(t)},
\end{equation}
with $U^{\rm F}(t+T)=U^{\rm F}(t)$ is the micromotion operator and $H^{\rm F}_{\rm eff}$ is a time-independent effective Floquet Hamiltonian acting in the physical Hilbert space. 

The matrix elements of the exact evolution operator in an arbitrary time-independent basis $\{\ket{j=i}\}$ follow directly from Eq.~\eqref{eq:U_Floquet}:
\begin{equation}
\bra{f}U(t,t_0)\ket{i}
=
\sum_\alpha
\ee^{-\ii \varepsilon_\alpha (t-t_0)}
\braket{f|\alpha(t)}
\braket{\alpha(t_0)|i}.
\label{eq:U_matrix_general}
\end{equation}
Using the Fourier decomposition of the Floquet modes, one may also write
\begin{equation}
\braket{f|\alpha(t)}
=
\sum_l \ee^{-\ii l\omega_d t}\braket{f|\alpha_l}.
\label{eq:projection_f_alpha_t}
\end{equation}

A particularly useful extended-space identity is
\begin{equation}
\bra{f}U(t,t_0)\ket{i}
=
\sum_{n\in\mathbb{Z}}
\ee^{-\ii n\omega_d t}
\Fbra{n,f}
\ee^{-\ii \mathcal{H}^{\rm F}_{\rm ex}(t-t_0)}
\Fket{0,i},
\label{eq:U_matrix_Sambe}
\end{equation}
which makes explicit the decomposition into processes involving different numbers of drive quanta.

\subsection{Transition amplitudes and transition probabilities}

The transition amplitude from an initial state $\ket{j=i}$ at time $t_0$, to a final state $\ket{j=f}$ at time $t$, is
\begin{equation}
U_{fi}(t,t_0)\equiv \bra{f}U(t,t_0)\ket{i},
\end{equation}
and the corresponding transition probability is
\begin{equation}
P_{f\leftarrow i}(t,t_0)=|U_{fi}(t,t_0)|^2.
\end{equation}

For periodically driven systems, it is natural to distinguish the elapsed time $\tau=t-t_0$ from the initial phase of the drive, encoded in $t_0$. The experimentally relevant quantity is often the transition probability averaged over the initial time $t_0$ over one drive period, at fixed $\tau$:
\begin{equation}
P_{f\leftarrow i}(\tau)
\equiv
\frac{1}{T}\int_0^T dt_0\, P_{f\leftarrow i}(t_0+\tau,t_0).
\label{eq:P_t0_avg}
\end{equation}

A further average over long elapsed times defines the long-time averaged transition probability,
\begin{equation}
\overline{P}_{f\leftarrow i}
\equiv
\lim_{\tau_{\rm max}\to\infty}
\frac{1}{\tau_{\rm max}}
\int_0^{\tau_{\rm max}} d\tau\, P_{f\leftarrow i}(\tau).
\label{eq:P_long_avg_def}
\end{equation}
In a periodically driven system, the observables in the steady-state are also periodic, so $\tau_{\rm max}$ can be thoughtfully chosen as a multiple of $T$.

Using the Floquet decomposition, one obtains the standard result~\cite{Shirley_Solution_1965,Chu_Quantum_1982,Chu_Beyond_2004,Chu_Recent_1985}
\begin{equation}
\overline{P}_{f\leftarrow i}
=
\sum_{\alpha}\sum_{l,l'}
\left|
\braket{f|\alpha_l}
\braket{\alpha_{l'}|i}
\right|^2.
\label{eq:Pbar_llprime}
\end{equation}
Note, this expression is obtained from the early Floquet studies of quantum mechanics, sometimes known as the Shirley limit, but it can also be obtained from the phenomenological approach as in Eq.~\eqref{eq:ShirleyLimit}.
It is useful to reorganize this expression according to the net sideband index
$n=l-l'$,
which labels the net number of drive quanta exchanged with the periodic field. We then define
\begin{equation}
\overline{P}_{f\leftarrow i}^{(n)}
\equiv
\sum_{\alpha}\sum_{l-l'=n}
\left|
\braket{f|\alpha_l}
\braket{\alpha_{l'}|i}
\right|^2,
\label{eq:Pbar_n}
\end{equation}
so that
\begin{equation}
\overline{P}_{f\leftarrow i}=\sum_n \overline{P}_{f\leftarrow i}^{(n)}.
\end{equation}
In this way, $n$ identifies the order of the multiphoton process induced by the periodic drive (in case of a coherent optical pump as a Floquet drive).

The same quantities admit a compact representation in the extended space. The probability associated with reaching the sector $\Fket{n,f}$ from $\Fket{0,i}$ after elapsed time $\tau$ is
\begin{equation}
P_{f\leftarrow i}^{(n)}(\tau)
=
\left|
\Fbra{n,f}
\ee^{-\ii {H}^{\rm F}_{\rm ex}\tau}
\Fket{0,i}
\right|^2.
\label{eq:Pn_Sambe_tau}
\end{equation}
After long-time averaging,
\begin{equation}
\overline{P}_{f\leftarrow i}^{(n)}
=
\sum_{\alpha,l}
\left|
\Fbraket{n,f}{l\alpha}\,\Fbraket{l\alpha}{0,i}
\right|^2,
\label{eq:Pn_Sambe_long}
\end{equation}
which is equivalent to Eq.~\eqref{eq:Pbar_n}, with $\Fket{l\alpha}$ being the eigenmodes of $H^{\rm F}_{\rm ex}$ in the extended space according to Eq.~\eqref{eq:Floquet_eigenproblem_ex_modified}.

\subsection{Density matrix in the Floquet basis for a closed system}

For a closed system, the density matrix obeys the 
 von-Neumann equation
\begin{equation}
\ii \partial_t \rho(t)=[{H}(t),\rho(t)].
\label{eq:vonNeumann}
\end{equation}
Expanding $\rho(t)$ in the instantaneous Floquet basis,
\begin{equation}
\rho(t)=\sum_{\alpha\beta}\rho_{\alpha\beta}(t)\,
\ket{\alpha(t)}\bra{\beta(t)},
\ 
\rho_{\alpha\beta}(t)\equiv
\bra{\alpha(t)}\rho(t)\ket{\beta(t)},
\label{eq:rho_Floquet_expand}
\end{equation}
and using Eq.~\eqref{eq:HF_t}, one finds~\cite{Hausinger_Dissipative_2010PhDThesis}
\begin{equation}
\partial_t \rho_{\alpha\beta}(t)
=
-\ii (\varepsilon_\alpha-\varepsilon_\beta)\rho_{\alpha\beta}(t).
\label{eq:rho_alpha_beta_closed}
\end{equation}
Thus,
\begin{equation}
\rho_{\alpha\beta}(t)
=
\rho_{\alpha\beta}(t_0)\,
\ee^{-\ii(\varepsilon_\alpha-\varepsilon_\beta)(t-t_0)}.
\label{eq:rho_alpha_beta_solution}
\end{equation}

The initial coefficients are determined by projection of $\rho(t_0)$ onto the Floquet basis,
\begin{equation}
\rho_{\alpha\beta}(t_0)=
\bra{\alpha(t_0)}\rho(t_0)\ket{\beta(t_0)}.
\end{equation}
For a pure initial state, $\rho(t_0)=\ket{i}\bra{i}$,
then
\begin{equation}
\rho_{\alpha\beta}(t_0)=
\braket{\alpha(t_0)|i}\braket{i|\beta(t_0)}.
\end{equation}

\subsection{Expectation values of observables}

For an arbitrary operator $O$, the expectation value is
\begin{equation}
\langle O(t)\rangle=\Tr{O\,\rho(t)},
\end{equation}
and using Eq.~\eqref{eq:rho_Floquet_expand}, this becomes
\begin{equation}
\langle O(t)\rangle
=
\sum_{\alpha\beta}
O_{\alpha\beta}(t)\,\rho_{\beta\alpha}(t),
\qquad
O_{\alpha\beta}(t)\equiv \bra{\alpha(t)}O\ket{\beta(t)}.
\label{eq:O_expect_general}
\end{equation}

Since $O_{\alpha\beta}(t)$ is $T$-periodic, it admits the Fourier expansion
\begin{equation}
O_{\alpha\beta}(t)=
\sum_l O_{\alpha\beta l}\,\ee^{-\ii l\omega_d t},
\quad
O_{\alpha\beta l}
=
\frac{1}{T}\int_0^T dt\, \ee^{\ii l\omega_d t} O_{\alpha\beta}(t).
\label{eq:Oab_fourier}
\end{equation}
Substituting Eq.~\eqref{eq:rho_alpha_beta_solution} into Eq.~\eqref{eq:O_expect_general} gives
\begin{equation}
\langle O(t)\rangle
=
\sum_{\alpha\beta l}
O_{\alpha\beta l}\,
\rho_{\beta\alpha}(t_0)\,
\ee^{\ii(\varepsilon_\alpha-\varepsilon_\beta-l\omega_d)(t-t_0)}
\ee^{-\ii l\omega_d t_0}.
\label{eq:O_expect_full}
\end{equation}

For a 
closed system, the off-diagonal Floquet coherences generally persist, and $\langle O(t)\rangle$ need not be periodic with period $T$. In contrast, in open driven systems the off-diagonal contributions are typically suppressed at long times, in which case only the diagonal Floquet populations remain relevant. Motivated by that long-time limit, one may define the diagonal (periodic) contribution
\begin{equation}
\langle O(t)\rangle_{\rm diag}
=
\sum_\alpha \rho_{\alpha\alpha}\, O_{\alpha\alpha}(t),
\label{eq:O_diag}
\end{equation}
where $\rho_{\alpha\alpha}$ are time independent for a closed system. The period average is then
\begin{equation}
\overline{O}
\equiv
\frac{1}{T}\int_0^T dt\, \langle O(t)\rangle_{\rm diag}
=
\sum_\alpha \rho_{\alpha\alpha}\, O_{\alpha\alpha 0}.
\label{eq:O_bar_diag}
\end{equation}

Using the Fourier components of the Floquet modes, one may write
\begin{equation}
O_{\alpha\alpha 0}
=
\sum_l \bra{\alpha_l}O\ket{\alpha_l},
\label{eq:Oaa0_sidebands}
\end{equation}
and therefore
\begin{equation}
\overline{O}
=
\sum_{\alpha,l}
\rho_{\alpha\alpha}\,
\bra{\alpha_l}O\ket{\alpha_l}.
\label{eq:O_bar_sidebands}
\end{equation}
In the extended-space language, Eq.~\eqref{eq:O_bar_sidebands} becomes
\begin{equation}
\overline{O}
=
\sum_\alpha
\rho_{\alpha\alpha}\,
\Fbra{\alpha}F_0\otimes O\Fket{\alpha}.
\label{eq:O_bar_Sambe}
\end{equation}
Equation~\eqref{eq:O_bar_Sambe} is particularly useful numerically because it allows one to evaluate time-averaged observables directly from the extended-space eigenvectors, without reconstructing the full time dependence of the physical state. In our calculations, this expression will be used extensively for long-time averaged observables such as the mean excitation number.

Alternatively, the expectation value of an observable, can be defined via $\langle O(t)\rangle\equiv\langle\psi(t)|O|\psi(t)\rangle$, and as before, this raw quantity is not necessarily time periodic due to the presence of off-diagonal terms ($\alpha\neq\beta$)
in the Floquet eigenbasis~\cite{Eckardt_Atomic_2017}, given by 
\begin{equation}
\langle O(t)\rangle=\sum_{\alpha\beta}c^*_\alpha c_\beta\,O_{\alpha\beta}(t)\,\mathrm{e}^{\ii(\varepsilon_\alpha-\varepsilon_\beta)t}.
\end{equation}
Again,
in real open systems,
 the off-diagonal terms are suppressed,
and the time evolution of observables
often becomes periodic with the same periodicity as the driving
field. 
This (prethermalization process) can be stimulated by
adding a phenomenological damping rate, $\Gamma(\omega)$ with $\Gamma(0)=0$, that yields to kill off 
the non-diagonal terms,
which are then damped out 
after a sufficiently long time~\cite{Chu_Beyond_2004,DeLiberato_Virtual_2017,Eckardt_Atomic_2017}.

Subsequently, we derive the expectation values,
\begin{equation}
\langle O(t)\rangle=\sum_{\alpha\beta}c^*_\alpha c_\beta\,O_{\alpha\beta}(t)\,\mathrm{e}^{\ii(\varepsilon_\alpha-\varepsilon_\beta)t-\Gamma(\varepsilon_\alpha-\varepsilon_\beta) t},
\end{equation}
and obtain the 
steady-state values 
\begin{equation}
    \langle O(t)\rangle_{\rm ss}\equiv\langle O(t> t_{\rm ss})\rangle=\sum_{\alpha}\lvert c_\alpha\rvert^2\,O_{\alpha\alpha}(t).
\end{equation}

 We then define the 
 the mean real excitation number 
 from
 $ \overline{O}=\frac{1}{T}\int_{t_\mathrm{ss}}^{t_\mathrm{ss}+T} dt\,\langle O(t)\rangle_{\rm ss}$, where $t_{\rm ss}$ and $T$ are long enough to yield a temporal average.
 For our case, the result is exactly $T$ periodic,
 so we only have to use one period. 
Hence, the long-time-averaged observable reads
\begin{equation}
\begin{split}
    \overline{O}&=\sum_{\alpha l}\lvert c_\alpha \rvert^2\,\langle \alpha_l\lvert O\rvert \alpha_l\rangle.
\end{split}
\label{eq:}
\end{equation}

Using the extended space formalism~\cite{Restrepo_Driven_2019,Restrepo_Quantum_2018},
\begin{equation}
\begin{split}
    \overline{O}&=\sum_{\alpha}\lvert c_\alpha \rvert^2\,\Fbra{\alpha} F_0\otimes O\Fket{\alpha}.
\end{split}
\label{eq:AveO_ex}
\end{equation}
Note because $c_\alpha=\braket{\alpha\vert \psi(0)}$, then $\lvert c_\alpha\rvert^2=\braket{\alpha\vert \psi(0)}\braket{\psi(0)\vert\alpha}=\rho_{\alpha\alpha}$, and hence, the two approaches yield equivalent result.
We call this approach as the \emph{phenomenological} approach as the dissipation is only considered phenomenologically, and the dissipative channels are not included properly through the master equation, but only through the 
imaginary part of the
quasienergies.

Finally, we note that when the observable projects onto the same channels that define the transition probability out of a chosen initial state, the averaged observable may be related to sums of $\overline{P}^{(n)}_{f\leftarrow i}$. The precise relation depends on the operator under consideration and on the choice of initial state.

\paragraph*{Proof of Equations \eqref{eq:O_bar_Sambe} and \eqref{eq:AveO_ex}.}
The Floquet extended space expression of the average number of excitation (population) has some benefits such as the efficiency in numerics as well as BZ identification-free procedure. Thus, here we show that Eq.~\eqref{eq:AveO_ex} is true algebraically and structurally. We expand the extended-space eigenvector in its sideband components,
Eq.~\eqref{eq:Floquet_extended_vector}, $\Fket{\alpha}=\sum_{l}\vert l)\otimes
\vert\alpha_{l}\rangle$. Since the temporal shift operators satisfy
$F_{m}\vert l)=\vert l+m)$ [Eq.~\eqref{eq:TemporalShift_Op}], the zeroth component is the
identity on the temporal space, $F_{0}=\textbf{1}_{\mathrm{temp}}$.
Using the orthonormality of the temporal basis, $(l'\vert l)
=\delta_{l'l}$ [Eq.~\eqref{eq:temporal_orthogonality}],
\begin{equation}
\Fbra{\alpha} F_{0}\otimes O\Fket{\alpha}
=\sum_{l',l}(l'\vert l)\,
\langle\alpha_{l'}\vert O\vert\alpha_{l}\rangle
=\sum_{l}\langle\alpha_{l}\vert O\vert\alpha_{l}\rangle
=O_{\alpha\alpha 0}.
\label{eq:proof_algebraic}
\end{equation}

Structurally, by the definition of the extended-space scalar product as the
period-averaged inner product, Eq.~\eqref{eq:Sambe_inner}, and using
$(t\Fket{\alpha}=\vert\alpha(t)\rangle$ [Eq.~\eqref{eq:alpha_t_from_extended}] together
with the time independence of $O$,
\begin{equation}
\Fbra{\alpha} F_{0}\otimes O\Fket{\alpha}
=\frac{1}{T}\int_{0}^{T}\!\mathrm{d}t\,
\langle\alpha(t)\vert O\vert\alpha(t)\rangle
=\overline{O_{\alpha\alpha}(t)}.
\label{eq:proof_structural}
\end{equation}
Equation~\eqref{eq:proof_structural} makes the content of the identity
transparent; \emph{the period average is not an additional
operation to be performed on top of the extended-space formalism--- it
is built into the Sambe-space inner product itself.} The diagonal
element of the lifted observable is, by construction of Eq.~\eqref{eq:Sambe_inner}, the
one-period average of the mode-diagonal expectation value.

\subsection{Direct extended-space evaluation of the phenomenological phase-locked and
phase-averaged averages}
\label{app:phen_extended}

The phase-locked result,
\begin{equation}
\overline{N}_{\Lambda}^{(\mathrm{phase-locked})}
=
\sum_{\alpha}|c_{\alpha}|^{2}
\Fbra{\alpha}
F_{0}\otimes
s_{\mathrm{B}}^{\Lambda-}s_{\mathrm{B}}^{\Lambda+}
\Fket{\alpha},
\label{eq:pl_start}
\end{equation}
is usually evaluated after selecting one Floquet representative
$\Fket{\alpha}$ from each quasienergy BZ. This selection
can become inconvenient near zone boundaries and multiphoton
resonances. We now show that the same quantity can instead be obtained
directly from the complete extended-space spectrum, without explicitly
identifying the Floquet replicas.

For a replica $\Fket{n\alpha}$, the diagonal matrix element of the
lifted number operator is
\begin{equation}
\begin{split}
n_{n\alpha}
&\equiv
\Fbra{n\alpha}
F_{0}\otimes
s_{\mathrm{B}}^{\Lambda-}s_{\mathrm{B}}^{\Lambda+}
\Fket{n\alpha}
\\
&=
\sum_l
\langle\alpha_{l+n}|
s_{\mathrm{B}}^{\Lambda-}s_{\mathrm{B}}^{\Lambda+}
|\alpha_{l+n}\rangle
=n_\alpha ,
\end{split}
\label{eq:n_class}
\end{equation}
where the last equality follows by shifting the summation index.
Thus, every replica of the same physical Floquet mode carries the same
cycle-averaged occupation.

To obtain the phase-locked weight, we define the temporal state
\begin{equation}
|\chi(t_0))
=
\sum_l e^{i l\omega_dt_0}|l),
\label{eq:chi_def}
\end{equation}
and
$\Fket{\chi(t_0),\psi_0}
\equiv|\chi(t_0))\otimes|\psi_0\rangle$.
Using
$|\alpha_l^{[n]}\rangle=|\alpha_{l+n}\rangle$,
\begin{equation}
\begin{split}
\Fbraket{n\alpha}{\chi(t_0),\psi_0}
&=
\sum_l
e^{il\omega_dt_0}
\langle\alpha_{l+n}|\psi_0\rangle
\\
&=
e^{-in\omega_dt_0}\,c_\alpha(t_0).
\end{split}
\label{eq:c_class}
\end{equation}
Hence,
\begin{equation*}
\left|
\Fbraket{n\alpha}{\chi(t_0),\psi_0}
\right|^2
=
|c_\alpha(t_0)|^2 ,
\end{equation*}
so that the physical phase-locked weight is also independent of the
replica index.

The replica overcounting is removed by projecting onto one temporal
block. We let
$\mathcal P_0=|0)(0|$ and define
\begin{equation}
P^{(0)}_{n\alpha}
\equiv
\Fbra{n\alpha}
\mathcal P_0\otimes\mathbf 1_{\rm dressed}
\Fket{n\alpha}
=
\langle\alpha_n|\alpha_n\rangle .
\label{eq:P0_def}
\end{equation}
For every physical Floquet mode,
\begin{equation}
\sum_n P^{(0)}_{n\alpha}
=
\sum_n\langle\alpha_n|\alpha_n\rangle
=
\Fbraket{\alpha}{\alpha}
=1 .
\label{eq:partition_unity}
\end{equation}
The central-block weights therefore provide a partition of unity over
the replicas.

Combining Eqs.~\eqref{eq:n_class}--\eqref{eq:partition_unity} gives
the unrestricted extended-space expression,
\begin{widetext}
\begin{equation}
\overline{N}_{\Lambda}^{(\mathrm{phase-locked})}
=
\sum_{n\alpha}
P^{(0)}_{n\alpha}
\left|
\Fbraket{n\alpha}{\chi(t_0),\psi_0}
\right|^2
\Fbra{n\alpha}
F_0\otimes
s_{\mathrm{B}}^{\Lambda-}s_{\mathrm{B}}^{\Lambda+}
\Fket{n\alpha}.
\label{eq:pl_extended}
\end{equation}
\end{widetext}
Note that no explicit BZ selection or replica identification is required.

For the phase-averaged result, the initial state is embedded only in
the zero temporal block,
$\Fket{0,\psi_0}=|0)\otimes|\psi_0\rangle$. In this case
\begin{equation*}
\left|
\Fbraket{n\alpha}{0,\psi_0}
\right|^2
=
|\langle\alpha_n|\psi_0\rangle|^2 ,
\end{equation*}
and summing directly over the replicas reconstructs
\begin{equation*}
p_\alpha
=
\sum_n|\langle\alpha_n|\psi_0\rangle|^2 .
\end{equation*}

Thus,
\begin{equation}
\overline{N}_{\Lambda}^{(\mathrm{phase-averaged})}
=
\sum_{n\alpha}
\left|
\Fbraket{n\alpha}{0,\psi_0}
\right|^2
\Fbra{n\alpha}
F_0\otimes
s_{\mathrm{B}}^{\Lambda-}s_{\mathrm{B}}^{\Lambda+}
\Fket{n\alpha}.
\label{eq:pa_extended}
\end{equation}
The phase-averaged expression therefore requires no additional
partition-of-unity factor, since the different replicas already carry
the individual sideband contributions to $p_\alpha$.

In numerical calculations the Sambe space is truncated at
$|l|\leq l_{\max}$. The central-block factor
$P^{(0)}_{n\alpha}$ suppresses replicas whose weight lies near the
truncation boundary, providing rapid convergence once the populated
sidebands are contained within the chosen cutoff. In our calculations
$l_{\max}\gtrsim30$ is sufficient at the strongest drives considered.
Useful convergence checks are
\begin{equation}
\sum_{n\alpha}
P^{(0)}_{n\alpha}
\left|
\Fbraket{n\alpha}{\chi,\psi_0}
\right|^2
\simeq
\sum_{n\alpha}
\left|
\Fbraket{n\alpha}{0,\psi_0}
\right|^2
\simeq1 .
\label{eq:sum_rules}
\end{equation}

\section{Floquet--Markov generalized master equation for the driven--dissipative quantum system}
\label{secS:FloquetTheory_OpenSystem}

We next derive the master equation for a weakly-open periodically driven system. Our goal is to obtain a nonsecular Floquet--Markov generalized master equation (GME) that retains the couplings between distinct Floquet transition frequencies. The derivation follows the standard Born--Markov procedure for periodically driven systems~\cite{Breuer_Theory_2002,Eckardt_High-frequency_2015,Restrepo_Driven_2019}, specialized here to a driven hybrid (with arbitrarily interacting subsystems) quantum system model and its generalizations~\cite{Settineri_Dissipation_2018}.

\subsection{Microscopic Hamiltonian and assumptions}

We consider the total system plus bath Hamiltonian,
\begin{equation}
H_{\rm tot}(t)=H(t)+H_{\rm B}+H_{\rm SB}(t),
\label{eq:Htot_open}
\end{equation}
where $H(t)=H_0+H_d(t)$ is the system Hamiltonian, with
\begin{equation*}
H(t+T)=H(t), \qquad T=\frac{2\pi}{\omega_d},
\end{equation*}
$H_{\rm B}$ is the environment (bath) Hamiltonian, and $H_{\rm SB}(t)$ describes the system--bath coupling. 

We allow $H_{\rm SB}(t)$ to be explicitly periodic,
\begin{equation}
H_{\rm SB}(t+T)=H_{\rm SB}(t),
\end{equation}
since this naturally arises, for example, after moving to a rotating or Floquet-engineered frame even when the laboratory-frame coupling is time independent.

We consider a generic multipartite system with subsystem index $\Lambda\in\{1,\dots,N_\Lambda\}$, each subsystem being weakly coupled to an independent bath. Thus,
\begin{equation}
H_{\rm B}=\sum_\Lambda H_{\rm B}^{(\Lambda)},
\end{equation}
with, for instance, bosonic baths
\begin{equation}
H_{\rm B}^{(\Lambda)}=\sum_k \omega_k^{(\Lambda)}\, b_k^{(\Lambda)\dagger} b_k^{(\Lambda)},
\end{equation}
or fermionic baths analogously.

We assume a bilinear coupling of the form
\begin{equation}
H_{\rm SB}(t)=\sum_\Lambda S_\Lambda(t)\otimes B_\Lambda,
\label{eq:HSB_bilinear}
\end{equation}
where $S_\Lambda(t)$ acts on the system Hilbert space and $B_\Lambda$ acts on bath $\Lambda$. For notational simplicity one may take the coupling operators Hermitian when appropriate, although the derivation below does not rely on this except when discussing symmetry relations.

The total density matrix $\rho_{\rm tot}(t)$ obeys the von Neumann equation,
\begin{equation}
\partial_t \rho_{\rm tot}(t)=-\ii[H_{\rm tot}(t),\rho_{\rm tot}(t)].
\end{equation}

We make the standard assumptions:
(i) weak system--bath coupling (Born approximation),
(ii) short bath memory compared to the relaxation timescale (Markov approximation),
and
(iii) an initially factorized state,
\begin{equation}
\rho_{\rm tot}(t_0)=\rho(t_0)\otimes \rho_{\rm B},
\label{eq:factorized_initial}
\end{equation}
where $\rho(t)=\mathrm{Tr}_{\rm B}[\rho_{\rm tot}(t)]$ is the reduced system density matrix. 

The bath state is assumed stationary,
\begin{equation}
[\rho_{\rm B},H_{\rm B}]=0,
\end{equation}
and centered,
\begin{equation}
\mathrm{Tr}_{\rm B}[\rho_{\rm B} B_\Lambda]=0.
\end{equation}
These assumptions ensure that bath correlations depend only on time differences.

\subsection{Born--Markov equation in the interaction picture}

We let $U(t,t_0)$ denote the exact unitary evolution operator of the driven \emph{closed} system,
\begin{equation}
\ii\partial_t U(t,t_0)=H(t)U(t,t_0),
\qquad
U(t_0,t_0)=\mathbf{1},
\end{equation}
and let
\begin{equation}
U_{\rm B}(t,t_0)=\ee^{-\ii H_{\rm B}(t-t_0)}.
\end{equation}

The interaction-picture reduced density matrix and coupling Hamiltonian are then
\begin{equation}
\widetilde{\rho}(t)=U^\dagger(t,t_0)U_{\rm B}^\dagger(t,t_0)\rho_{\rm tot}(t)U_{\rm B}(t,t_0)U(t,t_0),
\end{equation}
\begin{equation}
\widetilde{H}_{\rm SB}(t)
=
U^\dagger(t,t_0)U_{\rm B}^\dagger(t,t_0)H_{\rm SB}(t)U_{\rm B}(t,t_0)U(t,t_0).
\end{equation}

Writing
\begin{equation}
\widetilde{H}_{\rm SB}(t)=\sum_\Lambda \widetilde{S}_\Lambda(t)\otimes \widetilde{B}_\Lambda(t),
\end{equation}
with
\begin{equation}
\begin{split}
    \widetilde{S}_\Lambda(t)&=U^\dagger(t,t_0)S_\Lambda(t)U(t,t_0),
\\
\widetilde{B}_\Lambda(t)&=U_{\rm B}^\dagger(t,t_0)B_\Lambda U_{\rm B}(t,t_0),
\end{split}
\end{equation}
the Born--Markov equation takes the Redfield form
\begin{equation}
\partial_t \widetilde{\rho}(t)
=
-\int_0^\infty ds\,
\mathrm{Tr}_{\rm B}\!\left[
\widetilde{H}_{\rm SB}(t),
\left[
\widetilde{H}_{\rm SB}(t-s),\widetilde{\rho}(t)\otimes\rho_{\rm B}
\right]
\right].
\label{eq:Redfield_start}
\end{equation}

Introducing the bath correlation functions
\begin{equation}
C_{\Lambda\Lambda'}(s)
\equiv
\mathrm{Tr}_{\rm B}\!\left[
\widetilde{B}_\Lambda(s)\widetilde{B}_{\Lambda'}(0)\rho_{\rm B}
\right],
\label{eq:bath_corr}
\end{equation}
and using stationarity of the bath, Eq.~\eqref{eq:Redfield_start} becomes
\begin{equation}
\begin{aligned}
\partial_t \widetilde{\rho}(t)
&=
\sum_{\Lambda,\Lambda'}
\int_0^\infty ds\,
\Big\{
C_{\Lambda\Lambda'}(s)\,
\widetilde{S}_{\Lambda'}(t-s)\widetilde{\rho}(t)\widetilde{S}_{\Lambda}(t)
\\
&\qquad
-C_{\Lambda\Lambda'}(s)\,
\widetilde{S}_{\Lambda}(t)\widetilde{S}_{\Lambda'}(t-s)\widetilde{\rho}(t)
\\
&\qquad
+C_{\Lambda'\Lambda}^*(s)\,
\widetilde{S}_{\Lambda}(t)\widetilde{\rho}(t)\widetilde{S}_{\Lambda'}(t-s)
\\
&\qquad
-C_{\Lambda'\Lambda}^*(s)\,
\widetilde{\rho}(t)\widetilde{S}_{\Lambda'}(t-s)\widetilde{S}_{\Lambda}(t)
\Big\}.
\end{aligned}
\label{eq:Redfield_with_corr}
\end{equation}

If the baths are independent, then
\begin{equation}
C_{\Lambda\Lambda'}(s)=\delta_{\Lambda\Lambda'}\, C_\Lambda(s).
\label{eq:indep_baths_corr}
\end{equation}

\subsection{Floquet decomposition of the system coupling operators}

The periodicity of the system Hamiltonian allows one to decompose the interaction-picture system operators into Floquet transition components. 
Then any system coupling operator admits the decomposition
\begin{equation}
\widetilde{S}_\Lambda(t)
=
\sum_{\Delta}
\ee^{-\ii \Delta t}\,
S_\Lambda(\Delta),
\label{eq:S_tilde_Delta}
\end{equation}
where the Floquet transition frequencies are
\begin{equation}
\Delta\equiv \Delta_{\alpha\beta l}
=
\varepsilon_\beta-\varepsilon_\alpha+l\omega_d,
\qquad
l\in\mathbb{Z},
\label{eq:Floquet_transition_frequency}
\end{equation}
and the corresponding transition operators are
\begin{equation}
S_\Lambda(\Delta)
=
\sum_{\alpha,\beta,l:\,\Delta_{\alpha\beta l}=\Delta}
S_{\Lambda,\alpha\beta l}\,
\ket{\alpha}\bra{\beta}.
\label{eq:S_Delta_def}
\end{equation}

The coefficients, 
\begin{equation}
S_{\Lambda,\alpha\beta l}
=
\frac{1}{T}\int_0^T dt\, \ee^{\ii l\omega_d t}
\bra{\alpha(t)}S_\Lambda(t)\ket{\beta(t)}, 
\label{eq:S_alpha_beta_l}
\end{equation}
are the Fourier components of the operator matrix elements in the Floquet basis. In the extended-space notation introduced earlier, they can be written as
\begin{equation}
S_{\Lambda,\alpha\beta l}
=\sum_k \bra{\alpha_k}S_\Lambda\ket{\beta_{k+l}}
=\Fbra{\alpha}F_{-l}\otimes S_\Lambda\Fket{\beta},
\label{eq:S_alpha_beta_l_Sambe}
\end{equation}
when $S_\Lambda(t)$ is time independent in the physical Hilbert space. More generally, the same structure holds with the appropriate Fourier components of $S_\Lambda(t)$ included.

If the system coupling operators are Hermitian, then
\begin{equation}
S_\Lambda^\dagger(\Delta)=S_\Lambda(-\Delta),
\label{eq:S_Delta_hermitian}
\end{equation}
which follows from
\begin{equation}
S_{\Lambda,\alpha\beta l}=S_{\Lambda,\beta\alpha,-l}^*.
\end{equation}

Substituting Eq.~\eqref{eq:S_tilde_Delta} into Eq.~\eqref{eq:Redfield_with_corr} yields the {\it nonsecular Floquet--Redfield equation in the interaction picture}:
\begin{widetext}
    \begin{equation}
    \begin{split}
      \widetilde{\mathcal{L}}_\mathrm{diss}(t)\widetilde{\rho}(t)
      &=\sum_{\Lambda\Lambda'}\sum_{\Delta\Delta'}\mathrm{e}^{-\ii(\Delta-\Delta') t}\,\left[\Gamma_{\Lambda'\Lambda}(\Delta)\,\lcb S_{\rm B}^{\Lambda}(\Delta) \widetilde{\rho}(t)S_{\rm B}^{\Lambda'\dagger}(\Delta')
        -S_{\rm B}^{\Lambda'\dagger}(\Delta')S_{\rm B}^\Lambda(\Delta)\widetilde{\rho}(t)
        \rcb\right.
        \\
        &\hspace{2.0cm}\left. +\Gamma_{\Lambda\Lambda'}^*(\Delta')\lcb S_{\rm B}^\Lambda(\Delta)\widetilde{\rho}(t)S_{\rm B}^{\Lambda'\dagger}(\Delta')
- \widetilde{\rho}(t)S_{\rm B}^{\Lambda'\dagger}(\Delta')S_{\rm B}^\Lambda(\Delta)\rcb\right],
    \end{split}
    \label{}
\end{equation} 
\label{eq:interaction_nonsecular_FR}
\end{widetext}
where
\begin{equation}
\Gamma_{\Lambda\Lambda'}(\Delta)
=
\int_0^\infty ds\, \ee^{\ii \Delta s} C_{\Lambda\Lambda'}(s).
\label{eq:Gamma_def}
\end{equation}

The quantity $\Gamma_{\Lambda\Lambda'}(\Delta)$ is generally complex. Its real part determines dissipative transition rates, while its imaginary part gives Lamb-shift corrections. In the following, unless explicitly needed, the Lamb-shift contribution may be absorbed into an effective system Hamiltonian and we focus on the dissipative structure of the master equation.

\subsection{Schr\"{o}dinger-picture nonsecular Floquet--Markov GME}

Returning to the Schr\"odinger picture, via
\begin{equation}
\rho(t)=U(t,t_0)\widetilde{\rho}(t)U^\dagger(t,t_0),
\end{equation}
it is convenient to introduce the time-dependent Floquet operators:
\begin{equation}
S_\Lambda(\Delta;t)
\equiv
U(t,t_0)\,\ee^{-\ii \Delta t}S_\Lambda(\Delta)\,U^\dagger(t,t_0).
\label{eq:S_Delta_t_def}
\end{equation}
Using the Floquet decomposition of $U(t,t_0)$, one finds explicitly
\begin{equation}
S_\Lambda(\Delta;t)
=
\sum_{\alpha,\beta,l:\,\Delta_{\alpha\beta l}=\Delta}
S_{\Lambda,\alpha\beta l}\,
\ket{\alpha(t)}\bra{\beta(t)}.
\label{eq:S_Delta_t_explicit}
\end{equation}

The reduced density matrix then satisfies
\begin{equation}
\partial_t \rho(t)=\mathcal{L}(t)\rho(t),
\qquad
\mathcal{L}(t)=\mathcal{L}_{\rm S}(t)+\mathcal{L}_{\rm diss}(t),
\end{equation}
with
\begin{equation}
\mathcal{L}_{\rm S}(t)\rho=-\ii[H(t),\rho],
\label{eq:LS_open}
\end{equation}
and the {\it full nonsecular Floquet--Markov dissipator} is
\begin{widetext}
\begin{equation}
    \begin{split}
      {\mathcal{L}}_\mathrm{diss}(t){\rho}(t)  
&=\sum_{\Lambda\Lambda'}\sum_{\Delta\Delta'}\left\{\mathrm{e}^{-\ii l\omega_d t}\,\Gamma_{\Lambda'\Lambda}(\Delta)\,\lsb S_{\rm B}^{\Lambda}(\Delta;t) {\rho}(t)S_{\rm B}^{\Lambda'\dagger}(\Delta')
        -S_{\rm B}^{\Lambda'\dagger}(\Delta')S_{\rm B}^\Lambda(\Delta;t){\rho}(t)
        \rsb\right.
        \\
        &\hspace{2.0cm}\left. +\mathrm{e}^{\ii l'\omega_d t}\,\Gamma_{\Lambda\Lambda'}^*(\Delta')\lsb S_{\rm B}^\Lambda(\Delta){\rho}(t)S_{\rm B}^{\Lambda'\dagger}(\Delta';t)
- {\rho}(t)S_{\rm B}^{\Lambda'\dagger}(\Delta';t)S_{\rm B}^\Lambda(\Delta)\rsb\right\}.
    \end{split}
    \label{eq:FGME_nonsecular_Sch}
\end{equation}
\end{widetext}

Equation~\eqref{eq:FGME_nonsecular_Sch} is the Floquet--Markov generalized master equation used in this work. It is explicitly $T$-periodic in time, inherits the periodicity of the closed-system Floquet modes, and retains the couplings between distinct transition frequencies $\Delta$ and $\Delta'$. Thus, it goes beyond the full secular approximation and reduces to the usual FL form only after additional secularization.
For independent baths,
\begin{equation}
\Gamma_{\Lambda\Lambda'}(\Delta)=\delta_{\Lambda\Lambda'}\,\Gamma_\Lambda(\Delta),
\label{eq:indep_gamma}
\end{equation}
which considerably simplifies Eq.~\eqref{eq:FGME_nonsecular_Sch} and reduces it to Eq.~\eqref{eq:Ldiss_FGME_OriginalBasis} in the main text.

\subsection{Positive- and negative-frequency decomposition}

For later physical interpretation it is useful to separate positive- and negative-frequency parts
for the operators. For $\Delta>0$, we define
\begin{equation}
S_\Lambda^{+}(\Delta)\equiv S_\Lambda(\Delta),
\qquad
S_\Lambda^{-}(\Delta)\equiv S_\Lambda(-\Delta)=S_\Lambda^\dagger(\Delta).
\label{eq:S_pm_def}
\end{equation}
We also have 
\begin{equation}
S_\Lambda^{-}(\Delta)=\left[S_\Lambda^{+}(\Delta)\right]^\dagger.
\end{equation}
Accordingly,
\begin{equation}
S_\Lambda(\Delta)=S_\Lambda^{-}(\Delta)+S_\Lambda^{(0)}+S_\Lambda^{+}(\Delta).
\end{equation}
with $S_\Lambda^{(0)}=S_\Lambda(\Delta=0)$, and similarly for $S_\Lambda(\Delta;t)$.

At zero temperature, where $\Gamma(\Delta\leq0)=0$, or more generally when the bath supports only downward processes, one may keep only $\Gamma_\Lambda(\Delta)$ for $\Delta>0$ in the dissipator-RWA limit such that the relatively high oscillating terms with $\pm\Delta$, $\pm\Delta'$ and $\Delta+\Delta'$ are discarded and only terms with $\pm\Delta\mp\Delta'$ are retained, similar to the approach in Ref.~\onlinecite{Settineri_Dissipation_2018} but used for the TI-GME not Floquet channels. In that case, the dissipator is built from the positive-frequency transition operators $S_\Lambda^{(+)}(\Delta)$ and their Hermitian conjugates. Importantly, in the present nonsecular treatment the sums over $\Delta$ and $\Delta'$ are retained; that is, terms with $\Delta\neq \Delta'$ are not discarded unless one explicitly imposes a secular approximation.
With this, the derivation of the F-GME can be simplified as 
\begin{widetext}
\begin{equation}
    \begin{split}
      {\mathcal{L}}_\mathrm{diss}(t){\rho}(t)  
&=\sum_{\Lambda\Lambda'}\sum_{\Delta,\Delta'>0}\left\{\mathrm{e}^{-\ii l\omega_d t}\,\Gamma_{\Lambda'\Lambda}(\Delta)\,\lsb S_{\rm B}^{\Lambda+}(\Delta;t) {\rho}(t)S_{\rm B}^{\Lambda'-}(\Delta')
        -S_{\rm B}^{\Lambda'-}(\Delta')S_{\rm B}^{\Lambda+}(\Delta;t){\rho}(t)
        \rsb\right.
        \\
        &\hspace{2.0cm}\left. +\mathrm{e}^{\ii l'\omega_d t}\,\Gamma_{\Lambda\Lambda'}^*(\Delta')\lsb S_{\rm B}^{\Lambda+}(\Delta){\rho}(t)S_{\rm B}^{\Lambda'-}(\Delta';t)
- {\rho}(t)S_{\rm B}^{\Lambda'-}(\Delta';t)S_{\rm B}^{\Lambda+}(\Delta)\rsb\right\}.
    \end{split}
    \label{eq:FGME_nonsecular_Sch_RWA}
\end{equation}
\end{widetext}

\subsection{Fourier expansion of the Liouvillian and numerical implementation}

Since both the Hamiltonian and the Floquet dissipator are $T$-periodic, the full Liouvillian admits a Fourier expansion as well:
\begin{equation}
\mathcal{L}(t)=\sum_{n\in\mathbb{Z}} \ee^{-\ii n\omega_d t}\,\mathcal{L}_n.
\label{eq:L_fourier}
\end{equation}

For the coherent part,
\begin{equation}
H(t)=\sum_n \ee^{-\ii n\omega_d t} H_n,
\end{equation}
so that
\begin{equation}
\mathcal{L}_{{\rm S},n}\rho=-\ii[H_n,\rho].
\label{eq:LSn}
\end{equation}

Likewise, each operator $S_\Lambda(\Delta;t)$ is periodic and can be written as
\begin{equation}
S_\Lambda(\Delta;t)=\sum_n \ee^{-\ii n\omega_d t}\, S_{\Lambda,n}(\Delta),
\label{eq:SDelta_t_fourier}
\end{equation}
where
\begin{equation}
S_{\Lambda,n}(\Delta)=\frac{1}{T}\int_0^T dt\, \ee^{\ii n\omega_d t} S_\Lambda(\Delta;t).
\label{eq:SDelta_n_def}
\end{equation}
Using the Fourier expansion of the Floquet modes,
\begin{equation*}
\ket{\alpha(t)}=\sum_m \ee^{-\ii m\omega_d t}\ket{\alpha_m}.
\end{equation*}
Equation~\eqref{eq:S_Delta_t_explicit} gives
\begin{equation}
\begin{split}
S_{\Lambda,n}(\Delta)
&=
\sum_m
S_{\Lambda,\alpha\beta l}\,
\ket{\alpha_m}\bra{\beta_{m-n}}
\\
&=\sum_{m}S^\Lambda_{{\rm B},\alpha\beta l}\lvert\alpha_{m}\rangle\langle\beta_{m}^{[n]}\rvert=\sum_{m}S^\Lambda_{{\rm B},\alpha\beta l}\lvert\alpha_{m}^{[-n]}\rangle\langle\beta_{m}\rvert.
\end{split}
\label{eq:SDelta_n_sidebands}
\end{equation}
This expression is particularly convenient numerically because the operator Fourier components can be assembled directly from the Floquet sidebands.

Conveniently, Eq.~\eqref{eq:FGME_nonsecular_Sch} can be written
\begin{widetext}
\begin{equation}
    \begin{split}
      {\mathcal{L}}_\mathrm{diss}(t){\rho}(t)  
&=\sum_{\Lambda\Lambda'}\sum_{\Delta\Delta'}\sum_{nn'}\left\{\mathrm{e}^{-\ii n\omega_d t}\,\Gamma_{\Lambda'\Lambda}(\Delta)\,\lsb S_{{\rm B},n-l}^{\Lambda}(\Delta) {\rho}(t)S_{\rm B}^{\Lambda'\dagger}(\Delta')
        -S_{\rm B}^{\Lambda'\dagger}(\Delta')S_{{\rm B},n-l}^\Lambda(\Delta){\rho}(t)
        \rsb\right.
        \\
        &\hspace{2.0cm}\left. +\mathrm{e}^{\ii n'\omega_d t}\,\Gamma_{\Lambda\Lambda'}^*(\Delta')\lsb S_{\rm B}^\Lambda(\Delta){\rho}(t)S_{{\rm B},n'-l'}^{\Lambda'\dagger}(\Delta')
- {\rho}(t)S_{{\rm B},n'-l'}^{\Lambda'\dagger}(\Delta')S_{\rm B}^\Lambda(\Delta)\rsb\right\}.
    \end{split}
    \label{eq:FGME_nonsecular_Sch_1}
\end{equation}
\end{widetext}
Substituting Eq.~\eqref{eq:SDelta_t_fourier} into Eq.~\eqref{eq:FGME_nonsecular_Sch}, one obtains the Fourier components $\mathcal{L}_{{\rm diss},n}$ of the dissipator. The explicit result is
\begin{widetext}
\begin{equation}
    \begin{split}
      {\mathcal{L}}_{\mathrm{diss},n}\rho(t)
        &=\sum_{\Lambda\Lambda'}\sum_{\Delta\Delta'}
        \lcb\Gamma_{\Lambda'\Lambda}(\Delta)\lsb{S}_{{\rm B},n-l}^{\Lambda}(\Delta){\rho}(t){S}^{\Lambda'\dagger}_{{\rm B}}(\Delta')-{S}^{\Lambda'\dagger}_{\rm B}(\Delta'){S}_{{\rm B},n-l}^{\Lambda}(\Delta){\rho}(t)\rsb\right.
        \\
        &\hspace{5cm}\left.+\Gamma^*_{\Lambda\Lambda'}(\Delta')\lsb{S}^{\Lambda}_{\rm B}(\Delta){\rho}(t){S}_{{\rm B},-l'-n}^{\Lambda'\dagger}(\Delta')-{\rho}(t){S}_{{\rm B},-l'-n}^{\Lambda'\dagger}(\Delta'){S}^{\Lambda}_{\rm B}(\Delta)\rsb\rcb.
    \end{split}
    \label{eq:Ldiss_n}
\end{equation}
\end{widetext}

Equations~\eqref{eq:L_fourier}--\eqref{eq:Ldiss_n} provide the form used for numerical implementation in a truncated Fourier-Sambe representation. In practice, one truncates both the physical Hilbert space and the sideband index, constructs the matrices $S_{\Lambda,n}(\Delta)$ from the Floquet modes, and then builds the Liouvillian harmonics $\mathcal{L}_n$.

\subsection{Brillouin-zone-independent construction of Floquet channels
from the full extended spectrum}
\label{app:pair_channels}

The construction above uses one representative Floquet eigenpair per
physical mode and explicitly enumerates the sideband index $l$ in the
channel frequencies $\Delta_{\alpha\beta l}$ and matrix elements
$S^\Lambda_{\mathrm{B},\alpha\beta l}$. Since quasienergies are
defined modulo $\omega_d$, this requires a Brillouin-zone (BZ)
convention and can become numerically inconvenient near zone
boundaries, resonances, and avoided crossings
\cite{Shirley_Solution_1965,Sambe_Steady_1973,
Le_Defining_2020,Le_Missing_2022}.
Here we show that the same channel structure, and ultimately the
Fourier components of the F-GME dissipator, can be constructed
directly from the complete set of eigenpairs
\begin{equation}
    H_{\mathrm{ex}}^{\mathrm{F}}\Fket{r}
    =
    \varepsilon_r\Fket{r},
\end{equation}
without assigning BZ, replica, or physical-mode labels to the
eigenvectors.

\paragraph*{Channel-pairing identity.}
We consider any two extended-space eigenvectors $\Fket{r}$ and
$\Fket{r'}$, and  define
\begin{equation}
    \Delta_{rr'}
    \equiv
    \varepsilon_{r'}-\varepsilon_r,
    \qquad
    M_{rr'}^\Lambda
    \equiv
    \Fbra{r}
    F_0\otimes S^\Lambda_{\mathrm B}
    \Fket{r'} .
    \label{eq:pair_data}
\end{equation}
These two quantities already form a matched Floquet channel.
To see this, only for the purpose of establishing the equivalence,
write $\Fket{r}=\Fket{n\alpha}$ and
$\Fket{r'}=\Fket{n'\beta}$. With
$l=n'-n$ and the replica convention
$\varepsilon_{n\alpha}=\varepsilon_\alpha+n\omega_d$,
\begin{equation}
\begin{split}
    \Delta_{rr'}
    &=
    \varepsilon_\beta-\varepsilon_\alpha
    +(n'-n)\omega_d
    \\
    &=
    \Delta_{\alpha\beta l},
\end{split}
\label{eq:pair_freq}
\end{equation}
while, using
$\vert\alpha_m^{[n]}\rangle=\vert\alpha_{m+n}\rangle$,
\begin{equation}
\begin{split}
    M_{rr'}^\Lambda
    &=
    \sum_m
    \langle\alpha_{m+n}|
    S^\Lambda_{\mathrm B}
    |\beta_{m+n'}\rangle
    \\
    &=
    \sum_k
    \langle\alpha_k|
    S^\Lambda_{\mathrm B}
    |\beta_{k+l}\rangle
    =
    S^\Lambda_{\mathrm B,\alpha\beta l}.
\end{split}
\label{eq:pair_element}
\end{equation}

Thus, the quasienergy difference and matrix element of an unrestricted
pair reproduce exactly the frequency and amplitude of a physical
Floquet channel. The sideband index $l$ is encoded automatically in
the relative replica displacement and need not be identified
numerically.

\paragraph*{Removal of replica multiplicity.}
Each physical channel is represented repeatedly in the full extended
spectrum under a common translation of the two eigenvectors. As in
App.~\ref{app:phen_extended}, this redundancy can be removed with the
central-block weight
\begin{equation}
    w_r
    \equiv
    P_r^{(0)}
    =
    \Fbra{r}
    \mathcal P_0\otimes\mathbf 1_{\rm dressed}
    \Fket{r},
    \qquad
    \mathcal P_0=|0)(0|,
    \label{eq:pair_weight}
\end{equation}
which obeys
\begin{equation}
    \sum_n P_{n\alpha}^{(0)}=1
\end{equation}
for each physical Floquet class. Consequently, in the untruncated
Sambe space,
\begin{equation}
\begin{split}
&\sum_r w_r\sum_{r'}
f\!\left(
\Delta_{rr'},M_{rr'}^\Lambda
\right)
\\
&\hspace{15mm}
=
\sum_{\alpha\beta l}
f\!\left(
\Delta_{\alpha\beta l},
S^\Lambda_{\mathrm B,\alpha\beta l}
\right),
\end{split}
\label{eq:pair_sum_rule}
\end{equation}
for any function of the matched channel data. In a finite Sambe
truncation, Eq.~\eqref{eq:pair_sum_rule} is recovered as the sideband
cutoff is increased.

\paragraph*{BZ-free operator harmonics.}
The time-dependent operator associated with a channel pair can also
be obtained directly from the full extended-space eigenvectors.
For an integer harmonic $q$, define
\begin{equation}
    K_{rr'}(q)
    \equiv
    {\rm Tr}_{\rm temp}
    \left[
    (F_{-q}\otimes\mathbf 1_{\rm dressed})
    \Fket{r}\Fbra{r'}
    \right],
    \label{eq:pair_K}
\end{equation}
and
\begin{equation}
    X_{rr'}^\Lambda(q)
    \equiv
    M_{rr'}^\Lambda K_{rr'}(q).
    \label{eq:pair_X}
\end{equation}
Equation~\eqref{eq:pair_K} is evaluated directly from the extended
eigenvectors and therefore does not require extracting or identifying
their individual sideband components.

For completeness, expanding the extended vectors only algebraically
gives
\begin{equation}
    K_{rr'}(q)
    =
    \sum_m |r_m\rangle\langle r'_{m-q}|,
\end{equation}
and hence, for
$\Fket r=\Fket{n\alpha}$ and
$\Fket{r'}=\Fket{n'\beta}$,
\begin{equation}
    K_{rr'}(q)
    =
    \sum_k
    |\alpha_k\rangle
    \langle\beta_{k+l-q}|.
\end{equation}

Thus the Fourier expansion
\begin{equation}
    A_{rr'}^\Lambda(t)
    \equiv
    \sum_q e^{-iq\omega_dt}
    X_{rr'}^\Lambda(q)
    \label{eq:pair_A_time}
\end{equation}
reproduces the corresponding covariant Floquet transition,
\begin{equation}
    A_{rr'}^\Lambda(t)
    =
    e^{-il\omega_dt}
    S^\Lambda_{\mathrm B,\alpha\beta l}
    |\alpha(t)\rangle\langle\beta(t)|,
\end{equation}
without requiring $l$ to be determined in the numerical
implementation.

\paragraph*{Direct construction of the dissipator harmonics.}
For independent baths, it is convenient to collect the unrestricted
pair contributions into the unweighted and bath-weighted harmonics
\begin{align}
    \mathcal S_{\Lambda,q}
    &\equiv
    \sum_{r,r'}w_r
    X_{rr'}^\Lambda(q),
    \label{eq:pair_S}
    \\
    \mathcal G_{\Lambda,q}
    &\equiv
    \sum_{r,r'}w_r\,
    \Gamma^\Lambda(\Delta_{rr'})
    X_{rr'}^\Lambda(q).
    \label{eq:pair_G}
\end{align}
The first quantity contains the complete system-coupling contribution
at harmonic $q$, whereas the second contains the same channels
weighted by the bath response at their physical quasienergy
differences.

Writing
\begin{equation}
    \mathcal L_{\rm diss}(t)
    =
    \sum_n e^{-in\omega_dt}\,
    \mathcal L_{{\rm diss},n},
\end{equation}
the full nonsecular dissipator harmonics can then be assembled
directly as
\begin{equation}
\begin{split}
\mathcal L_{{\rm diss},n}\rho
=
\sum_{\Lambda,q}
\Big[
&
\mathcal G_{\Lambda,q}\rho
\mathcal S_{\Lambda,q-n}^{\dagger}
-
\mathcal S_{\Lambda,q-n}^{\dagger}
\mathcal G_{\Lambda,q}\rho
\\
&+
\mathcal S_{\Lambda,q}\rho
\mathcal G_{\Lambda,q-n}^{\dagger}
-
\rho\,
\mathcal G_{\Lambda,q-n}^{\dagger}
\mathcal S_{\Lambda,q}
\Big].
\end{split}
\label{eq:Ldiss_BZfree}
\end{equation}
Equation~\eqref{eq:Ldiss_BZfree} is the BZ-independent counterpart of
the representative-based Fourier construction above. All of its
ingredients are obtained directly from the raw eigenvalues and
eigenvectors of $H_{\mathrm{ex}}^{\mathrm F}$; no BZ selection,
replica identification, mode tracking, or explicit sideband-channel
label is required.

For a zero-temperature bath, negative-frequency decay processes may
be removed directly through the bath response,
$\Gamma^\Lambda(\Delta<0)=0$, so that no additional channel
classification is necessary. At finite temperature, both signs of
$\Delta_{rr'}$ are retained with the corresponding detailed-balance
relations.

In the numerical calculations presented here, the unrestricted
pair-sum construction and the representative-based implementation
give the same period-averaged populations within
$10^{-6}$--$10^{-9}$ over the investigated drive range, including
the channel-crossing regime, once the Sambe truncation is converged.
The pair construction therefore provides an equivalent formulation
while avoiding explicit BZ bookkeeping.

\subsection{Master equation in the Floquet basis}

It is also useful to write the GME directly in the Floquet basis. Expanding the reduced density matrix as
\begin{equation}
\rho(t)=\sum_{\alpha,\beta}\rho_{\alpha\beta}(t)\,
\ket{\alpha(t)}\bra{\beta(t)},
\label{eq:rho_floquet_open_expand}
\end{equation}
with
\begin{equation}
\rho_{\alpha\beta}(t)=\bra{\alpha(t)}\rho(t)\ket{\beta(t)},
\end{equation}
the master equation becomes
\begin{equation}
\partial_t \rho_{\alpha\beta}(t)
=
-\ii(\varepsilon_\alpha-\varepsilon_\beta)\rho_{\alpha\beta}(t)
+
\bra{\alpha(t)}\mathcal{L}_{\rm diss}(t)\rho(t)\ket{\beta(t)}.
\label{eq:rhoab_open_general}
\end{equation}

Substituting Eq.~\eqref{eq:FGME_nonsecular_Sch} and Eq.~\eqref{eq:rho_floquet_open_expand}, one obtains a linear system of the form
\begin{equation}
\partial_t \rho_{\alpha\beta}(t)
=
\sum_{\sigma,\sigma'}
\mathcal{L}_{\alpha\beta,\sigma\sigma'}(t)\,
\rho_{\sigma\sigma'}(t),
\label{eq:Floquet_basis_linear}
\end{equation}
where
\begin{equation}
\mathcal{L}_{\alpha\beta,\sigma\sigma'}(t)
=
-\ii(\varepsilon_\alpha-\varepsilon_\beta)\delta_{\alpha\sigma}\delta_{\beta\sigma'}
+
\mathcal{D}_{\alpha\beta,\sigma\sigma'}(t),
\label{eq:L_super_floquet}
\end{equation}
and $\mathcal{D}(t)$ is the dissipative kernel obtained from the operator products in Eq.~\eqref{eq:FGME_nonsecular_Sch}, with
\begin{widetext}
\begin{equation}
    \begin{split}
       \mathcal{D}_{\lambda\lambda',\nu\nu'}(t)
&=\sum_{\Lambda\Lambda}\sum_{\alpha\beta l}\sum_{\alpha'\beta'l}\sum_{\sigma\sigma'}\Big\{
\mathrm{e}^{-\ii l\omega_{d}t}\Gamma^{\Lambda'\Lambda*}(\Delta)\lsb U^{{\rm F}}_{\lambda\alpha}(t){S}^{\Lambda+}_{\alpha\beta l} U^{{\rm F}\dagger}_{\beta\sigma}(t) \delta_{\sigma\nu}\delta_{\nu'\alpha'} {S}^{\Lambda'-}_{\alpha'\beta' l'}\delta_{\beta'\lambda'}-\delta_{\lambda\alpha'}{S}^{\Lambda'-}_{\alpha'\beta' l'}U^{{\rm F}}_{\beta'\alpha}(t){S}_{\alpha\beta l}^{\Lambda+}U^{{\rm F}\dagger}_{\beta\sigma}(t)\delta_{\sigma\nu}\delta_{\nu'\lambda'}\rsb
\\
&\hspace{0.5cm}+\mathrm{e}^{\ii l'\omega_{d}t}
\Gamma^{\Lambda\Lambda'}(\Delta')\lsb \delta_{\lambda\alpha}{S}^{\Lambda+}_{\alpha\beta l}\delta_{\beta\nu}\delta_{\nu'\sigma}
U^{{\rm F}\dagger}_{\sigma\alpha'}(t){S}^{\Lambda'-}_{\alpha'\beta' l'}U^{{\rm F}}_{\beta'\lambda'}(t)- \delta_{\lambda\nu}\delta_{\nu'\sigma} U^{{\rm F}\dagger}_{\sigma\alpha'}(t){S}^{\Lambda'-}_{\alpha'\beta' l'}U^{{\rm F}}_{\beta'\alpha}(t) {S}^{\Lambda+}_{\alpha\beta l} \delta_{\beta\lambda'}\rsb\Big\}.
    \end{split}
    \label{eq:D_abcd}
\end{equation}
\end{widetext}

This form is convenient conceptually, but for numerical work it is often simpler to use the time-independent basis of $H_0$ or the Fourier-expanded Liouvillian representation described above.

\subsection{Secular approximation}

When the full secular approximation is applied, it is advantageous to
work in the interaction picture, since the resulting dissipative
generator becomes time independent. Starting from
Eq.~\eqref{eq:interaction_nonsecular_FR}, the rapidly oscillating terms
with $\Delta\neq\Delta'$ are neglected, so that only resonant
contributions satisfying $\Delta=\Delta'$ are retained. The
interaction-picture dissipator then reduces to
\begin{widetext}
\begin{equation}
\begin{split}
\widetilde{\mathcal L}_{\rm diss}^{\rm sec}
\widetilde{\rho}(t)
=
\sum_{\Lambda\Lambda'}
\sum_{\Delta}
\Bigl\{
&
\Gamma_{\Lambda'\Lambda}(\Delta)
\Bigl[
S_{\rm B}^{\Lambda}(\Delta)
\widetilde{\rho}(t)
S_{\rm B}^{\Lambda'\dagger}(\Delta)
-
S_{\rm B}^{\Lambda'\dagger}(\Delta)
S_{\rm B}^{\Lambda}(\Delta)
\widetilde{\rho}(t)
\Bigr]
\\
&+
\Gamma_{\Lambda\Lambda'}^{*}(\Delta)
\Bigl[
S_{\rm B}^{\Lambda}(\Delta)
\widetilde{\rho}(t)
S_{\rm B}^{\Lambda'\dagger}(\Delta)
-
\widetilde{\rho}(t)
S_{\rm B}^{\Lambda'\dagger}(\Delta)
S_{\rm B}^{\Lambda}(\Delta)
\Bigr]
\Bigr\},
\end{split}
\label{eq:SecFGME_IP}
\end{equation}
\end{widetext}
where $S_{\rm B}^{\Lambda}(\Delta)$ is the total transition
operator associated with the physical transition frequency $\Delta$.
In particular, it is not necessarily associated with only one
Floquet-channel label $(\alpha,\beta,l)$, as explained below.

\subsubsection*{Frequency grouping of degenerate channels}

An important point in implementing the secular approximation is that
secularization is performed with respect to physical transition
frequencies rather than transition labels. Distinct Floquet channels
\begin{equation}
a\equiv(\alpha,\beta,l),
\qquad
a'\equiv(\alpha',\beta',l'),
\end{equation}
must therefore remain coherently combined whenever
\begin{equation}
\Delta_a=\Delta_{a'},
\qquad
\Delta_a
\equiv
\varepsilon_\beta-\varepsilon_\alpha+l\omega_d.
\end{equation}

For every distinct transition frequency $\Omega$, we define the
frequency-degenerate channel set
\begin{equation}
\mathcal C_{\Omega}
=
\left\{
(\alpha,\beta,l):
\Delta_{\alpha\beta l}=\Omega
\right\},
\label{eq:frequency_group_set}
\end{equation}
and construct the corresponding grouped transition operator
\begin{equation}
S_{\rm B}^{\Lambda}(\Omega)
=
\sum_{(\alpha,\beta,l)\in\mathcal C_{\Omega}}
S_{{\rm B},\alpha\beta l}^{\Lambda}
\ket{\alpha}\bra{\beta}.
\label{eq:frequency_grouped_operator}
\end{equation}
The operator $S_{\rm B}^{\Lambda}(\Omega)$ in
Eq.~\eqref{eq:SecFGME_IP} must therefore contain the coherent sum of
all channels belonging to the same frequency group. Constructing a
separate dissipator for every label $(\alpha,\beta,l)$ would generally
oversecularize the generator by discarding interference terms between
frequency-degenerate channels.

For independent baths,
\begin{equation}
\Gamma_{\Lambda\Lambda'}(\Omega)
=
\delta_{\Lambda\Lambda'}\Gamma_{\Lambda}(\Omega),
\end{equation}
then Eq.~\eqref{eq:SecFGME_IP} becomes
\begin{widetext}
\begin{equation}
\begin{split}
\widetilde{\mathcal L}_{\rm sec}
\widetilde{\rho}
=
\sum_{\Lambda}
\sum_{\Omega}
\Bigl\{
&
\Gamma_{\Lambda}(\Omega)
\Bigl[
S_{\rm B}^{\Lambda}(\Omega)
\widetilde{\rho}
S_{\rm B}^{\Lambda\dagger}(\Omega)
-
S_{\rm B}^{\Lambda\dagger}(\Omega)
S_{\rm B}^{\Lambda}(\Omega)
\widetilde{\rho}
\Bigr]
\\
&+
\Gamma_{\Lambda}^{*}(\Omega)
\Bigl[
S_{\rm B}^{\Lambda}(\Omega)
\widetilde{\rho}
S_{\rm B}^{\Lambda\dagger}(\Omega)
-
\widetilde{\rho}
S_{\rm B}^{\Lambda\dagger}(\Omega)
S_{\rm B}^{\Lambda}(\Omega)
\Bigr]
\Bigr\}.
\end{split}
\label{eq:SecFGME_grouped}
\end{equation}
\end{widetext}

At zero temperature, the frequency sum may be restricted to the
positive-frequency decay groups, $\Omega>0$. At finite temperature,
both positive- and negative-frequency groups must be retained together
with the corresponding detailed-balance relations.

When the bath response is real and nonnegative, and its
principal-value Lamb-shift contribution is neglected, one may define
one grouped collapse operator for each bath channel and distinct
transition frequency,
\begin{equation}
C_{\Lambda,\Omega}
=
\sqrt{\Gamma_{\Lambda}(\Omega)}
S_{\rm B}^{\Lambda}(\Omega).
\label{eq:grouped_collapse_operator}
\end{equation}
Using the dissipator convention
\begin{equation}
\mathcal D[C]\rho
=
2C\rho C^{\dagger}
-
C^{\dagger}C\rho
-
\rho C^{\dagger}C,
\label{eq:dissipator_convention_sec}
\end{equation}
the secular Liouvillian can then be written as
\begin{equation}
\widetilde{\mathcal L}_{\rm sec}
=
\sum_{\Lambda,\Omega}
\mathcal D[C_{\Lambda,\Omega}].
\label{eq:grouped_secular_liouvillian}
\end{equation}

In numerical calculations, equality of transition frequencies is
identified using a small frequency-grouping tolerance
$\delta_{\rm freq}$,
\begin{equation}
\left|
\Delta_{\alpha\beta l}
-
\Delta_{\alpha'\beta'l'}
\right|
<
\delta_{\rm freq}.
\label{eq:frequency_grouping_tolerance}
\end{equation}
The tolerance $\delta_{\rm freq}$ is chosen only to identify
frequencies that are equal up to numerical precision. It is not a
physical coarse-graining scale. Grouping physically distinct but
nearby frequencies over a finite frequency window would instead
constitute a partial-secular or coarse-grained approximation.

This frequency-grouped construction retains all interference terms
between degenerate channels. For example, distinct labels such as
$(g,e,l)$ and $(e,g,l+1)$ must be combined whenever they correspond to
the same physical transition frequency. Likewise, whenever they are
retained, all population-type channels with $\alpha=\beta$ and the
same sideband index $l$ belong to the same frequency group because
\begin{equation}
\Delta_{\alpha\alpha l}
=
l\omega_d
\end{equation}
independently of $\alpha$.

\paragraph*{Steady-state observables in the secular interaction picture.}

After the full secular approximation, the interaction-picture master
equation takes the time-independent form
\begin{equation}
\frac{d\widetilde{\rho}(t)}{dt}
=
\widetilde{\mathcal L}_{\rm sec}
\widetilde{\rho}(t).
\label{eq:secular_IP_master_equation}
\end{equation}
The interaction-picture steady state is therefore obtained from
\begin{equation}
\widetilde{\mathcal L}_{\rm sec}
\widetilde{\rho}_{\rm ss}
=
0,
\qquad
{\rm Tr}
\left[
\widetilde{\rho}_{\rm ss}
\right]
=
1.
\label{eq:secular_IP_steady_state}
\end{equation}

We let $\ket{\alpha(t)}$ denote the $T$-periodic Floquet mode and
$\varepsilon_\alpha$ its quasienergy. The Floquet propagator can be
written as
\begin{equation}
U_{\rm F}(t)
=
\sum_{\alpha}
e^{-\ii\varepsilon_\alpha t}
\ket{\alpha(t)}
\bra{\alpha},
\label{eq:Floquet_propagator_secular}
\end{equation}
where $\ket{\alpha}$ denotes the corresponding time-independent
reference state in the Floquet interaction picture. The
Schr\"odinger-picture asymptotic state is reconstructed as
\begin{equation}
\rho_{\rm ss}(t)
=
U_{\rm F}(t)
\widetilde{\rho}_{\rm ss}
U_{\rm F}^{\dagger}(t).
\label{eq:secular_ss_reconstruction_operator}
\end{equation}

Equivalently,
\begin{equation}
\rho_{\rm ss}(t)
=
\sum_{\alpha\beta}
\widetilde{\rho}_{\alpha\beta}^{\rm ss}
e^{-\ii(\varepsilon_\alpha-\varepsilon_\beta)t}
\ket{\alpha(t)}
\bra{\beta(t)}.
\label{eq:secular_ss_reconstruction}
\end{equation}
For a nondegenerate fully secular steady state,
$\widetilde{\rho}_{\rm ss}$ is normally diagonal in the Floquet basis,
and the explicit quasienergy phases in
Eq.~\eqref{eq:secular_ss_reconstruction} consequently cancel.
Equation~\eqref{eq:secular_ss_reconstruction} also covers cases with
degenerate sectors in which stationary coherences may remain.

For any observable $O$, the long-time periodic average is
\begin{equation}
\overline{O}
=
\frac{1}{T}
\int_0^T
dt\,
{\rm Tr}
\left[
O\rho_{\rm ss}(t)
\right].
\label{eq:secular_observable_average}
\end{equation}
In particular, for the excitation number of subsystem $\Lambda$,
\begin{equation}
\overline{N}_{\Lambda}
=
\frac{1}{T}
\int_0^T
dt\,
{\rm Tr}
\left[
s_{\rm B}^{\Lambda-}
s_{\rm B}^{\Lambda+}
\rho_{\rm ss}(t)
\right].
\label{eq:secular_excitation_average}
\end{equation}

Equivalently, one may transform the observable to the Floquet
interaction picture,
\begin{equation}
\widetilde{O}(t)
=
U_{\rm F}^{\dagger}(t)
O
U_{\rm F}(t),
\label{eq:secular_transformed_observable}
\end{equation}
and evaluate
\begin{equation}
\overline{O}
=
\frac{1}{T}
\int_0^T
dt\,
{\rm Tr}
\left[
\widetilde{O}(t)
\widetilde{\rho}_{\rm ss}
\right].
\label{eq:secular_observable_average_IP}
\end{equation}
Thus, in the secular interaction-picture formulation, the density
matrix is time independent while the physical time dependence is
carried by the transformed observables.

\paragraph*{Spectrum in the secular interaction picture.}

The incoherent first-order spectrum is obtained from the
period-averaged two-time correlation function
\begin{equation}
\begin{split}
\overline{G}_{\Lambda}^{(1)}(\tau)
=
\frac{1}{T}
\int_0^T
dt\,
\Bigl[
&
\left\langle
\widetilde{s}_{\rm B}^{\Lambda-}(t+\tau)
\widetilde{s}_{\rm B}^{\Lambda+}(t)
\right\rangle_{\rm ss}
\\
&-
\left\langle
\widetilde{s}_{\rm B}^{\Lambda-}(t+\tau)
\right\rangle_{\rm ss}
\left\langle
\widetilde{s}_{\rm B}^{\Lambda+}(t)
\right\rangle_{\rm ss}
\Bigr],
\end{split}
\label{eq:secular_incoherent_correlator}
\end{equation}
where
\begin{equation}
\widetilde{s}_{\rm B}^{\Lambda\pm}(t)
=
U_{\rm F}^{\dagger}(t)
s_{\rm B}^{\Lambda\pm}
U_{\rm F}(t),
\label{eq:secular_transformed_ladder_operator}
\end{equation}
and all expectation values are evaluated with
$\widetilde{\rho}_{\rm ss}$.

Using the quantum regression theorem,
\begin{equation}
\begin{split}
&
\left\langle
\widetilde{s}_{\rm B}^{\Lambda-}(t+\tau)
\widetilde{s}_{\rm B}^{\Lambda+}(t)
\right\rangle_{\rm ss}
\\
&\qquad =
{\rm Tr}
\left[
\widetilde{s}_{\rm B}^{\Lambda-}(t+\tau)
e^{\widetilde{\mathcal L}_{\rm sec}\tau}
\left(
\widetilde{s}_{\rm B}^{\Lambda+}(t)
\widetilde{\rho}_{\rm ss}
\right)
\right].
\end{split}
\label{eq:secular_QRT}
\end{equation}
The incoherent spectrum is then obtained from
Eq.~\eqref{eq:Spectrum_def}. This procedure is numerically convenient
because the propagator
$e^{\widetilde{\mathcal L}_{\rm sec}\tau}$ is generated by a
time-independent Liouvillian.

\paragraph*{Numerical implementation.}

The full-secular calculation can be implemented as follows:

\begin{enumerate}

\item
Diagonalize the closed periodically driven Hamiltonian in Floquet
space and obtain the quasienergies $\varepsilon_\alpha$ and Floquet
modes $\ket{\alpha(t)}$.

\item
Construct all transition frequencies,
\begin{equation}
\Delta_{\alpha\beta l}
=
\varepsilon_\beta-\varepsilon_\alpha+l\omega_d,
\end{equation}
and the corresponding Floquet matrix elements
$S_{{\rm B},\alpha\beta l}^{\Lambda}$.

\item
At zero temperature, retain the positive-frequency decay channels
$\Delta_{\alpha\beta l}>0$. At finite temperature, retain both signs
of $\Delta_{\alpha\beta l}$ together with the appropriate
detailed-balance-related rates.

\item
Partition the retained channels into frequency-degenerate sets
$\mathcal C_{\Omega}$ according to
Eq.~\eqref{eq:frequency_grouping_tolerance}. The tolerance
$\delta_{\rm freq}$ is used only to identify equal frequencies within
numerical precision.

\item
For every bath channel $\Lambda$ and distinct transition frequency
$\Omega$, construct the grouped transition operator,
\begin{equation}
S_{\rm B}^{\Lambda}(\Omega)
=
\sum_{(\alpha,\beta,l)\in\mathcal C_{\Omega}}
S_{{\rm B},\alpha\beta l}^{\Lambda}
\ket{\alpha}\bra{\beta}.
\end{equation}

\item
For independent baths with real and nonnegative response functions,
construct the grouped collapse operators
\begin{equation}
C_{\Lambda,\Omega}
=
\sqrt{\Gamma_{\Lambda}(\Omega)}
S_{\rm B}^{\Lambda}(\Omega),
\end{equation}
and build the time-independent secular Liouvillian:
\begin{equation}
\widetilde{\mathcal L}_{\rm sec}
=
\sum_{\Lambda,\Omega}
\mathcal D[C_{\Lambda,\Omega}].
\end{equation}
If the imaginary principal-value part of the bath response is
retained, the corresponding Lamb-shift Hamiltonian must also be
included. Alternatively, the generator may be assembled directly from
Eq.~\eqref{eq:SecFGME_grouped}.

\item
If working in the Floquet frame rather than the fully rotating
interaction picture, include the quasienergy Hamiltonian contribution
\begin{equation}
-\ii[H_{\rm F},\,\cdot\,].
\end{equation}

\item
Solve
\begin{equation}
\widetilde{\mathcal L}_{\rm sec}
\widetilde{\rho}_{\rm ss}
=
0,
\end{equation}
subject to
${\rm Tr}[\widetilde{\rho}_{\rm ss}]=1$, and calculate observables by
period-averaging either the reconstructed Schr\"odinger-picture state
or the corresponding interaction-picture observables.

\item
Compute spectra using the quantum regression theorem with the same
time-independent Liouvillian
$\widetilde{\mathcal L}_{\rm sec}$, and average the resulting
two-time correlation function over one drive period.

\end{enumerate}

\paragraph*{Partial secular approximation.}

A less restrictive approximation may be introduced by discarding only
terms that oscillate rapidly on a prescribed physical
coarse-graining timescale. One commonly used Floquet partial-secular
scheme first requires equal Floquet harmonics,
\begin{equation}
l=l',
\end{equation}
while retaining couplings between sufficiently close quasienergy
transitions satisfying
\begin{equation}
\left|
(\varepsilon_\beta-\varepsilon_\alpha)
-
(\varepsilon_{\beta'}-\varepsilon_{\alpha'})
\right|
\lesssim
\delta_{\rm ps},
\label{eq:partial_secular_window}
\end{equation}
where $\delta_{\rm ps}$ is a physical coarse-graining scale rather
than a numerical tolerance. This construction preserves selected
near-resonant channel couplings while removing rapidly oscillating
terms associated with well-separated Floquet sidebands. Additional
conditions on the resulting coefficient matrix are required if a
GKSL generator is desired.

The partial-secular scale $\delta_{\rm ps}$ should therefore be
distinguished from the numerical grouping tolerance
$\delta_{\rm freq}$ in
Eq.~\eqref{eq:frequency_grouping_tolerance}: the former defines a
physical approximation, whereas the latter only identifies exactly
degenerate frequencies within finite numerical precision.

\subsection{General remarks}

First, Eq.~\eqref{eq:FGME_nonsecular_Sch} is a nonsecular
Floquet--Markov master equation. It retains couplings between distinct
Floquet transition frequencies $\Delta$ and $\Delta'$ and therefore
generally differs from the fully secular Floquet master equation. This
distinction is essential whenever near-degenerate transition
frequencies or small dissipative splittings are relevant.

Second, complete positivity is not guaranteed at the
Redfield/nonsecular level in general. The present approach is
therefore best viewed as a weak-coupling Born--Markov generalized
master equation rather than a
Gorini--Kossakowski--Sudarshan--Lindblad (GKSL) form. This is precisely
the regime relevant to our comparison between the Floquet generalized
master equation and time-independent-basis approaches.

Third, at zero temperature, the bath rates satisfy
$\Gamma_{\Lambda}(\Delta)\approx0$ for $\Delta<0$, so that the bath
induces only downward net-energy transitions. At finite temperature,
both signs of $\Delta$ contribute, and the corresponding
detailed-balance relations must be retained.

\section{Floquet--Liouville formalism for periodically driven open quantum systems}
\label{seS:FL}
In this Appendix, we present the Floquet--Liouville (FL)
formalism employed in the main text.
The central idea is that the dynamics of a periodically
driven open quantum system can be understood through
the eigenmodes of a time-independent superoperator
defined in an enlarged FL extended space.

Just as ordinary Floquet theory transforms the
time-periodic Schr\"{o}dinger equation into a
time-independent quasienergy eigenvalue problem,
the FL formalism transforms the time-periodic
master equation into a time-independent
eigenvalue problem for a FL
supermatrix.

The resulting eigenvalues determine the
oscillation frequencies and decay rates of the
driven dissipative system, while the associated
eigenvectors determine the dynamical modes,
steady state, and spectral response.
\label{secS:FL}

In the previous section, we derived the nonsecular F-GME for a periodically driven open quantum system, of the general form
\begin{equation}
\partial_t \rho(t)=\mathcal{L}(t)\rho(t),
\qquad
\mathcal{L}(t+T)=\mathcal{L}(t),
\label{eq:FGME_periodic}
\end{equation}
where $\mathcal{L}(t)$ is the full time-periodic Liouvillian superoperator, including both the coherent $\mathcal{L}_{\rm S}(t)$ and dissipative $\mathcal{L}_{\rm diss}(t)$ contributions. 

We now reformulate this problem in the Floquet--Liouville (FL) extended space. This construction is the Liouville-space analogue of the Sambe-space formalism introduced for the closed system. Its main advantage is that it converts the time-periodic master equation into a time-independent eigenvalue problem in an enlarged space, thereby providing a unified and numerically convenient framework for the steady state, the propagator, and two-time correlation functions.

\subsection{Liouville-space notation and periodic Liouvillian}

We first vectorize the system density operator and all system operators in Liouville space. We denote vectors in Liouville space by rounded kets and bras,
\begin{equation}
\Lket{\rho} \in \mathbb{H}_{\rm L},
\qquad
\Lbraket{A}{\rho}=\Tr{A^\dagger \rho},
\label{eq:Liouville_notation}
\end{equation}
where $\mathbb{H}_{\rm L}$ is the Liouville space associated with the physical Hilbert space. Note $\rho$ is an operator (a matrix in a linear representation) in the physical Hilbert space, but it transforms into a vector $\Lket{\rho}$ in the Liouville space. 

This Liouville space here is span by the exteriour product of the dressed basis set, $\mathsf{
B}_\mathrm{L}=\lcb\Lket{jk}\equiv\ket{j}\bra{k},\,\ket{j},\ket{k}\in\mathbb{H}_{\rm dressed}\rcb$.
The scalar product is the Hilbert--Schmidt inner product,
\begin{equation}
\Lbraket{A}{B}=\Tr{A^\dagger B}.
\end{equation}

In this notation, the master equation \eqref{eq:FGME_periodic} becomes
\begin{equation}
\partial_t \Lket{\rho(t)}=\mathscr{L}(t)\,\Lket{\rho(t)},
\qquad
\mathscr{L}(t+T)=\mathscr{L}(t),
\label{eq:ME_Liouville}
\end{equation}
where $\mathscr{L}(t)$ is the Liouvillian supermatrix representing $\mathcal{L}(t)$ in Liouville space.

Since the Liouvillian is periodic, it admits a Fourier expansion,
\begin{equation}
\mathscr{L}(t)=\sum_{n\in\mathbb{Z}} \ee^{-\ii n\omega_d t}\,\mathscr{L}_n.
\label{eq:L_t_Fourier}
\end{equation}
The formal analogy with the closed-system Floquet problem is now immediate. In particular, Eq.~\eqref{eq:ME_Liouville} is a linear differential equation with periodic coefficients, so Floquet's theorem applies directly in Liouville space.

\subsection{Floquet theorem in Liouville space}

Using Floquet's theorem, solutions of Eq.~\eqref{eq:ME_Liouville} can be written as
\begin{equation}
\Lket{\rho_\mu(t)}=\ee^{\lambda_\mu t}\,\Lket{R_\mu(t)},
\qquad
\Lket{R_\mu(t+T)}=\Lket{R_\mu(t)},
\label{eq:FL_right_time}
\end{equation}
where $\lambda_\mu\in\mathbb{C}$ are the Floquet--Liouville exponents and $\Lket{R_\mu(t)}$ are the corresponding $T$-periodic right Floquet--Liouville modes. 
Substituting Eq.~\eqref{eq:FL_right_time} into Eq.~\eqref{eq:ME_Liouville} yields
\begin{equation}
\mathscr{L}^{\rm F}(t)\Lket{R_\mu(t)}=\lambda_\mu \Lket{R_\mu(t)}.
\label{eq:FL_time_eigenproblem}
\end{equation}
where $\mathscr{L}^{\rm F}(t)\equiv\mathscr{L}(t)-\partial_t$. 

This is the Liouville-space analogue of the quasienergy equation for the closed system. Since $\mathscr{L}(t)$ is generally non-Hermitian, one must also introduce left FL modes $\Lbra{L_\mu(t)}$, defined by
\begin{equation}
\Lbra{L_\mu(t)}\mathscr{L}^{\rm F}(t)=\lambda_\mu \Lbra{L_\mu(t)},
\qquad
\Lbra{L_\mu(t+T)}=\Lbra{L_\mu(t)}.
\label{eq:FL_left_time}
\end{equation}
The left and right modes may be chosen biorthonormal over one period:
\begin{equation}
\frac{1}{T}\int_T dt\, \Lbraket{L_\mu(t)}{R_\nu(t)}=\delta_{\mu\nu}.
\label{eq:FL_biorth_time}
\end{equation}

\subsection{Floquet--Liouville extended space}

To make the eigenvalue problem time independent, we introduce the temporal Fourier space $\mathbb{H}_{\rm temp}$, exactly as in the closed-system Sambe construction, with basis $\{|n)\}_{n\in\mathbb{Z}}$ satisfying
\begin{equation}
(t|n)=\ee^{-\ii n\omega_d t},
\qquad
(n|m)=\delta_{nm}.
\end{equation}
The Floquet--Liouville extended space is then
\begin{equation}
\mathbb{H}^{\rm FL}_{\rm ex}
=
\mathbb{H}_{\rm temp}\otimes \mathbb{H}_{\rm L}.
\end{equation}
Vectors in $\mathbb{H}^{\rm FL}_{\rm ex}$ will be denoted by triple bracket notation,
\begin{equation}
\FLket{R_\mu}\in \mathbb{H}^{\rm FL}_{\rm ex},
\qquad
\FLbra{L_\mu} \in \mathbb{H}^{{\rm FL}\ast}_{\rm ex}.
\end{equation}

A convenient basis is
\begin{equation}
\FLket{n,jk} \equiv |n)\otimes \ket{j}\bra{k},
\end{equation}
where $\{\ket{jk}\equiv\ket{j}\bra{k}\}$ is a basis of Liouville space.

The time-periodic right mode $\Lket{R_\mu(t)}$ is expanded as
\begin{equation}
\Lket{R_\mu(t)}=\sum_{n\in\mathbb{Z}} \ee^{-\ii n\omega_d t}\Lket{R_{\mu,n}},
\label{eq:Rmu_fourier}
\end{equation}
and similarly for the left mode,
\begin{equation}
\Lbra{L_\mu(t)}=\sum_{n\in\mathbb{Z}} \ee^{+\ii n\omega_d t} \Lbra{L_{\mu,n}}.
\label{eq:Lmu_fourier}
\end{equation}
We then define the corresponding extended-space vectors
\begin{equation}
\FLket{R_\mu}
=
\sum_n |n)\otimes \Lket{R_{\mu,n}},
\qquad
\FLbra{L_\mu}
=
\sum_n (n|\otimes \Lbra{L_{\mu,n}}.
\label{eq:R_L_extended}
\end{equation}

Introducing the temporal shift operators $F_m$ and the temporal number operator $N_{\rm temp}$ through
\begin{equation}
F_m|n)=|n+m),
\qquad
N_{\rm temp}|n)=n|n),
\end{equation}
the time-independent FL supermatrix is
\begin{equation}
\mathscr{L}^{\rm F}(t)\to\mathscr{L}^{\rm F}_{\rm ex}
=
\sum_{m\in\mathbb{Z}} F_m\otimes \mathscr{L}_m
+\ii\omega_d\, N_{\rm temp}\otimes \mathbf{1}_{\rm L},
\label{eq:LFL_def}
\end{equation}
where $\mathbf{1}_{\rm L}$ is the identity in Liouville space. 

The eigenvalue problems in the extended space are then
\begin{equation}
\mathscr{L}^{\rm F}_{\rm ex}\FLket{R_\mu}
=
\lambda_\mu \FLket{R_\mu},
\label{eq:FL_right_extended}
\end{equation}
\begin{equation}
\FLbra{L_\mu}\,\mathscr{L}^{\rm F}_{\rm ex}
=
\lambda_\mu \FLbra{L_\mu}.
\label{eq:FL_left_extended}
\end{equation}
The left and right eigenvectors satisfy the biorthonormality and completeness relations
\begin{equation}
\FLbraket{L_\mu}{R_\nu}=\delta_{\mu\nu},
\qquad
\sum_\mu\FLket{R_\mu}\FLbra{L_\mu}
=
\mathbf{1}_{\rm F-L}.
\label{eq:FL_biorth_extended}
\end{equation}

As in ordinary Floquet theory, the spectrum is replicated in strips shifted by integer multiples of $\ii\omega_d$:
\begin{equation}
\lambda_{\mu}^{[k]}=\lambda_\mu+\ii k\omega_d,
\qquad
k\in\mathbb{Z}.
\label{eq:FL_zone_replica}
\end{equation}
It is therefore sufficient to keep one representative from a single FL Brillouin strip.
Hence, Eqs.~\eqref{eq:FL_right_extended} and \eqref{eq:FL_left_extended} actually has more eigenvalues and eigenvectors so that the extended-space matrix eigenvalue problem posseses a larger size and must be correctly rewritten as
\begin{equation}
\mathscr{L}^{\rm F}_{\rm ex}\FLket{R_{n\mu}}
=
\lambda_{n\mu} \FLket{R_{n\mu}}.
\label{eq:FL_eigenproblem_ex_modified}
\end{equation}

By extended-space redundancy convention, then, $\lambda_{\mu}\equiv\lambda_{0\mu}$ and $\FLket{R_{n\mu}}=\sum_l \tket{l}\otimes\Lket{R_{\mu,l}^{[n]}}$, so that $\FLket{R_\mu}=\FLket{R_{0\mu}}$, and $\mu$ indexes the FL extended space modes, so that these modes can be written as superpositions of basis states in the Liouville space in terms of the dressed states
\begin{equation}
    \begin{split}
        \Lket{R_\mu(t)}&=\sum_{jk} R^{jk}_{\mu}(t)\,\ket{j}\bra{k},
        \\
        \Lbra{L_\mu(t)}&=\sum_{jk} L_{\mu}^{jk}(t)\,\ket{j}\bra{k},
    \end{split}
\end{equation}
or, in terms of the Floquet modes
\begin{equation}
    \begin{split}
        \Lket{R_\mu(t)}&=\sum_{\alpha\beta} R_{\mu}^{\alpha\beta}(t)\,\ket{\alpha(t)}\bra{\beta(t)},
        \\
        \Lbra{L_\mu(t)}&=\sum_{\alpha\beta} L_{\mu}^{\alpha\beta}(t)\,\ket{\alpha(t)}\bra{\beta(t)}.
    \end{split}
\end{equation}

Likewise the FL extended-space modes are written via
in terms of the dressed states:
\begin{equation}
    \begin{split}
        \FLket{R_\mu}&=\sum_{njk} R_{\mu}^{njk}\,\tket{n}\otimes\ket{j}\bra{k},
        \\
        \FLbra{L_\mu}&=\sum_{njk} L_{\mu}^{njk}\,\tket{n}\otimes\ket{j}\bra{k},
    \end{split}
\end{equation}
or, in terms of the Floquet modes
\begin{equation}
    \begin{split}
        \FLket{R_\mu}&=\sum_{n\alpha\beta} R_{\mu}^{n\alpha\beta}\,\tket{n}\otimes\ket{\alpha}\bra{\beta}
        \\
        \FLbra{L_\mu}&=\sum_{n\alpha\beta} L_{\mu}^{n\alpha\beta}\,\tket{n}\otimes\ket{\alpha}\bra{\beta}.
    \end{split}
\end{equation}

Moreover, $\lambda_{\mu}=-\gamma_{\mu}-\ii\Delta_{\mu}$, where $\gamma_{\mu}$ represents the linewidth  (inverse lifetime) and $\Delta_{\mu}$ denotes the frequency of each Floquet-Liouville mode in the spectral decomposition in the extended space. Note, as seen above that $\mu$ cannot be necessarily mapped into a single dressed state or Floquet mode, thus, $\gamma_{\mu}$ and $\Delta_\mu$ cannot be individually mapped by a dressed states or Floquet mode characters, but a combination of several of them, unless a Floquet mode is dominant enough and well separated.

\subsection{Trace mode and periodic steady state}

For a trace-preserving master equation, the identity operator defines a left zero mode of the Liouvillian,
\begin{equation}
\Lbraket{\mathbf{1}}{\mathscr{L}(t)}=0
\qquad
\Longrightarrow
\qquad
\FLbra{\mathbf{1}}\mathscr{L}^{\rm F}_{\rm ex}=0,
\label{eq:trace_left_mode}
\end{equation}
where the corresponding extended-space trace bra is
\begin{equation}
\FLbra{\mathbf{1}}
\equiv
(0|\otimes \Lbra{\mathbf{1}}.
\label{eq:trace_FL_bra}
\end{equation}

The asymptotic periodic steady state is associated with the right Floquet--Liouville mode whose exponent is zero,
\begin{equation}
\mathscr{L}^{\rm F}_{\rm ex}\FLket{R_{\rm ss}}=0.
\label{eq:ss_right_mode}
\end{equation}
The corresponding physical periodic steady state is recovered as
\begin{equation}
\Lket{\rho_{\rm ss}(t)}
=
\sum_n \ee^{-\ii n\omega_d t}\,\Lket{R_{{\rm ss},n}},
\label{eq:rho_ss_reconstruct}
\end{equation}
with normalization fixed by
\begin{equation}
\Lbraket{\mathbf{1}}{\rho_{\rm ss}(t)}=1
\qquad
\Leftrightarrow
\qquad
\FLbraket{\mathbf{1}}{R_{\rm ss}}=1.
\label{eq:ss_normalization}
\end{equation}

\paragraph*{Connection to the steady-state formulation in Fourier and extended space.}
Since the Liouvillian is periodic, the asymptotic periodic steady state $\rho_{\rm ss}(t)$ can be expanded in a Fourier series,
\begin{equation}
\rho_{\rm ss}(t)=\sum_{n\in\mathbb{Z}} \rho_n\, \ee^{-\ii n\omega_d t},
\label{eq:rho_ss_fourier}
\end{equation}
where the operators $\rho_n$ are time independent. Inserting Eq.~\eqref{eq:rho_ss_fourier} together with Eq.~\eqref{eq:L_fourier} into
\begin{equation}
\partial_t \rho_{\rm ss}(t)=\mathcal{L}(t)\rho_{\rm ss}(t),
\end{equation}
and matching equal Fourier harmonics yields
\begin{equation}
-\ii n\omega_d\, \rho_n
=
\sum_{m\in\mathbb{Z}} \mathcal{L}_m\, \rho_{n-m}.
\label{eq:ss_fourier_recursion}
\end{equation}
This is the direct analogue, for the driven Liouvillian, of the Sambe-space Floquet eigenvalue problem for the driven Hamiltonian.

Introducing the vector of Fourier components
\begin{equation}
\vec{\rho}_{\rm ss}=\big(\dots,\rho_{-1},\rho_0,\rho_{1},\dots\big)^T,
\end{equation}
Eq.~\eqref{eq:ss_fourier_recursion} can be written compactly in the Liouville--Sambe space as
\begin{equation}
\mathscr{L}^{\rm F}_{\rm ex}\,\vec{\rho}_{\rm ss}=0.
\label{eq:ss_extended}
\end{equation}

Equation~\eqref{eq:ss_extended} provides a time-independent linear problem for the asymptotic periodic steady state. In practice, after truncation in Fourier space, the physical steady-state solution is obtained as the normalized null vector of the extended Liouvillian,
$\mathscr{L}^{\rm F}_{\rm ex}$.
The normalization condition $\Tr{\rho_{\rm ss}(t)}=1$ fixes the overall scale of the null vector. Once $\rho_{\rm ss}(t)$ is known, observables, time-averaged populations, and two-time correlation functions may be computed straightforwardly.

\subsection{Propagator in Floquet--Liouville space}

Let $\mathscr{U}(t,t_0)$ be the Liouville-space propagator generated by $\mathscr{L}(t)$,
\begin{equation}
\partial_t \mathscr{U}(t,t_0)=\mathscr{L}(t)\mathscr{U}(t,t_0),
\qquad
\mathscr{U}(t_0,t_0)=\mathbf{1}_{\rm L}.
\label{eq:Liouville_propagator}
\end{equation}
Using the FL modes, one obtains the spectral decomposition
\begin{equation}
\mathscr{U}(t,t_0)
=
\sum_\mu
\ee^{\lambda_\mu (t-t_0)}
\Lket{R_\mu(t)}\Lbra{L_\mu(t_0)}.
\label{eq:Liouville_propagator_FL}
\end{equation}

Equivalently, in the extended space, we have
\begin{equation}
\Lket{R_\mu(t)}=(t\FLket{R_\mu},
\qquad
\Lbra{L_\mu(t)}=\FLbra{L_\mu}t).
\end{equation}
Thus, 
\begin{equation}
\mathscr{U}^{\rm F}_{\rm ex}(t,t_0)
=
\ee^{\mathscr{L}^{\rm F}_{\rm ex} (t-t_0)}.
\label{eq:Liouville_propagator_FL_ex}
\end{equation}

\subsection{Two-time correlator in Liouville space}

Let $A$ and $B$ be system operators. The two-time correlation function in the Schr\"odinger picture is
\begin{equation}
G_{AB}(t,\tau)
\equiv
\langle A(t)B(t+\tau)\rangle
=
\Tr{A\,\rho_B(t,\tau)},
\label{eq:GAB_def}
\end{equation}
where
\begin{equation}
\Lket{\rho_B(t,\tau)}
\equiv
\mathscr{U}(t+\tau,t)\, B\, \Lket{\rho(t)}
\label{eq:rhoB_tau_def}
\end{equation}
is the auxiliary state obtained via the quantum regression theorem by applying $B$ at time $t$ and propagating forward by delay $\tau$. In the Liouville-space notation, this becomes
\begin{equation}
G_{AB}(t,\tau)=\Lbra{A}\mathscr{U}(t+\tau,t)\,B\Lket{\rho(t)}.
\label{eq:GAB_Liouville}
\end{equation}
In the long-time regime, $\rho(t)\to \rho_{\rm ss}(t)$, and the correlator becomes
\begin{equation}
G_{AB}^{\rm ss}(t,\tau)
=
\Lbra{A}\mathscr{U}(t+\tau,t)\,B\Lket{\rho_{\rm ss}(t)}.
\label{eq:GAB_ss}
\end{equation}

Since the steady state is periodic, $G_{AB}^{\rm ss}(t,\tau)$ is periodic in the first time argument:
\begin{equation}
G_{AB}^{\rm ss}(t+T,\tau)=G_{AB}^{\rm ss}(t,\tau).
\end{equation}
The experimentally relevant object is therefore the average over one drive period,
\begin{equation}
\overline{G}_{AB}(\tau)
\equiv
\frac{1}{T}\int_T dt\, G_{AB}^{\rm ss}(t,\tau).
\label{eq:Gbar_def}
\end{equation}

\subsection{Correlator in the Floquet--Liouville extended space}

We now express Eq.~\eqref{eq:Gbar_def} directly in the FL eigenbasis. Using Eq.~\eqref{eq:Liouville_propagator_FL} together with the periodic steady state, we obtain
\begin{equation}
G_{AB}^{\rm ss}(t,\tau)
=
\sum_\mu
\ee^{\lambda_\mu \tau}
\Lbraket{A}{R_\mu(t)}\Lbraket{L_\mu(t)}{B|\rho_{\rm ss}(t)}.
\label{eq:GAB_ss_modes}
\end{equation}

Averaging over one period yields
\begin{equation}
\overline{G}_{AB}(\tau)
=
\sum_\mu
\ee^{\lambda_\mu \tau}\,
\mathcal{W}_{AB,\mu},
\label{eq:Gbar_mode_sum}
\end{equation}
with the mode weights
\begin{equation}
\mathcal{W}_{AB,\mu}
\equiv
\frac{1}{T}\int_T dt\,
\Lbraket{A}{R_\mu(t)}\Lbraket{L_\mu(t)}{B|\rho_{\rm ss}(t)}.
\label{eq:Wmu_time}
\end{equation}

To write this compactly in the extended space, define the lifted FL operators
\begin{equation}
\mathcal{A}_{\rm FL}\equiv F_0\otimes A,
\qquad
\mathcal{B}_{\rm FL}\equiv F_0\otimes B,
\label{eq:A_B_lifted}
\end{equation}
or, lifted FL bra states
\begin{equation}
\FLbra{\mathcal{A}_{\rm FL}}
\equiv
(0|\otimes \Lbra{A}
\quad
\FLbra{\mathcal{B}_{\rm FL}}
\equiv
(0|\otimes \Lbra{B},
\label{eq:A_B_FL_bra}
\end{equation}
and the extended steady-state vector
\begin{equation}
\FLket{R_{\rm ss}}
=
\sum_n |n)\otimes \Lket{R_{{\rm ss},n}}.
\end{equation}
Then the mode weight \eqref{eq:Wmu_time} is simply
\begin{equation}
\mathcal{W}_{AB,\mu}
=
\FLbraket{\mathcal{A}_{\rm FL}}{R_\mu}\,
\FLbraket{L_\mu}{\mathcal{B}_{\rm FL}\vert R_{\rm ss}},
\label{eq:Wmu_extended}
\end{equation}
where
$\FLbra{\mathcal{A}_{\rm FL}}$
is given in Eq.~\eqref{eq:A_B_FL_bra} and $\mathcal{B}_{\rm FL}$ is given in Eq.~\eqref{eq:A_B_lifted}.
Therefore, the period-averaged steady-state correlator takes the compact FL form
\begin{equation}
\overline{G}_{AB}(\tau)
=
\sum_\mu
\ee^{\lambda_\mu \tau}\,
\FLbraket{\mathcal{A}_{\rm FL}}{R_\mu}\,
\FLbraket{L_\mu}{\mathcal{B}_{\rm FL}\vert R_{\rm ss}}.
\label{eq:Gbar_final_FL}
\end{equation}

For the commonly used first-order correlation function, one chooses
\begin{equation}
A=s^-,
\qquad
B=s^+,
\end{equation}
or, for the incoherent part,
\begin{equation}
A=\delta s^- \equiv s^- - \langle s^-(t)\rangle_{\rm ss},
\qquad
B=\delta s^+ \equiv s^+ - \langle s^+(t)\rangle_{\rm ss}.
\end{equation}
Then Eq.~\eqref{eq:Gbar_final_FL} directly gives the time-averaged steady-state first-order correlator.

\subsection{Spectrum in the Floquet--Liouville representation}

The emission spectrum associated with the period-averaged correlator is defined by the one-sided Fourier transform
\begin{equation}
\mathsf{S}_{AB}(\omega)
\propto
\mathrm{Re}\int_0^\infty d\tau\, \ee^{\ii\omega \tau}\,\overline{G}_{AB}(\tau).
\label{eq:spectrum_def}
\end{equation}
Substituting Eq.~\eqref{eq:Gbar_final_FL}, and assuming $\mathrm{Re}[\lambda_\mu]<0$ for all non-steady modes, we obtain
\begin{equation}
\mathsf{S}_{AB}(\omega)
\propto
\mathrm{Re}\sum_\mu
\frac{
\FLbraket{\mathcal{A}_{\rm FL}}{R_\mu}\,
\FLbraket{L_\mu}{\mathcal{B}_{\rm FL}\vert R_{\rm ss}}
}{
-\lambda_\mu-\ii\omega
}.
\label{eq:spectrum_FL_final}
\end{equation}

Equivalently, writing
\begin{equation}
\lambda_\mu=-\gamma_\mu-\ii \Delta_\mu,
\qquad
\gamma_\mu\ge 0,
\label{eq:Omega_gamma_delta}
\end{equation}
the spectrum is expressed as a sum over FL resonances,
\begin{equation}
\mathsf{S}_{AB}(\omega)
\propto
\mathrm{Re}\sum_\mu
\frac{
\mathcal{W}_{AB,\mu}
}{
\gamma_\mu-\ii(\omega-\Delta_\mu)
}
.
\label{eq:spectrum_resonance_sum}
\end{equation}
Thus, each FL eigenmode contributes a resonance centered at frequency $\Delta_\mu$
with linewidth $\gamma_\mu$ and complex residue $\mathcal{W}_{AB,\mu}$.

\subsection{Population modes, coherent contributions, and incoherent emission spectra}

An important practical aspect of the FL spectrum is the distinction between modes that are predominantly population-like and those that are predominantly coherence-like. This distinction is particularly useful when analyzing time-averaged excitation numbers and emission spectra. The coherent (elastic) contribution originates from the periodic steady state itself, whereas the incoherent (inelastic) spectrum is generated by the decaying FL modes. In many situations, population-dominated modes primarily govern relaxation toward the periodic steady state, while coherence-dominated modes are responsible for oscillatory spectral features. However, in the presence of nonsecular couplings, FL eigenmodes generally become hybridized mixtures of populations and coherences, and the separation is no longer exact.

\subsubsection{Definition of population modes}

Recall that each FL eigenmode satisfies
\begin{equation}
\mathscr{L}^{\rm F}_{\rm ex}\FLket{R_\mu}
=
\lambda_\mu\FLket{R_\mu},
\qquad
\lambda_\mu=-\gamma_\mu-\ii \Delta_\mu,
\label{eq:FL_mode_population_start}
\end{equation}
where, for convenience, we merged the double index into one, $n\mu\to\mu$ as we work in the full extended space without identifying the BZ strips so that the index runs through the whole sidebands, $\mu=0,1,\dots,\mathrm{dim}(\mathbb{H}^{\rm FL}_{\rm ex})-1$.
Modes whose imaginary part lies at an integer harmonic of the drive frequency,
\begin{equation}
\Delta_\mu = m\omega_d,
\qquad
m\in\mathbb{Z},
\label{eq:population_condition}
\end{equation}
and whose real part is zero or asymptotically smallest,
\begin{equation}
\gamma_\mu \approx 0,
\label{eq:population_decay}
\end{equation}
correspond to the periodic steady-state manifold and its harmonic population oscillations. 

These are referred to here as \emph{population modes}. They originate from diagonal components of the density operator in the Floquet basis, i.e. from occupations rather than coherences.
Note this argument also confirms the derivation of the phenomenological approach~\cite{Akbari_Floquet_2025} for the population in Eqs.~\eqref{eq:Nss_t_phen} and \eqref{eq:Nbar_phen}.

The most important member of this family is the true steady-state mode,
\begin{equation}
\lambda_{\rm ss}=0,
\end{equation}
which reconstructs the periodic asymptotic density matrix,
\begin{equation}
\Lket{\rho_{\rm ss}(t)}
=
\sum_n \ee^{-\ii n\omega_d t}\Lket{R_{{\rm ss},n}}.
\end{equation}

In contrast, modes with
\begin{equation}
\Delta_\mu \neq m\omega_d
\end{equation}
are associated with coherences between distinct Floquet sectors and produce genuine inelastic resonances.

\subsubsection{Numerical detection of population modes}

After diagonalizing $\mathscr{L}^{\rm F}_{\rm ex}$, the population modes are identified numerically through the following criteria:

\begin{enumerate}
\item {Near-zero decay rate:}
\begin{equation}
\lvert\mathrm{Re}[\lambda_\mu]\rvert < \epsilon_\gamma .
\end{equation}

\item {Harmonic frequency locking:}
\begin{equation}
\min_{m\in\mathbb{Z}}
\lvert\mathrm{Im}[\lambda_\mu - m\omega_d]\rvert < \epsilon_\omega .
\end{equation}

\item {Large diagonal weight (optional diagnostic):}  One can reconstruct the physical Liouville component of $\FLket{R_\mu}$ and evaluate its projection onto operators diagonal in the Floquet basis.
\end{enumerate}
Here, $\epsilon_\gamma$ and $\epsilon_\omega$ are numerical tolerances set by truncation accuracy and machine precision.

In practical computations, one first isolates the exact zero mode $\lambda_{\rm ss}=0$, then identifies additional harmonically shifted replicas or slowly decaying population harmonics.

\subsubsection{Contributions to time-averaged excitation number}

Let $N$ be an excitation-number operator. The periodic expectation value is
\begin{equation}
\langle N(t)\rangle=\Lbraket{N}{\rho_{\rm ss}(t)},
\end{equation}
and its period average is
\begin{equation}
\overline{N}
=
\frac{1}{T}\int_T dt\, \Lbraket{N}{\rho_{\rm ss}(t)}.
\end{equation}

Using the Fourier expansion of the steady state, only the zero-harmonic population component survives:
\begin{equation}
\overline{N}
=
\Lbraket{N}{R_{{\rm ss},0}}.
\label{eq:AveN_population}
\end{equation}
Hence, the averaged excitation number is determined entirely by the steady-state population sector. \emph{Coherence modes do not contribute after period averaging unless they are phase-locked to zero net harmonic}.
Therefore, $\overline{N}$ is fundamentally a population observable.

\subsubsection{Contributions to the correlator}

Consider the averaged first-order correlator in Eq.~\eqref{eq:Gbar_mode_sum},
which remain long-lived and correspond to coherent oscillations synchronized with the drive. In the strict steady-state limit ($\gamma_\mu=0$), these yield nondecaying contributions.

Thus, population modes generate the coherent part of the emitted field, including:
\begin{itemize}
\item Elastic Rayleigh peak at $\omega=0$ in the rotating frame,
\item Harmonics at $\omega=\pm m\omega_d$,
\item Deterministic modulation inherited from the periodic steady state.
\end{itemize}
In contrast, decaying coherence modes ($\gamma_\mu>0$) generate Lorentzian-like inelastic sidebands.

\subsubsection{Removing population modes: incoherent spectrum}

The incoherent emission spectrum is defined from fluctuation operators:
\begin{equation}
\delta B(t)=B-\langle B(t)\rangle_{\rm ss},
\qquad
\delta A(t)=A-\langle A(t)\rangle_{\rm ss},
\end{equation}
and the correlator
\begin{equation}
\overline{G}^{\,{\rm inc}}_{AB}(\tau)
=
\frac{1}{T}\int_T dt\,
\langle \delta A(t+\tau)\,\delta B(t)\rangle .
\label{eq:Ginc_def}
\end{equation}
Equivalently,
\begin{equation}
\overline{G}^{\,{\rm inc}}_{AB}(\tau)
=
\overline{G}_{AB}(\tau)
-
\frac{1}{T}\int_T dt\,
\langle A(t+\tau)\rangle_{\rm ss}\langle B(t)\rangle_{\rm ss}.
\label{eq:Ginc_subtract}
\end{equation}

In the FL modal expansion, this subtraction exactly removes the population modes. Therefore,
\begin{equation}
\overline{G}^{\,{\rm inc}}_{AB}(\tau)
=
\sum_{\mu\notin\mathcal{P}}\ee^{\lambda_\mu \tau}\,
\mathcal{W}_{AB,\mu}
\label{eq:Ginc_population_removed}
\end{equation}
where $\mathcal{P}$ denotes the set of detected population modes.
Hence, the incoherent spectrum becomes
\begin{equation}
\mathsf{S}_{AB}(\omega)
\propto
\mathrm{Re}
\sum_{\mu\notin\mathcal{P}}
\frac{\mathcal{W}_{AB,\mu}}
{-\lambda_\mu-\ii\omega},
\label{eq:Sinc_population_removed}
\end{equation}
which contains only decaying coherent resonances.

\subsubsection{Numerical implementation steps}

In numerical calculations, a robust procedure is
as follows:
\begin{enumerate}
\item Diagonalize $\mathscr{L}^{\rm F}_{\rm ex}$.
\item Detect population modes using Eqs.~\eqref{eq:population_condition}--\eqref{eq:population_decay}.
\item Compute residues $\mathcal{W}_{AB,\mu}$.
\item Exclude $\mu\in\mathcal{P}$ when constructing the spectrum.
\end{enumerate}
This avoids spurious narrow peaks or delta-like harmonics in the incoherent spectrum and isolates the physically relevant fluorescence signal.

\subsubsection{Physical interpretations}

Population modes encode how the bath populates Floquet states and therefore determine:

\begin{itemize}
\item Steady-state occupations,
\item $\overline{N}$ and related averaged observables,
\item Coherent elastic scattering.
\end{itemize}

Coherence modes encode transitions between populated Floquet sectors and therefore determine:

\begin{itemize}
\item Linewidths,
\item Fluorescence sidebands,
\item Mollow-type triplets,
\item Interference dips and resonance structures.
\end{itemize}

Separating these two classes of Floquet--Liouville modes is therefore essential for interpreting driven-dissipative spectra.

\subsection{Period-averaged excitation number in the Floquet--Liouville formalism}
In addition to two-time correlators and spectra, the Floquet--Liouville construction provides a natural and efficient route to the long-time average of one-time observables. Let \(N\) denote the excitation-number operator of interest, for example the atomic excitation operator \(N=\sigma^+\sigma^-\) for the driven TLS in the bare states representation, or generally, \(N=s^{\Lambda-}s^{\Lambda+}\) in the dressed states representation.

In the Schr\"odinger picture, its instantaneous expectation value is
\begin{equation}
\langle N(t)\rangle = \Tr{N\,\rho(t)}.
\end{equation}
For a periodically driven open system, the asymptotic state is not time independent in the Schr\"odinger picture, but rather approaches a \(T\)-periodic steady state,
\begin{equation}
\rho_{\rm ss}(t+T)=\rho_{\rm ss}(t).
\end{equation}

Accordingly, the experimentally relevant stationary quantity is the average over one driving period,
\begin{equation}
\overline{N}
\equiv
\frac{1}{T}\int_T dt\, \Tr{N\,\rho_{\rm ss}(t)}.
\label{eq:AveN_def}
\end{equation}

\paragraph*{Liouville-space expression.}
In Liouville notation, Eq.~\eqref{eq:AveN_def} becomes
\begin{equation}
\overline{N}
=
\frac{1}{T}\int_T dt\, \Lbraket{N}{\rho_{\rm ss}(t)},
\label{eq:AveN_Liouville}
\end{equation}
where \(\Lbra{N}\) is the Hilbert--Schmidt bra associated with the operator \(N\). Expanding the periodic steady state into its Fourier components,
\begin{equation}
\Lket{\rho_{\rm ss}(t)}
=
\sum_{n\in\mathbb Z} e^{-\ii n\omega_d t}\,\Lket{\rho_{{\rm ss},n}},
\label{eq:rho_ss_fourier}
\end{equation}
one obtains
\begin{equation}
\overline{N}
=
\sum_{n}
\left[
\frac{1}{T}\int_T dt\, e^{-\ii n\omega_d t}
\right]\Lbraket{N}{\rho_{{\rm ss},n}}
=
\Lbraket{N}{\rho_{{\rm ss},0}}.
\label{eq:AveN_zero_component}
\end{equation}
Thus, only the zeroth Fourier component of the periodic steady state contributes to the period-averaged excitation number.

\paragraph*{Floquet--Liouville extended-space form.}
This result can be written particularly compactly in the Floquet--Liouville extended space. Let
\begin{equation}
\FLket{R_{\rm ss}}
=
\sum_{n\in\mathbb Z}\tket{n}\otimes \Lket{\rho_{{\rm ss},n}},
\end{equation}
denote the normalized steady-state right eigenvector of the Floquet--Liouville supermatrix,
\begin{equation}
\mathscr{L}^{\rm F}_{\rm ex}\FLket{R_{\rm ss}}=0.
\end{equation}

Then, the period-average operation simply projects onto the temporal zeroth harmonic. Defining
\begin{equation}
\FLbra{N_{\rm FL}}
\equiv
\tbra{0}\otimes \Lbra{N},
\label{eq:N_FL_bra}
\end{equation}
one finds
\begin{equation}
\overline{N}
=
\FLbraket{N_{\rm FL}}{R_{\rm ss}}
=
\tbra{0}\otimes \Lbra{N}\,R_{\rm ss})\!\rrangle.
\label{eq:AveN_FL}
\end{equation}
Equivalently, since the zeroth temporal basis vector \(\tket{0}\) selects the \(n=0\) Fourier block of the steady state, Eq.~\eqref{eq:AveN_FL} is exactly the extended-space representation of Eq.~\eqref{eq:AveN_zero_component}.

\paragraph*{Trace normalization and practical implementation.}
The steady-state mode must be normalized by the trace condition. In the Floquet--Liouville extended space, the corresponding left trace bra is
\begin{equation}
\FLbra{\mathbf{1}_{\rm FL}}
=
\tbra{0}\otimes \Lbra{\mathbf{1}},
\end{equation}
so that the physical periodic steady state is fixed by
\begin{equation}
\FLbraket{\mathbf{1}_{\rm FL}}{R_{\rm ss}} = 1.
\label{eq:AveN_trace_norm}
\end{equation}

Once the steady-state right eigenvector has been identified and normalized according to Eq.~\eqref{eq:AveN_trace_norm}, the evaluation of \(\overline{N}\) requires only a single overlap, Eq.~\eqref{eq:AveN_FL}. In numerical implementations, this is especially convenient because no explicit time propagation over one period is needed after the steady-state eigenmode has been obtained.

\paragraph*{Mode interpretation of \(\overline{N}\).}
Equation~\eqref{eq:AveN_FL} also clarifies the physical meaning of \(\overline{N}\) in the Floquet--Liouville language. Unlike the incoherent spectrum, which depends on the full family of decaying Floquet--Liouville modes and their residues, the period-averaged excitation number depends only on the steady-state mode \(R_{\rm ss}\). Therefore, \(\overline{N}\) probes the population content of the asymptotic periodic state rather than the transient coherence modes responsible for fluorescence sidebands. This is precisely why the population-like Floquet--Liouville sector is central for averaged observables, whereas incoherent spectra require the exclusion of the steady-family modes and the inclusion of the decaying coherence modes.

\paragraph*{Connection to the decomposition in Floquet components.}
If the steady state is decomposed into the Floquet basis as
\begin{equation}
\rho_{\rm ss}(t)
=
\sum_{\alpha\beta}\rho_{\alpha\beta}^{\rm ss}(t)\,
|\alpha(t)\rangle\langle\beta(t)|,
\end{equation}
then the average excitation number is controlled by the zeroth harmonic of the diagonal and off-diagonal steady-state components through the operator matrix elements of \(N\). In the extended-space representation, all such contributions are already encoded in the overlap \(\FLbraket{N_{\rm FL}}{R_{\rm ss}}\). Hence the Floquet--Liouville formalism provides a unified framework in which both the stationary one-time observable \(\overline{N}\) and the two-time quantities such as \(\overline{G}^{(1)}(\tau)\) and \(\mathsf{S}(\omega)\) are obtained directly from the same extended-space eigenstructure.

\subsection{Practical remarks}

Several points are worth emphasizing.
First, the Floquet--Liouville eigenvalues $\lambda_\mu$ encode both decay and frequency oscillations. Their real parts determine relaxation or decoherence rates, while their imaginary parts determine the spectral positions of the resonances.

Second, because the spectrum of $\mathscr{L}^{\rm F-L}_{\rm ex}$ is replicated in vertical strips separated by $\ii\omega_d$, one must select one FL Brillouin strip when listing the physically distinct modes. This is fully analogous to choosing a single quasienergy Brillouin zone in the closed-system Floquet problem.

Third, Eq.~\eqref{eq:Gbar_final_FL} provides the exact relation between the period-averaged steady-state correlator and the FL modes for the FGME derived in the previous section. It is therefore the natural starting point for computing incoherent spectra in periodically driven open quantum systems.

Moreover, the formalism is completely general: once the periodic Liouvillian $\mathscr{L}(t)$ is known, one can construct $\mathscr{L}^{\rm F}_{\rm ex}$, solve the biorthogonal eigenvalue problem \eqref{eq:FL_right_extended}--\eqref{eq:FL_left_extended}, identify the normalized steady state $\FLket{R_{\rm ss}}$, and then evaluate correlators and spectra from Eqs.~\eqref{eq:Gbar_final_FL} and \eqref{eq:spectrum_FL_final}.

Finally, Eq.~\eqref{eq:Sinc_population_removed} may be written explicitly in terms of the real and imaginary parts of the residue. Writing
\begin{equation}
\mathcal{W}_{AB,\mu}=\mathcal{W}^{(\mathrm r)}_\mu+\ii \mathcal{W}^{(\mathrm i)}_\mu,
\end{equation}
we obtain
\begin{equation}
\mathsf{S}_{AB}(\omega)
\propto
\sum_{\mu\notin\mathcal{P}}
\frac{
\mathcal{W}^{(\mathrm r)}_\mu \gamma_\mu
+\mathcal{W}^{(\mathrm i)}_\mu(\omega_\mu-\omega)
}{
(\omega-\omega_\mu)^2+\gamma_\mu^2
}.
\label{eq:Sinc_population_removed_explicit}
\end{equation}

The real part of the residue produces the symmetric Lorentzian contribution, while the imaginary part produces a dispersive correction that distorts the line shape. In the limit of a well-isolated resonance, where neighboring modes are negligible and the dispersive correction is weak, the height of the $\mu$th peak at $\omega=\omega_\mu$ is approximately
\begin{equation}
\mathsf{S}_{AB}(\omega_\mu)\propto \mathcal{R}_\mu\equiv\frac{\mathcal{W}^{(\mathrm r)}_\mu}{\gamma_\mu}.
\end{equation}

Thus, for isolated peaks, $\mathcal{W}^{(\mathrm r)}_\mu/\gamma_\mu$ provides a useful estimate of the peak height, whereas $\mathcal{W}^{(\mathrm r)}_\mu$ controls the integrated symmetric spectral weight of the mode. When several modes overlap strongly, however, the observed spectral feature is determined by the full sum in Eq.~\eqref{eq:Sinc_population_removed_explicit}, and no unique peak strength can in general be assigned to a single mode.

Also, if the intermixing of the FL modes are not strong the the extended FL mode frequency is sufficiently close to the Floquet mode frequencies (that is the zeroth order perturbation without correction), $\Delta_\mu\approx\Delta_\mu^{(0)}=\Delta_{\alpha\beta l}$. 

In addition to the spectral positions and linewidths encoded in the Floquet--Liouville eigenvalues, it is often useful to identify the microscopic Floquet transitions that predominantly compose a given Floquet--Liouville mode. To this end, each right eigenmode may be expanded in the operator basis generated by the Floquet states and temporal harmonics as
\begin{equation}
\FLket{R_\mu}
=
\sum_{\alpha\beta l}
R_{\mu}^{\alpha\beta l}\,
\tket{l}\otimes \ket{\alpha}\bra{\beta},
\label{eq:Rmu_basis_expand}
\end{equation}
where $\{|\alpha\rangle\}$ denotes the Floquet basis chosen within one quasienergy Brillouin zone, and $l\in\mathbb{Z}$ labels the harmonic (sideband) sector. The complex amplitudes
\begin{equation}
R_{\mu}^{\alpha\beta l}
=
\FLbraket{l,\alpha\beta}{R_\mu}
\label{eq:rmu_projection}
\end{equation}
are obtained by projection onto the orthonormal extended basis states
$\FLket{l,\alpha\beta}
\equiv
\tket{l}\otimes \ket{\alpha}\bra{\beta}$.

A natural measure of the contribution of a given Floquet transition $(\alpha,\beta,l)$ to the mode $\mu$ is the normalized modal weight
\begin{equation}
r_{\mu}^{\alpha\beta l}
=
\frac{|R_{\mu}^{\alpha\beta l}|^2}
{\sum_{\alpha'\beta' l'} |R_{\mu}^{\alpha'\beta' l'}|^2},
\qquad
\sum_{\alpha\beta l} r_{\mu}^{\alpha\beta l}=1.
\label{eq:weight_modal}
\end{equation}

Sorting the set $\{r_{\mu}^{\alpha\beta l}\}$ in descending order provides a direct decomposition of the Floquet--Liouville mode into its dominant Floquet components. In practice, to avoid overinterpreting numerically insignificant admixtures, one may retain only those components satisfying
\begin{equation}
r_{\mu}^{\alpha\beta l}\ge \eta_{\rm cut},
\label{eq:weight_cut}
\end{equation}
with a typical choice $\eta_{\rm cut}=0.1$, corresponding to contributions carrying at least $10\%$ of the modal norm.

This decomposition is particularly useful for assigning physical meaning to spectral resonances. If a mode is dominated by a single component $(\alpha,\beta,l)$, then the associated resonance may be approximately identified with the Floquet transition frequency
\begin{equation}
\Delta_\mu^{(0)}=\Delta_{\alpha\beta l}
=
\varepsilon_\beta-\varepsilon_\alpha+l\omega_d.
\end{equation}

Furthermore, the structure of the dominant coefficients distinguishes population-like and coherence-like sectors. Components with $\alpha=\beta$ correspond to diagonal Floquet occupations and are therefore associated with population harmonics and coherent elastic emission. By contrast, components with $\alpha\neq\beta$ correspond to off-diagonal Floquet coherences and primarily generate inelastic fluorescence features. Consequently, inspecting the dominant set $\{R_{\mu}^{\alpha\beta l}\}$ provides a practical diagnostic for determining whether a given FL mode mainly contributes to averaged observables such as $\overline{N}$, to coherent scattering, or to incoherent spectral sidebands.

\section{Additional Results}
\label{secS:AdditionalResults}
In this appendix, we provide some additional results and discussion for more ranges of the drive parameters and comparison with the limits of some common applied approximations.
Our calculation is performed with a minimum of $l_{\rm max}=30$. Therefore, for the single-TLS case, $\mathrm{dim}(\mathbb{H}_{\rm ex}^{\rm FL})=244$ is the number of FL modes, so that $\mu=0,1,\dots,243$. Likewise, for the coupled-TLS model, $\mathrm{dim}(\mathbb{H}_{\rm ex}^{\rm FL})=976$ is the number of FL modes, so that $\mu=0,1,\dots,975$.

\subsection{Driven-dissipative TLS: incoherent spectra when the drive is detuned}

\begin{figure*}[!htpb]
\centering
\includegraphics[width=.9\linewidth]
{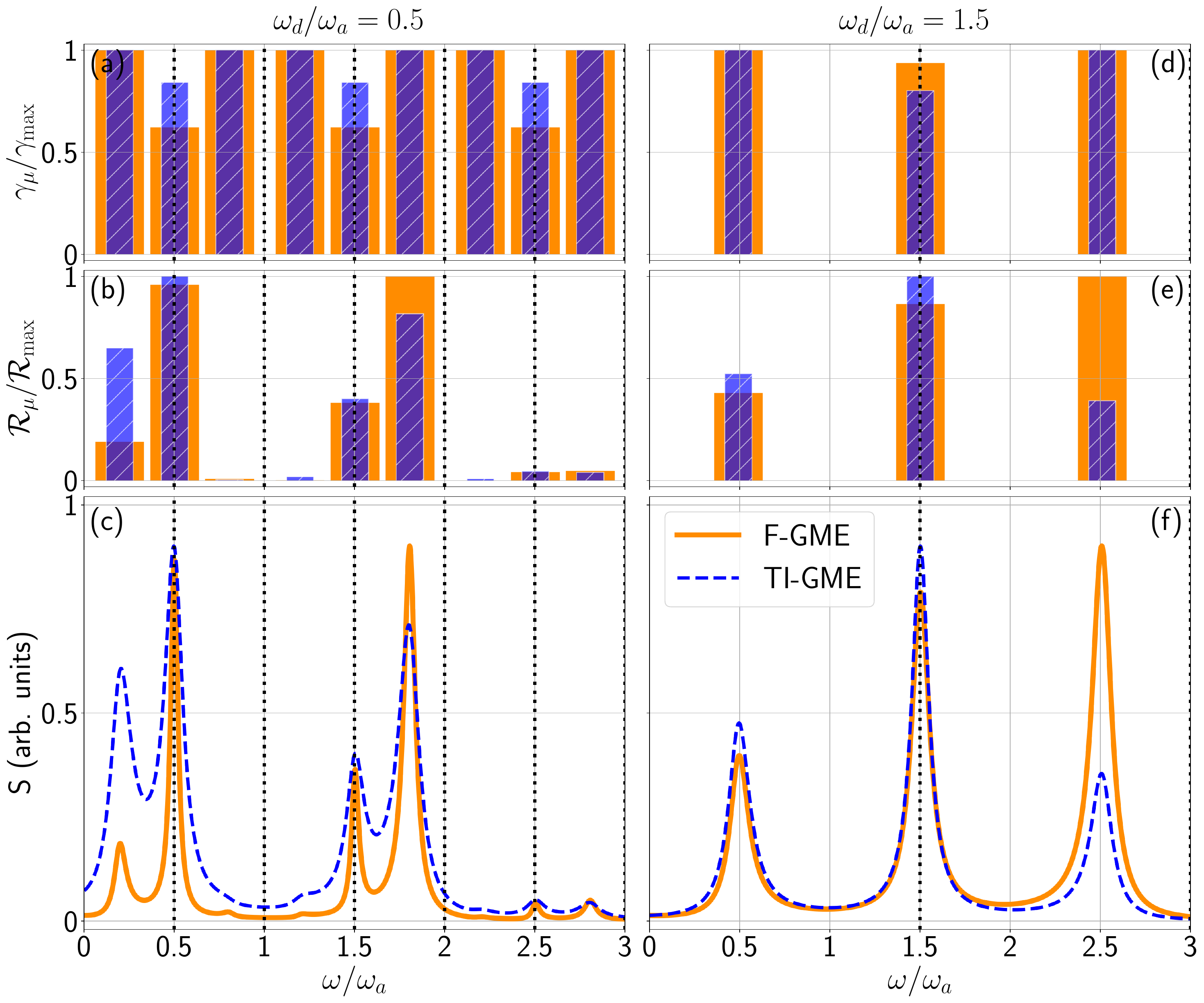} 
\caption{\textbf{Driven-dissipative TLS: incoherent spectrum for $\eta_d=1$ with drive-atom detuning.}
We show the dominant decay rates $\gamma_\mu$ [panels (a),(d)], the relative peak-height factors $\mathcal{R}_\mu\propto \mathrm{Re}(W_\mu)/\gamma_\mu$ [panels (b),(e)], and the resulting incoherent emission spectra [panels (c),(f)] for a driven two-level system at $\omega_d/\omega_a=0.5$ (left column) and $\omega_d/\omega_a=1.5$ (right column). In all panels, $\gamma=0.1\,\omega_a$ and $\eta_d=1$. The vertical dotted black lines mark the multiple frequencies of the drive. The dominant Floquet--Liouville resonances, $\omega=\Delta_\mu/\omega_a$. The bar plots compare the exact Floquet--Liouville mode frequencies $\Delta_\mu$ and linewidths $\gamma_\mu$ with the dominant underlying coherent Floquet transition frequencies $\Delta_{\alpha\beta l}=\varepsilon_\beta-\varepsilon_\alpha+l\omega_d$ extracted from the closed-system Floquet Hamiltonian. Pure peaks correspond to modes dominated by a single Floquet transition, whereas hybrid peaks arise when one Floquet--Liouville mode carries comparable weights from multiple Floquet transitions. The spectra in the bottom panels are the incoherent spectra obtained from the Floquet--Liouville residues and linewidths.
}
\label{figS:Spectra_etad1_Flat}
\end{figure*}

In the main text, we mainly investigated the flat bath spectra for the resonantly driven systems. Here, we look at the spectra where the drive frequency is detuned, $\omega_d\neq\omega_a$. 

In Fig.~\ref{figS:Spectra_etad1_Flat}, we analyze the (incoherent) spectra of the driven-dissipative TLS for a fixed value of the drive amplitude $\eta_d=1$ and dissipation rate of $\kappa=0.1\omega_a$ for a flat bath. Panels (c) and (f) show the spectra for $\omega_d=0.5\omega_a$ and $\omega_d=1.5\omega_a$, respectively. The corresponding dominant FL modes of the spectra then are shown in panels (b) and (e), as well as the corresponding modal effective dissipative rates in panels (a) and (b).
It is expected that the spectra are normally composed of the Mollow-triplet-like structrure centrally peaked at odd drive frequencies.
Thus, one already expects the discrepancy between the TI-GME and the F-GME as soon as the Floquet sidebands appear.

In Fig.~\ref{figS:Spectra_etad1_Flat}(c), the spectra show some Mollow-triplet structures formed about the odd multiples of the drive frequency, strongest at $\omega=(0.5,1.5,2.5)\omega_a$, but with nonuniform sidebands. Note these Mollow-triplet structures also interfere when are close to each other, thus can deform each other. We clearly observe that the TI-GME predict the corret peaks' positions, but with incorrect widths and strengths, as expected because this theory has a time-independent dissipator so that all the dissipative effects are collected in a single dominant channel of $n=0$, the primary Floquet sideband. However, in a general multiphoton process the dissipation can also exist in the nonprimary sidebands and hence contribute in the dissipative process.

Figure~\ref{figS:Spectra_etad1_Flat} resolves the incoherent emission spectrum of the driven TLS into its underlying FL modes. For each exact FL eigenmode $\mu$, the spectrum is characterized by three central quantities: the linewidth $\gamma_\mu=-\mathrm{Re}\,\lambda_\mu$, the resonance frequency $\Delta_\mu=-\mathrm{Im}\,\lambda_\mu$, and the residue factor $W_\mu$, whose real part controls the isolated-mode peak height through $\mathcal{R}_\mu\equiv\mathrm{Re}(W_\mu)/\gamma_\mu$. At the same time, each exact dissipative mode can be decomposed into dominant coherent Floquet transition components $(\alpha,\beta,l)$, with frequencies
$\Delta_{\alpha\beta l}$
obtained from the closed-system Floquet Hamiltonian. This decomposition is crucial because the observed spectral peaks are governed by the exact dissipative frequencies $\Delta_\mu$, whereas the dominant microscopic content of each mode is encoded in the set of underlying $\Delta_{\alpha\beta l}$.

\paragraph*{Weak-mixing versus hybrid regimes.}
A central message of Fig.~\ref{figS:Spectra_etad1_Flat} is that the relation between $\Delta_\mu$ and $\Delta_{\alpha\beta l}$ is not always one-to-one. In the weak-mixing regime, a FL mode is almost entirely supported on a single Floquet transition, i.e.,
\begin{equation*}
r_\mu^{(\alpha\beta l)}\simeq 1,
\end{equation*}
and then
\begin{equation*}
\Delta_\mu \approx \Delta_{\alpha\beta l}.
\end{equation*}
These are the \emph{pure} peaks. In contrast, in the hybrid regime a single FL mode contains comparable contributions from two or more coherent Floquet transitions. Then the actual dissipative resonance frequency $\Delta_\mu$ need not coincide with any individual $\Delta_{\alpha\beta l}$, and instead emerges as the collective oscillation frequency of the hybrid dissipative mode. These are the \emph{hybrid} peaks.

\paragraph*{Detuning case with  $\omega_d/\omega_a=0.5$.}
For the lower drive frequency, $\omega_d/\omega_a=0.5$, the spectrum contains both pure and hybrid resonances. The strongest mode is $\mu=59$, with
\begin{equation*}
\lambda_{59}/\omega_a\approx -0.0387 - \ii\,1.805,
\end{equation*}
so that
\begin{equation*}
\gamma_{59}\approx 0.0387\,\omega_a,
\qquad
\Delta_{59}\approx 1.805\,\omega_a,
\end{equation*}
and this mode is essentially pure, being dominated by the single component $(a,b,l)=(g,e,4)$ with weight
\begin{equation*}
r_{59}^{(ge4)}\approx 0.9996,
\end{equation*}
and coherent transition frequency
\begin{equation*}
\Delta_{ge4}\approx 1.805\,\omega_a.
\end{equation*}

Likewise, the mode $\mu=51$ is almost purely $(e,g,0)$ with
\begin{equation*}
\Delta_{51}\approx 0.195\,\omega_a
\approx \Delta_{eg0},
\end{equation*}
and the negative-frequency partners $\mu=41$ and $\mu=49$ are similarly pure, corresponding respectively to $(g,e,0)$ and $(e,g,-4)$. However, the negative modes are not physical in the emission spectrum. These resonances therefore behave as nearly unmixed Floquet sideband transitions.

In contrast, the modes $\mu=141$ and $\mu=150$ are genuinely hybrid. For $\mu=141$ one finds
\begin{equation*}
\Delta_{141}\approx0.5\,\omega_a=\omega_d,
\qquad
\gamma_{141}\approx 0.024\,\omega_a,
\end{equation*}
while its dominant Floquet contents are
\begin{equation*}
(g,e,2),\ (e,g,0),\ (e,g,2),\ (g,e,0),
\end{equation*}
with weights $r_\mu^{\alpha\beta l}$, approximately,
\begin{equation*}
0.3112,\ 0.3112,\ 0.1567,\ 0.1567,
\end{equation*}
respectively. 

The corresponding coherent Floquet transition frequencies are
\begin{equation*}
\Delta_{ge2}\approx 0.805\,\omega_a,\quad
\Delta_{eg0}\approx 0.195\,\omega_a,\quad
\Delta_{eg2}\approx 1.195\,\omega_a,
\end{equation*}
hybridized also with a negative channel,
\begin{equation*}
\Delta_{ge0}\approx -0.195\,\omega_a.
\end{equation*}
Thus, the exact dissipative resonance sits at
$\Delta_{141}=0.5\,\omega_a$,
even though none of the dominant coherent transitions is individually located at that frequency. The same phenomenon occurs for $\mu=150$, whose exact resonance appears at
\begin{equation*}
\Delta_{150}\approx1.5\,\omega_a=3\omega_d,
\end{equation*}
while its dominant coherent components are centered around
\begin{equation*}
1.195\,\omega_a,\quad 1.805\,\omega_a,\quad 2.195\,\omega_a,\quad 0.805\,\omega_a.
\end{equation*}

Thus, the peaks at $\omega/\omega_a=0.5$ and $1.5$ are not direct single-transition lines; rather, they are collective FL resonances generated by dissipative mixing among several nearby Floquet sideband transitions. This is why panels (a)--(c) contain strong peaks exactly at the drive frequency and its first harmonic, even though the dominant closed-system Floquet transitions lie around them rather than on top of them.

\paragraph*{Physical meaning of the hybrid peaks for $\omega_d/\omega_a=0.5$.}
The hybrid peaks at $\omega/\omega_a=0.5$ and $1.5$ are especially important physically. They show that in the open periodically driven system, the dissipator does more than broaden pre-existing Floquet transitions: it also couples them and reorganizes them into collective decay/oscillation modes. In this case, the exact Liouvillian produces resonances centered at the drive harmonics, while the underlying coherent transitions remain split around those harmonics. 

Hence the observed spectral line is a dissipative collective object rather than a simple imprint of one closed-system transition. The relatively small linewidth
\begin{equation*}
\gamma_{141}=\gamma_{150}\approx 0.0241\,\omega_a
\end{equation*}
compared to the pure-mode linewidth
\begin{equation*}
\gamma_{51}=\gamma_{59}\approx 0.0387\,\omega_a
\end{equation*}
also explains why these hybrid peaks are tall and sharp in the spectrum.

\paragraph*{Detuning case with $\omega_d/\omega_a=1.5$.}
For the higher drive frequency, $\omega_d/\omega_a=1.5$, the same structure appears but in a much simpler form. The dominant peaks are centered near
\begin{equation*}
\omega/\omega_a \approx 0.492,\ 1.5,\ 2.508,
\end{equation*}
together with weaker higher-order sidebands. The strongest mode is $\mu=64$ with
\begin{equation*}
\Delta_{64}\approx 2.508\,\omega_a,
\qquad
\gamma_{64}\approx 0.0639\,\omega_a,
\end{equation*}
and it is almost purely the transition $(e,g,2)$ with
\begin{equation*}
r_{64}^{(eg2)}\approx 0.9998,
\qquad
\Delta_{eg2}\approx 2.508\,\omega_a.
\end{equation*}

Similarly, the mode $\mu=65$ is almost purely $(g,e,0)$ and produces the peak at
\begin{equation*}
\Delta_{65}\approx 0.492\,\omega_a,
\end{equation*}
while the negative-frequency counterpart $\mu=42$ is almost purely $(e,g,0)$ at
\begin{equation*}
\Delta_{42}\approx -0.492\,\omega_a.
\end{equation*}
These are therefore conventional pure Floquet sideband peaks.

The central resonance at $\omega/\omega_a=1.5$ is again hybrid. The mode $\mu=140$ has
\begin{equation*}
\Delta_{140}=1.5\,\omega_a,
\qquad
\gamma_{140}\approx 0.06\,\omega_a,
\end{equation*}
but it is composed almost equally of
\begin{equation*}
(g,e,0),\qquad (e,g,2),
\end{equation*}
with weights
\begin{equation}
r_{140}^{(ge0)}\approx r_{140}^{(eg2)}\approx 0.4898.
\end{equation}
The two associated coherent Floquet transitions lie at
\begin{equation*}
\Delta_{ge0}\approx 0.492\,\omega_a,
\qquad
\Delta_{eg2}\approx 2.508\,\omega_a,
\end{equation*}
whose average is precisely the drive frequency,
\begin{equation*}
\frac{\Delta_{ge0}+\Delta_{eg2}}{2}\approx 1.5\,\omega_a.
\end{equation*}

\begin{figure*}[!htpb]
\centering
\includegraphics[width=.75\linewidth]
{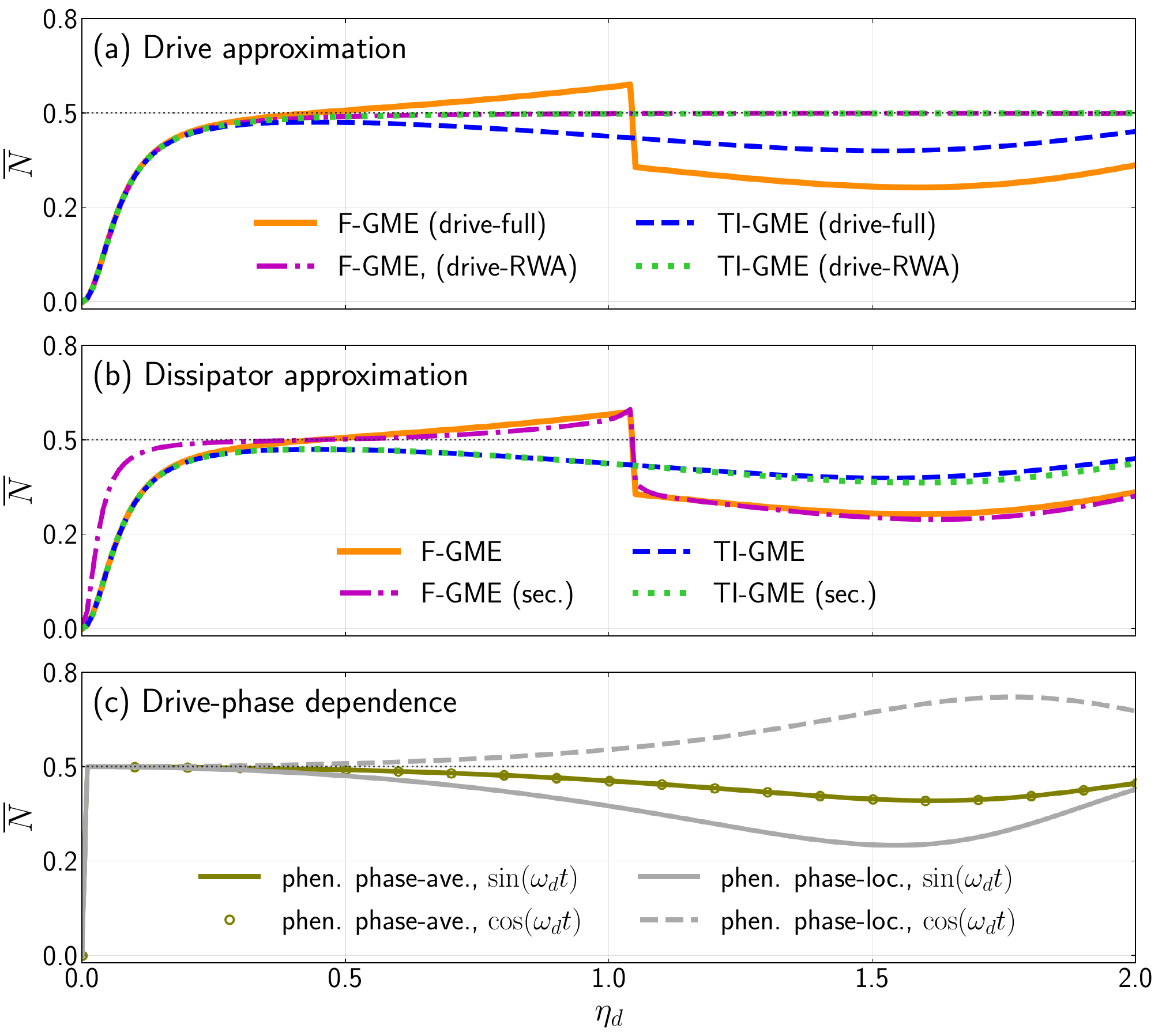} 
\caption[]{\textbf{Driven-dissipative TLS: the effect of drive-RWA and bath-RWA (secular) approximations.}
Long-time averaged excitation number $\overline{N}$ for $\gamma=0.1\omega_a$, comparing the F-GME, TI-GME, and phenomenological Floquet treatment, with drive-RWA (a) and with bath-RWA (secular) (b). 
In both panels, $\omega_d=\omega_a$. The dissipators in the GMEs' results of panel (a) are considered in the nonsecular and full form.
}
\label{figS:TLSRWA_etad}
\end{figure*}

Thus, the resonance at $\omega=\omega_d$ is a hybrid FL mode formed by dissipative coupling of the lower and upper Floquet sidebands. This explains why the spectrum in panel (f) contains a strong central peak exactly at $\omega_d$, even though no single dominant coherent Floquet transition is located there. The same pattern repeats for the weaker higher-frequency modes, such as $\mu=150$ and $\mu=143$, which hybridize the neighboring sidebands around $\omega/\omega_a=\pm 4.5$, but negligible here.

\paragraph*{Comparison between the two driving frequencies.}
Comparing the left and right columns, one sees that the lower-frequency drive $\omega_d/\omega_a=0.5$ generates a richer set of visible resonances, including several strong hybrid peaks at $\omega/\omega_a=0.5$ and $1.5$, as well as sidebands at $\omega/\omega_a\approx 0.195,\ 0.805,\ 1.195,\ 1.805,\dots$. In contrast, for $\omega_d/\omega_a=1.5$ the dominant incoherent spectrum is concentrated into three main peaks at
$\omega/\omega_a \approx 0.492,\ 1.5,\ 2.508$,
with the central peak again being hybrid and the side peaks being nearly pure. Thus, increasing the drive frequency simplifies the mode content of the observable spectrum, but does not eliminate the essential dissipative hybridization mechanism.

Figure~\ref{figS:Spectra_etad1_Flat}, therefore, illustrates two complementary facts. First, many peaks can indeed be interpreted as broadened Floquet sideband transitions, with
\begin{equation*}
\Delta_\mu \approx \Delta_{\alpha\beta l},
\qquad
r_\mu^{\alpha\beta l}\approx 1.
\end{equation*}
Second, and more importantly, periodically driven open systems also support genuinely dissipative hybrid resonances, for which
\begin{equation}
\Delta_\mu \neq \Delta_{\alpha\beta l}
\end{equation}
for any single dominant component, and instead the mode carries comparable weights from multiple Floquet transitions. These hybrid peaks are not artifacts of fitting; they are direct eigenmodes of the FL superoperator. In this sense, the FL decomposition provides a sharper physical interpretation of the spectrum than a purely closed-system Floquet picture, because it identifies not only the coherent transition channels but also how dissipation reorganizes them into the actual radiative resonances seen in the emission spectrum.

\subsection{Driven-dissipative TLS: The effects of the drive- and dissipator-RWA and secular approximation in populations}

We examine the effects of the drive rotating-wave approximation (drive-RWA) in Fig.~\ref{figS:TLSRWA_etad}(a).
We observe that in the drive-RWA limit of both approaches, the cycle-average population deflects from the correct results when the drive becomes sufficiently large, $\eta_d\sim\omega_a$, toward the RWA-saturation limit of 0.5. 
Importantly, the TI-GME and F-GME results become nearly indistinguishable when the drive-RWA is imposed.
For both of the phenomenological approaches (not shown here), both curves immediately after $\eta_d=0^+$ stick to the saturation limit of 0.5. 
This demonstrates that, for the cycle-average population of a driven TLS considered here, the discrepancies between the two theories originate primarily from the Floquet sideband structure generated by the non-RWA drive terms.
 When these counter-rotating processes are removed, the dissipative dynamics become effectively describable in terms of a single dominant transition channel, and the distinction between the time-independent and Floquet dissipative treatments largely disappears.
Indeed, these results, with drive-RWA, when the TLS is highly pumped, are invalid. 

Note we have also checked the effect of the dissipator-RWA for both TI-GME and F-GME for the cycle-averaged population, and the effect is negligible.

Next, we investigate the effect of secularization for both GMEs in the average number of excitations, in Fig.~\ref{figS:TLSRWA_etad}(b).
Expectedly, the effect is negligible for the TI-GME
as the dissipator is governed by a single transition frequency, $\omega_{ge}$, so that there are no distinct dissipative channels available for nonsecular coupling.
However, it affects the result of the F-GME mainly in the lower to moderately strong regime of driving $\eta_d\lesssim\omega_a$.
Secularization therefore removes genuine Floquet-channel interactions and can substantially modify the steady-state populations.

Figure~\ref{figS:TLSRWA_etad}(a,b) generally highlights two distinct physical mechanisms; the differences between TI-GME and F-GME originate primarily from the drive-induced Floquet sidebands generated by the non-RWA terms of the Hamiltonian, whereas the differences between secular and nonsecular F-GME results originate from dissipative couplings between multiple Floquet relaxation channels. Thus, both the coherent non-RWA drive processes and the nonsecular Floquet dissipative couplings play essential and complementary roles in accurately describing strongly driven open quantum systems.

\begin{figure}[thpb]
\centering 
\includegraphics[width=.99\linewidth]
{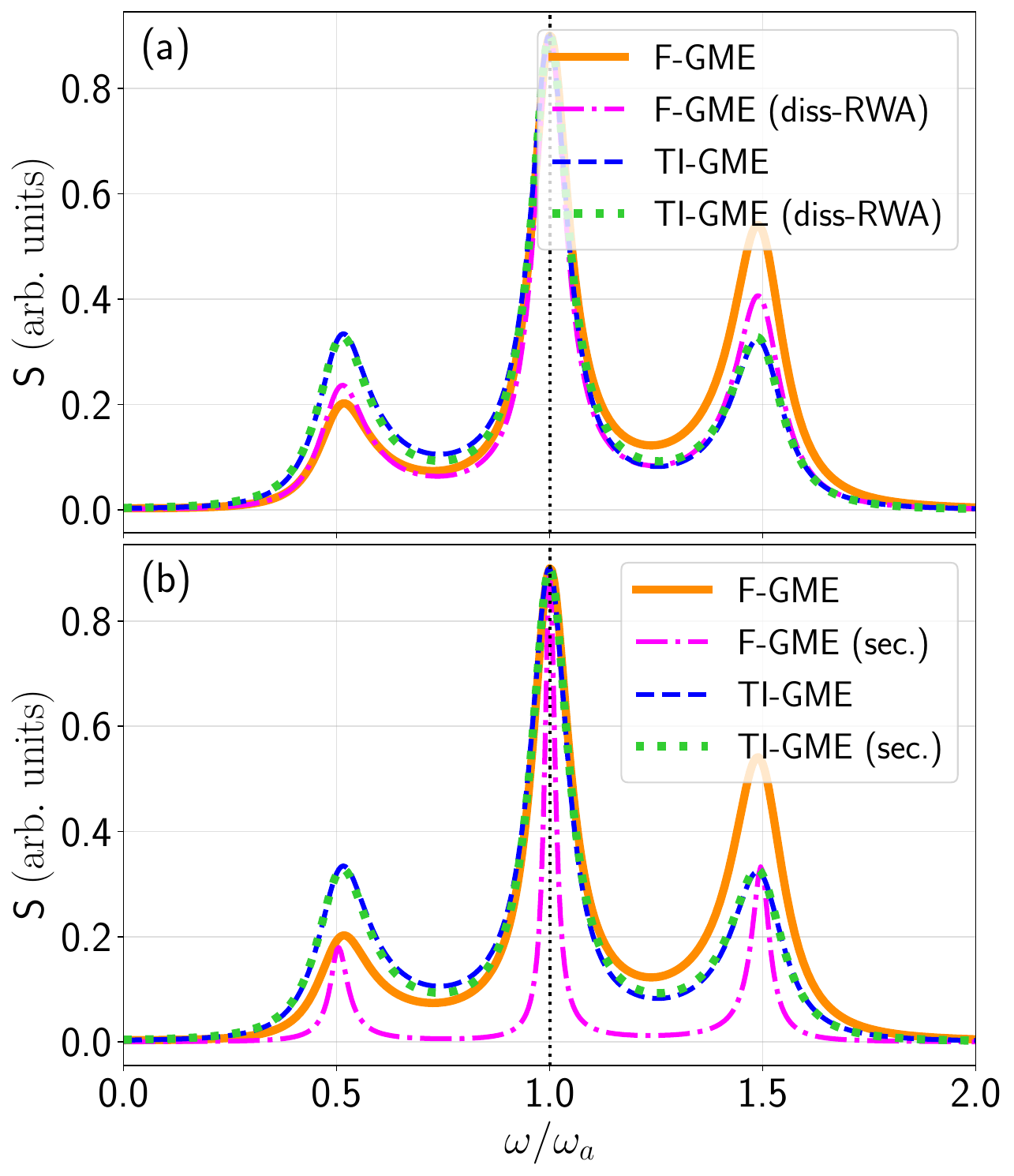} 
\caption[]{\textbf{Driven-dissipative TLS: the effect of dissipator-seculaerization and dissipator-RWA approximation in spectrally resolved results.}
Incoherent emission spectra as in Fig.~\ref{fig:TLSSpectra_wd1_Flat}(c), to check the optional dissipator-related approximations, i.e., the dissipator-RWA investigation (a), where the high frequency terms in the dissipator operator decomposition are discarded; and the dissipator-secularization investigation (b), where only the resonant terms in the dissipator operator decomposition are kept.
In all panels, $\omega_d=\omega_a$, $\eta_d=0.5$ and the TLS is attached to a flat bath with $\gamma=0.1\omega_a$.
}
\label{figS:TLS_etad_DissipatorApproxs}
\end{figure}

As mentioned, the phenomenological Floquet result obtained within the drive-RWA saturates close to $\overline{N}=0.5$ and it happens as quickly as possible, corresponding to the familiar resonantly driven TLS in which the counter-rotating drive processes are neglected. In contrast, the full-drive phenomenological result can exceed $0.5$ at large drive amplitudes due to the additional excitation pathways generated by the counter-rotating terms.  
Panel (c) of Fig.~\ref{figS:TLSRWA_etad} demonstrates the dependence of the phase-locked phenomenological prediction on the relative phase between the initial state and the drive. Changing the drive from sine to cosine time-function, corresponding to $\phi_d:0\to\pi/2$, modifies the phase-locked result because it changes the coherent interference between Floquet sidebands at the initial time. In contrast, the phase-averaged result is invariant under this shift because the sideband phases disappear after phase averaging.

 \subsection{Driven-dissipative TLS: The effects of different approximations in the dissipator in spectra}

Panel (a) of Fig.~\ref{figS:TLS_etad_DissipatorApproxs} showcases the effect of operator decomposition in the dissipator (RWA). The results confirm that this can influence the F-GME prediction more than that of the TI-GME. Expectedly, it does not affect the results from TI-GME, but manily affect the higher-energy portion of the spectrally-resolved results because the faster oscillation portion is dropped in the dissipator-RWA approximation.
Next, we present, in Fig.~\ref{figS:TLS_etad_DissipatorApproxs}(b), the effect of secular approximation in the dissipator for both GMEs. Expectedly again, for the flat bath single-channel TI-GME, secularization is not effective, while it dreadfully decreases the precision of the F-GME predictions as the dissipative drive-assisted-ladder channels are multiples. This effect can also be more alarming as the drive becomes stronger, particularly entering into the deep-strong drive, which can also destroy the complete positivity of the master equation.

\clearpage
\newpage
\bibliography{main}

\end{document}